\documentclass[a4paper, aps,prd,reprint,twocolumn,superscriptaddress,showpacs,nofootinbib,floatfix,longbibliography,preprintnumbers]{revtex4-2}
\usepackage[a4paper,top=2.25cm,bottom=2.25cm,left=2.25cm,right=2.25cm]{geometry}

\usepackage{amsfonts,mathtools,bm,braket,mathrsfs}
\usepackage{siunitx}
\DeclareSIUnit\parsec{pc}

\usepackage{graphicx}
\graphicspath{{./plot/}}

\usepackage{multirow,booktabs,arydshln}

\usepackage[normalem]{ulem}         
\usepackage[final]{microtype}       
\usepackage[dvipsnames]{xcolor}
\definecolor{refblue1}{rgb}{0.0, 0.19, 0.45}
\definecolor{refblue2}{rgb}{0.0, 0.32, 0.65}
\usepackage[unicode]{hyperref}
\hypersetup{
  colorlinks=true,
  citecolor=refblue2,
  linkcolor=red,
  urlcolor=refblue2
}

\usepackage{calligra}

\usepackage{orcidlink}

\DeclareMathAlphabet{\mathcalligra}{T1}{calligra}{m}{n}
\DeclareFontShape{T1}{calligra}{m}{n}{<->s*[2.0]callig15}{}

\newcommand\inp[2]{\langle #1 \,|\, #2 \rangle}

\DeclareMathAlphabet{\mathpzc}{OT1}{pzc}{m}{it}
	
\definecolor{LightCyan}{rgb}{0.88,1,1}
\definecolor{lightgray}{gray}{0.9}

\def \match     {\mathscr{M} }
\def \Hz        {\,\mathrm{Hz} }
\def \intrinsic {\vec{\Lambda}_\mathrm{int}}
\def \extrinsic {\vec{\Lambda}_\mathrm{ext}}
\def \dL        {d_{\rm L}}
\def \tc        {t_{\rm c}}

\def \msun      {\rm{M}_\odot}
\def \mchirp    {\mathcal{M}}

\def \imrpxhm     {\mathtt{IMRPhenomXHM}}
\def \imrphm     {\mathtt{IMRPhenomHM}}
\def \aLIGOZDHP     {\mathtt{aLIGOZeroDetHighPower}}

\def \flow      {f_{\rm low} }

\def \fhigh     {f_{\rm high}}
\def \Mpc       {\mathrm{Mpc}}
\def \mchirp    {\mathcal{M}}

\def \dynesty {\mathtt{dynesty}}

\def \IITGn     {Department of Physics, Indian Institute of Technology Gandhinagar, Gujarat 382055, India.\vspace*{3pt}}

\def \TIFR    {\mbox{Department of Astronomy \& Astrophysics, Tata Institute of Fundamental Research}, 1, Homi Bhabha Road, Mumbai- 400005, Maharashtra, India.\vspace*{3pt}}

\newcommand{\UCLouvain}{Centre for Cosmology, Particle Physics and Phenomenology - CP3, Universit\'{e} Catholique de Louvain, \\ Louvain-La-Neuve, B-1348, Belgium}

\newcommand{\ROB}{Royal Observatory of Belgium, Avenue Circulaire, 3, 1180 Uccle, Belgium}

\begin{document}

\title{Rapid parameter estimation with the full symphony of compact binary mergers \mbox{using meshfree approximation}}

\author{\textsc{Abhishek~Sharma}\orcidlink{0009-0007-2194-8633}}
\email{sharma.abhishek@iitgn.ac.in }
\affiliation{\IITGn}

\author{\textsc{Lalit~Pathak}\orcidlink{https://orcid.org/0000-0002-9523-7945}}
\email{lalit.pathak@tifr.res.in}
\affiliation{\TIFR}

\author{\textsc{Soumen~Roy}\orcidlink{0000-0003-2147-5411}}
\email{soumen.roy@uclouvain.be}
\affiliation{\UCLouvain}
\affiliation{\ROB}

\author{\textsc{Anand~S.~Sengupta}\orcidlink{0000-0002-3212-0475}
\vspace*{3pt}} 
\email{asengupta@iitgn.ac.in}
\affiliation{\IITGn}

\begin{abstract}
We present a fast Bayesian inference framework to address the growing computational cost of gravitational-wave parameter estimation. The increased cost is driven by improved broadband detector sensitivity, particularly at low frequencies due to advances in detector commissioning, resulting in longer in-band signals and a higher detection rate. 
Waveform models now incorporate features like higher-order modes, further increasing the complexity of standard inference methods.
Our framework employs meshfree likelihood interpolation with \mbox{radial basis functions} to accelerate Bayesian inference using the IMRPhenomXHM waveform model that incorporates higher modes of the gravitational wave signal. 
In the initial start-up stage, interpolation nodes are placed within a constant-match metric ellipsoid in the intrinsic parameter space. 
During sampling, likelihood is evaluated directly using the precomputed interpolants, bypassing the costly steps of on-the-fly waveform generation and overlap-integral computation.
We improve efficiency by sampling in a rotated parameter space aligned with the eigenbasis of the metric ellipsoid, where parameters are uncorrelated by construction. This speeds up sampler convergence.
This method yields unbiased parameter recovery when applied to 100 simulated neutron-star–black-hole signals (NSBH) in LIGO–Virgo data, while reducing computational cost by up to an order of magnitude for the longest-duration signal.
The meshfree framework equally applies to symmetric compact binary systems dominated by the quadrupole mode, supporting parameter estimation across a broad range of sources.
Applied to a simulated NSBH signal in Einstein Telescope data where the effects of Earth’s rotation are neglected for simplicity, our method achieves an $\mathcal{O}(10^4)$ speed-up, demonstrating its potential use in the 3G era.
\end{abstract}


\maketitle

\section{Introduction}
\label{sec:intro}
The direct detection of gravitational waves (GWs) from compact binary coalescences (CBCs) by the LIGO–Virgo–KAGRA (LVK) collaboration~\cite{LIGOScientific:2016aoc, Abbott_2019, Abbott_2021, gwtc_2.1, gwtc_3, LIGOScientific:2014pky,VIRGO:2014yos,Akutsu2021_KAGRA_overview} has opened a new observational window onto the universe. These detections, comprising mergers of binary black holes (BBHs), binary neutron stars (BNSs), and neutron star–black hole (NSBH) systems, have enabled unprecedented studies of compact objects and the environments in which they form and evolve.

By the end of the third observing run (O3), the LVK network had detected nearly 100 CBC events~\cite{Abbott_2021,gwtc_2.1,gwtc_3,Mehta:2023zlk}, providing a statistically significant sample for population studies, precision tests of general relativity, and constraints on the neutron star equation of state~\cite{TGR_2021, Abbott2023GWTC3pop, PhysRevLett.119.161101}. The fourth observing run (O4) is currently ongoing and is expected to substantially increase the number of detections, as well as improve the sensitivity to a wider range of source parameters, including lower-mass binaries and more distant systems~\cite{LVK2024GW230529,LVK2025GW231123}.

The expanding gravitational wave catalog offers critical insights into the mass and spin distributions of compact objects, the rate of binary mergers across cosmic time~\cite{Abbott2016RateBBH, Abbott2023GWTC3pop, gwtc_3, Abbott2021PopPropsGWTC2, Fairhurst2023MassParam}, and potential signatures of beyond-standard-model physics~\cite{LIGO2021CosmicStrings,Abbott2022DarkPhoton,Aurrekoetxea2024CosmicStringGW190521}. As the number of detected events grows, the development of efficient and reliable methods for parameter estimation, model selection, and population inference becomes essential for fully exploiting the scientific potential of these observations.

The improved sensitivity of ground-based interferometers is expected to increase the detection rate in upcoming observing runs. 
The inference of source parameters through full Bayesian inference is computationally expensive, often requiring several hours to days of CPU runtime per event~\cite{Christensen2022RMP_PE1}. This presents a significant bottleneck, particularly in the context of limited computational resources available to the community. However, recent advances such as relative binning, reduced-order quadrature, and machine-learned posterior estimation, have enabled accurate inference in minutes or even seconds for many events~\cite{Krishna:2023bug,Morisaki:2023kuq,Dax2025RealTimeBNS,Green_2021,Mushkin2025dotPE}. Accelerated inference methods that retain accuracy while reducing computational burden are therefore essential to fully capitalize on the scientific potential of current and future detectors.

Beyond individual event characterization, fast Bayesian inference also plays a pivotal role in enabling hierarchical population analyses~\cite{Mandel_2019}, which require re-sampling or re-weighting posterior distributions from large ensembles of events. Additionally, rapid estimation of source parameters facilitates timely electromagnetic follow-up in multi-messenger astrophysics, supports real-time decision-making in low-latency detection pipelines, and allows more extensive inclusion of waveform complexities such as higher-order multipoles. Furthermore, by lowering the computational barriers, fast inference methods democratize access to advanced analysis tools across the community. 

Bayesian parameter estimation involves sampling the posterior distribution which inturn requires computation of the expensive likelihood function $\mathcal{O}(10^6)$ times in a typical parameter estimation (PE) run~\cite{Veitch2015LALInference}. The brute-force evaluation of the likelihood function is computationally expensive due to two main steps: waveform generation and subsequent overlap integral computation.
Several techniques have been proposed to reduce the computational cost of likelihood evaluations in GW parameter estimation. 

One such method is the heterodyned likelihood or relative binning~\cite{Cornish:2010kf,Zackay:2018qdy,Cornish:2021lje,Leslie:2021ssu,Krishna:2023bug,Narola:2023men} which speeds up likelihood evaluation by using a reference waveform to approximate the ratio of nearby waveforms as a smooth, piece-wise linear function within frequency bins, allowing efficient reuse of precomputed overlap integrals.
Another widely used approach is Reduced Order Quadratures (ROQ)~\cite{Canizares:2014fya,Smith:2016qas,Morisaki:2020oqk,Morisaki:2023kuq,Smith:2021bqc}, in which waveforms are expressed as a linear combination of a precomputed basis. This basis needs only to be evaluated at a small set of carefully chosen frequency points known as empirical interpolation nodes  that are identified using a greedy algorithm. The overlap integral is then computed at these reduced frequency points. 

Another technique is multibanding, which exploits the chirping nature of compact binary signals to sample the waveform sparsely in frequency regions where high resolution is unnecessary~\cite{Vinciguerra:2017ngf,Morisaki:2021ngj}. Other approaches involve approximating the likelihood function itself using Gaussian process interpolation~\cite{Pankow:2015cra,Lange:2018pyp,Wagner:2025bih}, providing a smooth and fast surrogate for the true likelihood. In addition, several methods based on likelihood marginalization have been developed to integrate out nuisance parameters analytically or semi-analytically~\cite{Veitch2013_T1300326,Farr2014_T1400460, Singer:2015ema,Islam:2022afg,Roulet:2024hwz,Thrane2019PASA}. 
Another method leverages the intuitive understanding of how various binary parameters affect the observed GW signal to construct computationally efficient PE algorithm~\cite{Fairhurst:2023idl}. To further accelerate inference, numerous advanced sampling algorithms have been proposed that improve the efficiency of posterior sampling~\cite{Williams:2021qyt,Karamanis:2022ksp, Wong:2023lgb, Tiwari:2023mzf, Williams:2023ppp, Tiwari:2024qzr, Nitz:2024nhj, Wouters:2024oxj,Williams:2025szm, Vretinaris:2025wdu}. Finally, several likelihood-free inference methods have emerged. These approaches use machine learning models, trained on simulated data, to approximate the likelihood or posterior distributions in real-time~\cite{Gabbard:2019rde,Green:2020hst,Chua:2019wwt,Dax:2021tsq,Dax:2022pxd,Bhardwaj:2023xph,Kolmus:2024scm,Dax:2024mcn}.

Our previous works introduced efficient methods for fast likelihood evaluation utilizing interpolation with radial basis functions (RBFs) \cite{Pathak:2022iar, Pathak:2023ixb}, demonstrating significant computational acceleration while maintaining high accuracy in parameter estimation. However, previous works were limited to the quadrupole-only waveform models and were impeded by narrow prior range over the intrinsic parameters. In this work, we present an extension of our rapid-PE framework to incorporate higher multipoles of GW radiation and also mitigate the limitation of narrow priors by incorporating metric guided node placement strategy in the intrinsic parameter space.

Instead of using the chirp mass, mass ratio, and component spins, to represent the intrinsic parameter space, we demonstrate the use of completely uncorrelated parameters composed of the eigen-directions of the metric (represented later as $\Delta e$ coordinates) as sampling coordinates. Sampling in these coordinates results in faster convergence of the sampler with fewer likelihood calls compared to the conventional set of sampling coordinates. We perform a percentile-percentile test by performing analysis over 104 simulated NSBH systems and demonstrate that our method produces unbiased estimates of the source parameters. We also reanalyzed asymmetric real event \texttt{GW190814}~\cite{LIGOScientific:2020zkf} observed during the third observing run (O3) and show that the posteriors obtained are consistent with those obtained by the LVK analysis.

Although the dominant GW emission from a CBC system is quadrupolar in nature, the higher order multipoles contribute significantly for systems with  asymmetric component masses and orbits that are inclined with respect to the line of sight. 
Neglecting higher modes can lead to biased inference of source properties~\cite{Varma:2014jxa, PhysRevD.87.104003, CalderonBustillo:2015lrt, Chatziioannou:2019dsz, Kalaghatgi2020IMRPhenomHM, LIGOScientific:2020stg, LIGOScientific:2020zkf}. On the other hand, incorporating them in the signal model could help break degeneracies in the parameter space such as those between orbital inclination and luminosity distance, or between mass ratio and spins of black holes as shown in several studies~\cite{Usman:2018imj, Hannam:2013uu, Ohme:2013nsa}.

The rest of the paper is organized as follows. Section~\ref{sec:likelihood} consists of four subsections. Subsection~\ref{subsec:bayesian_inference} introduces the basics of Bayesian inference, followed by the decomposition of the likelihood function in terms of spherical harmonic multipoles in Subsection~\ref{subsec:likelihood_function}. Subsection~\ref{subsec:start_up} describes the start-up stage of our meshfree interpolation strategy, which is further divided into two parts: the metric-based node placement method in Subsection~\ref{subsubsec:node_placement}, and the construction of interpolants using singular value decomposition (SVD) and RBFs in Subsection~\ref{subsec:svd_interpolants_gen}. The online stage, where the interpolated likelihood is evaluated during the sampling, is
described in Subsection~\ref{subsec:online_stage}. In Section~\ref{sec:results}, we demonstrate the application of the meshfree interpolation method for parameter estimation of both simulated signals (Subsection~\ref{subsec:sim_results}) and previously detected GW events (Subsection~\ref{subsec:lvk_gw_events}). Finally, Section~\ref{sec:conclusion} summarizes our work and outlines future directions.

\section{Meshfree likelihood interpolation}
\label{sec:likelihood}
\subsection{Bayesian inference}
\label{subsec:bayesian_inference}

A gravitational-wave signal is characterized by parameters \( \vec{\Lambda} \) that lie in a product space,
\[
\vec{\Lambda} \in \intrinsic \times \extrinsic,
\]
where \( \intrinsic \) denotes the space of intrinsic parameters that govern the dynamics and phase evolution of the waveform (such as the component masses and spins), and \( \extrinsic \) denotes the space of extrinsic parameters, including the source’s sky location (right ascension \( \alpha \), declination \( \delta \)), polarization angle \( \psi \), inclination angle \( \iota \), coalescence phase \( \varphi \), geocentric coalescence time \( \tc \), and luminosity distance \( \dL \). These extrinsic parameters affect the amplitude, arrival time, and modulation of the signal in the detectors.

Given strain data \( d(t) \) recorded by a network of detectors, containing a signal \( h(\vec{\Lambda}_\mathrm{true}; t) \) from a compact binary coalescence and additive noise \( n(t) \), the goal is to infer the posterior distribution \( p(\vec{\Lambda} \mid d) \) over the source parameters. The posterior is related to the likelihood \( p(d \mid \vec{\Lambda}) \) and the prior \( p(\vec{\Lambda}) \) via Bayes’ theorem:
\begin{equation}
    p(\vec{\Lambda} \mid d) = \frac{p(d \mid \vec\Lambda)\, p(\vec \Lambda)}{p(d)},
    \label{eq:bayes_theorem}
\end{equation}
where \( p(d) \) is the evidence, which acts as a normalization constant. For aligned-spin systems, the parameter space is typically 11-dimensional, making direct evaluation of the posterior intractable. Stochastic sampling algorithms such as Markov chain Monte Carlo (MCMC)~\cite{Foreman_Mackey_2013} and nested sampling~\cite{skilling2006nested} are therefore employed to explore \( p(\vec{\Lambda} \mid d) \).

\subsection{Likelihood function}
\label{subsec:likelihood_function}
For gravitational-wave data $d^{(k)}(t) = h^{(k)}(\vec{\Lambda}_\mathrm{true}; t) + n^{(k)}(t)$ recorded at the $k^\text{th}$ detector, under the assumption of Gaussian and stationary noise $n^{(k)}(t)$, the log-likelihood ratio of the signal hypothesis to the null hypothesis is given by
\begin{equation}
    \ln \mathcal{L}(\vec \Lambda) = \sum_{k=1}^{N_{\text{d}}} \langle h^{(k)}(\vec \Lambda) \mid\boldsymbol{d}^{(k)}\rangle - \frac{1}{2} \sum_{k=1}^{N_{\text{d}}} ||h^{(k)}(\vec \Lambda)||^2,
    \label{eq:logl_basic}
\end{equation}
where $N_d$ is the number of detectors and $\inp{a}{b}$ is the noise-weighted inner product of two time domain signals $a(t)$ and $b(t)$, 
\begin{equation}
    \displaystyle \inp{a}{b} = 4 \:\Re\int_{\flow}^{\fhigh} \frac{\tilde{a}^{\ast}(f) \: \tilde{b}(f)}{S_n(f)}\: df,
    \label{eq:innerProduct}
\end{equation}
and $||a||^2 = \inp{a}{a}$. The quantity $S_n(f)$ is the one-sided noise power spectral density and $\tilde{a}(f)$ denotes the Fourier transform of $a(t)$. 

The gravitational wave signal $h(t)$, traveling in an arbitrary direction $(\iota, \varphi)$ in the source frame, can be expressed in terms of the spin-weighted spherical harmonics with spin weight -2:

\begin{equation}
    h(t;\iota, \varphi) \equiv h_+ - ih_\times= \sum_{\ell=2}^{\infty}  \sum_{m = -\ell}^{\ell} {}_{-2}Y_{\ell m}(\iota, \varphi) \, h_{\ell m}.
    \label{eq:h_spherical_harmonic_basis}
\end{equation}
In the following expressions we will drop the symbol $-2$ from ${}_{-2}Y_{\ell m}$. The $\imrpxhm$ supports the $(\ell, |m|) = (2, 2), (2, 1), (3, 2), (3, 3)$ and $(4, 4)$ modes. Following the convention in~\cite{Garcia-Quiros:2020qpx}, the polarizations $\tilde{h}_+(f)$ and $\tilde{h}_\times(f)$ in frequency domain in terms of positive and negative modes can be written as,
\begin{align}
\label{eq:hplus_hcross_decomposed1}
\tilde{h}_{+}(f) &= \frac{1}{2} \sum_{\ell, m} \left( Y_{\ell - m} + (-1)^{\ell} Y_{\ell m}^{*} \right) \tilde{h}_{\ell - m}(f), \\
\tilde{h}_{\times}(f) &= \frac{i}{2} \sum_{\ell, m} \left( Y_{\ell - m} - (-1)^{\ell} Y_{\ell m}^{*} \right) \tilde{h}_{\ell - m}(f).
\label{eq:hplus_hcross_decomposed2}
\end{align}
Note that in these expressions, both the polarizations are expressed only in terms of the negative spherical harmonic modes, $\tilde{h}_{\ell - m}$. And, in the above summations, $Y_{\ell - m}$ and $Y_{\ell m}$ contributes only for the negative and positive modes, respectively. Using such a decomposition, the log-likelihood function given by Eq.~\eqref{eq:logl_basic} can be expressed in terms of spherical harmonic modes. 

The projected GW signal at the $k^\text{th}$ detector is given by,
\begin{align}
\nonumber
    h^{(k)}(\vec\Lambda; t) =& F^{(k)}_+(\alpha, \delta, \psi, t)\; h_+(\intrinsic;t) \\
    &+ F^{(k)}_\times(\alpha, \delta, \psi, t)\; h_\times(\intrinsic;t),
    \label{eq:projected_h_td}
\end{align}
where, $F^{(k)}_+$ and $F^{(k)}_\times$ are antenna pattern functions that depend on the detector's geometry, position of the source in the sky and polarization angle. 
The time dependence in antenna patterns arises from changing orientation of the detector arms as the signal accumulates over time. 
For the systems and detectors considered in this analysis, the gravitational-wave signals remain in band for a much shorter duration than the timescale of Earth's rotation. This allows us to treat the antenna patterns as effectively constant (frozen in time) during the signal and evaluate them at the geocentric coalescence time.
Note that this assumption would not be applicable for third-generation (3G) detectors such as CE and ET, where the signal remains in band for about an hour. 
For example, for a NSBH system with component masses $m_1 = 5.31\,\msun$, $m_2=1.22\,\msun$, and spins, $\chi_{1z} = -0.69$, $\chi_{2z}=-0.02$, the in-band duration in second generation detectors assuming a lower cut-off frequency of $15\Hz$ is $\sim 150$ seconds. The in-band duration of the same signal will be $\sim 50$ minutes in ET, assuming a lower cut-off frequency of $5\Hz$. 

Assuming frozen antenna patterns, the projected GW signal in frequency domain is given by
\begin{align}
    \label{eq:projected_h_fd}
    \nonumber
    \tilde{h}^{(k)}(f) = &[F^{(k)}_+(\alpha, \delta, \psi)\; \tilde{h}_+(f) \\
    &+ F^{(k)}_\times(\alpha, \delta, \psi)\; \tilde{h}_\times(f)]e^{-2\pi i f \Delta t^{(k)}},
\end{align}
where $\Delta t^{(k)}$ is the time taken by the GW signal to travel from geocenter to the $k^\text{th}$ detector's location. On substituting Eq.~\eqref{eq:projected_h_fd} into Eq.~\eqref{eq:logl_basic} and using Eq.~\eqref{eq:hplus_hcross_decomposed1} and Eq.~\eqref{eq:hplus_hcross_decomposed2}, the log-likelihood function given by Eq.~\eqref{eq:logl_basic} takes the following form,
\begin{widetext}
    \begin{align}
\ln \mathcal{L}(\vec{\Lambda}) = \sum_{k=1}^{N_d} \sum_{\ell,m} &\frac{1}{2} \left\{ (-1)^\ell Y_{\ell m} (F^{(k)}_{+} + i F^{(k)}_{\times}) + Y_{\ell m}^{*} (F^{(k)}_{+} - i F^{(k)}_{\times}) \right\} \langle e^{2\pi i f \Delta t^{(k)}} h_{\ell -m} | d^{(k)} \rangle \nonumber \\
&\quad + \sum_{\ell,m,\ell',m'} \frac{1}{4}\Big[ \Big\{ (-1)^\ell (-1)^{\ell'} Y_{\ell m} Y_{\ell' m'}^{*} + Y_{\ell -m}^{*} Y_{\ell' -m'} \Big\} \Big\{\left(F_{+}^{(k)}\right)^2 + \left(F_{\times}^{(k)}\right)^2\Big\} \nonumber \\
&\quad + \Big\{ (-1)^\ell Y_{\ell m} Y_{\ell' -m'} + (-1)^{\ell'} Y_{\ell -m}^{*} Y_{\ell' m'}^{*} \Big\} \Big\{\left(F_{+}^{(k)}\right)^2 - \left(F_{\times}^{(k)}\right)^2\Big\} \nonumber \\
&\quad + 2i F^{(k)}_{+} F^{(k)}_{\times} \Big\{ (-1)^\ell Y_{\ell m} Y_{\ell' -m'} - (-1)^{\ell'} Y_{\ell -m}^{*} Y_{\ell' m'}^{*} \Big\} \Big]\inp{h_{\ell -m}}{h_{\ell' -m'}},
\label{eq:logl_complete}
\end{align}
\end{widetext}
where we have assumed antenna pattern functions to be real, which is a valid assumption when the GW wavelength is larger than the detector arms so that the transfer functions are unity (under the long-wavelength limit)~\cite{Rakhmanov:2008is}, which is actually the case for the current generation GW observatories. For time or frequency dependent antenna pattern functions, strategy described in Ref.~\cite{Sharma:2024sfb} can be adopted.

The computation of the log-likelihood function requires waveform generation followed by the evaluation of the overlap integrals.
These computations increase the overall runtime of the parameter estimation analysis even with efficient sampling algorithms. Thus, our primary aim is to mitigate the cost of computing the log-likelihood by bypassing these computations and directly evaluating the inner-products using various interpolants generated during the start-up stage as explained in next subsection.

The purpose of writing the log-likelhood function in this form is to factor out the dependence of all extrinsic parameters from the inner-products $\inp{\cdot}{\cdot}$, so that the inner-products only depend on the intrinsic parameters. However, there is a time-shift factor that depends on the extrinsic parameters inside the inner-product in first term of Eq.~\eqref{eq:logl_complete}, this can be easily dealt with as described in the later subsection. This factorization of the likelihood function in terms of extrinsic and intrinsic parameters is the crucial element that our method leverages to accelerate its computation. The intrinsic parameters dependent part, namely, the overlap integrals are the most expensive quantities to evaluate in the log-likelihood expression. Therefore, a meshfree interpolation is carried out only in the intrinsic parameter space, $\intrinsic$ to directly evaluate them during the sampling stage. On the other hand, the quantities that depend on extrinsic parameters are computationally inexpensive to evaluate and appear as overall amplitude factors in the inner products. The stage in which the interpolants are constructed is referred to as the {\it{start-up}} stage, while the subsequent stage, where the interpolants are utilized to evaluate the likelihood in real-time, is called the {\it{online}} stage. 

We now describe the these two stages in detail in the next subsection.
\begin{figure*}[!hbt]
    \centering
    \includegraphics[width=1\linewidth]{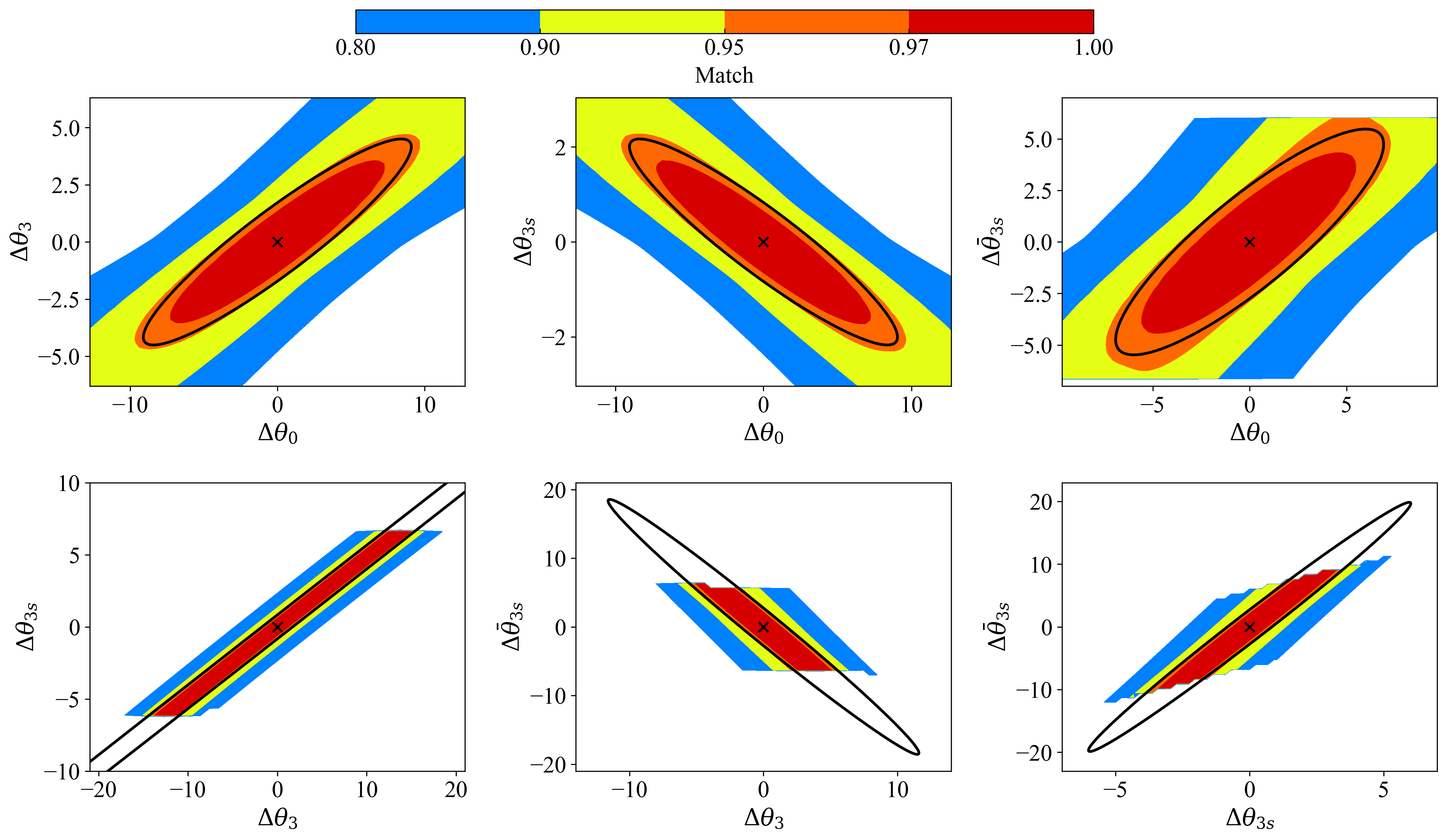}
    \caption{Comparison between the match and the metric approximation of the match across all possible combinations of 2-dimensional orthogonal slices over $\{\theta_0, \theta_3, \theta_{3s}, \bar{\theta}_{3s}\}$ of the 4-dimensional ellipsoid constructed using the semi-analytic metric evaluated at the injection. Note that in each plot, the remaining two coordinates are fixed to the value corresponding to the reference point where metric is evaluated. The black contour corresponds to $\mathscr{M} = 0.95$ ellipse as obtained from the metric while the filled colors show the exact match. The reference point at which metric is computed corresponds to the following parameters: $m_1 = 15 \msun, \; m_2 = 3 \msun, \; \chi_{1z} = 0.5$, and $\chi_{2z} = 0.05$.}
    \label{fig:2d_metric_slices}
\end{figure*}

\subsection{Start-up stage: Interpolating the likelihood}
\label{subsec:start_up}
By factorizing the log-likelihood function into parts dependent on extrinsic and intrinsic parameters, we isolate the computationally expensive component namely, overlap integrals—which depend only on the intrinsic parameters. This reduces the dimensionality of the space over which expensive computations are required. We directly evaluate these overlap integrals using an interpolation scheme in the intrinsic parameter space, thereby bypassing the costly step of waveform generation during the online stage.

The specific inner-products for which interpolants are constructed are:
\begin{equation}
    \left\langle e^{2\pi i f \Delta t^{(k)}} h_{\ell -m} \middle| d^{(k)} \right\rangle = z^{(k)}_{\ell m},
\end{equation}
and
\begin{equation}
    \left\langle h_{\ell -m} \middle| h_{\ell' -m'} \right\rangle = \sigma_{\ell m \ell' m'}.
\end{equation}

Due to the unknown time-shift factor in $z^{(k)}_{\ell m}$, it is evaluated at discrete values of circular time-shifts uniformly spaced over a specified range centered around the trigger time. As a result, $z^{(k)}_{\ell m}$ becomes a vector quantity, denoted as $\vec{z}^{(k)}_{\ell m}$. Since directly interpolating vector-valued quantities is challenging, we first express $\vec{z}^{(k)}_{\ell m}$ in a suitable basis and then construct interpolants for the corresponding basis coefficients.

The region in intrinsic parameter space where the interpolants are to be generated is guided by the best-matched template point as obtained from an upstream search pipeline~\cite{GstLAL_2010, usman2016pycbc} and the points where these interpolants are generated in this region are called {\it{nodes}}. In our previous work, we used a 4-dimensional hyper-rectangle centered around the best-matched template  to identify the region where we place the nodes. 
The placement of interpolation nodes within the intrinsic parameter space significantly influences the quality of the resulting basis vectors and, in turn, the faithfulness of the interpolant. Using a hyper-rectangular boundary yields accurate interpolants only when the region is very narrow. This effectively imposes tight priors on the intrinsic parameters during the online stage.
A key limitation of rectangular boundaries is that they do not account for correlations among parameters. As a result, many interpolation nodes sampled uniformly within the rectangle fall well outside the region of high likelihood, landing instead in noise-dominated areas. This issue is addressed by a new node placement strategy described in the next subsection.

\subsubsection{Metric guided placement of nodes}
\label{subsubsec:node_placement}
The goal is to place interpolation nodes in the intrinsic parameter space so that they capture the structure of the likelihood near its peak. Nodes located far from this high-likelihood region contribute little to interpolation accuracy. Our node placement strategy is therefore tailored to enable accurate likelihood interpolation suitable for full Bayesian inference.

The match between two normalized waveforms $a$ and $b$ measures their closeness and is defined as inner product between $a$ and $b$ maximized over an overall amplitude, phase ($\varphi_{\rm ref}$), and time ($t_{\rm ref}$),
\begin{equation}
\match(a, b) = \max_{t_{\rm ref}, \varphi_{\rm ref}} \: \inp{\hat{a} }{\hat{b} } 
\label{eq:Match}
\end{equation}
where $\hat{a}$ is the normalized waveform such that  $\hat{a} = a/\sqrt{\inp{a}{a}}$. The match function gives an approximate measure of the distribution of the likelihood function in terms of intrinsic parameters. The {\it{mismatch}}, ($1 - \mathscr{M}$), between two neighboring points in the parameter space can be approximated as the metric distance~\cite{Owen:1995tm}, i.e,
\begin{equation}
1 - \match \simeq \sum_{ij}  g_{ij} \, \Delta \Lambda_{\rm{int}}^i \, \Delta \Lambda_{\rm{int}}^j, 
\label{eq:MisMatch}
\end{equation}
where, $\Delta \intrinsic$ is the separation between the two neighboring points. It is a general equation of an hyper-ellipsoid in $\intrinsic$ space. Therefore, a constant match ellipsoid region centered around a point guided by the best-matched template is the most suitable region to place the interpolation nodes. 

Because of the discreteness of the template bank, the choice of best-matched template itself as the ellipsoid center is sub-optimal. Therefore, we find the optimal centre by performing a gradient descent search and minimizing the negative of the network signal-to-noise ratio (SNR). 
We use the $\texttt{IMRPhenomXHM}$ waveform approximant to compute the SNR during the optimization process, varying not only the intrinsic parameters but also over extrinsic parameters such as the inclination angle ($\iota$) and coalescence phase ($\varphi$). The initial guess for the optimization routine is taken to be the best-matched template for the intrinsic parameters from an upstream search pipeline, with both the inclination angle and coalescence phase initialized to zero.

This SNR optimization step resembles procedures used in simple-PE~\cite{Fairhurst2023simplePE}, which maximizes likelihood over a reduced parameter space to guide fast posterior sampling, and online search pipelines like PyCBC Live~\cite{DalCanton2021_PyCBCLive}, which apply differential evolution to refine the coarse initial result from the template bank by optimizing the network SNR. In contrast, we use the  location of the optimized network SNR at which the parameter space metric is constructed.

In approximating the mismatch as metric distance, a suitable coordinate system is crucial where the metric components vary slowly across the parameter space. Therefore, we construct metric in our 4-dimensional intrinsic parameter space using the following coordinates,
\begin{align}
    \label{eq:metric_coords}
    \nonumber
    \theta_0 &= \frac{5}{128 \eta} (\pi M f_0)^{-5/3} \\ \nonumber
    \theta_3 &= \frac{\pi}{4\eta} (\pi M f_0)^{-2/3} \\ \nonumber
    \theta_{3s} &= \frac{113 \; \chi_{\rm{r}}}{192 \eta} (\pi M f_0)^{-2/3} \\
    \bar{\theta}_{3s} &=  \frac{113 \; \chi_{\rm{eff}}}{192 \eta} (\pi M f_0)^{-2/3}
\end{align}
where, $f_0$ is the lower-cut-off frequency, $M$ is the total mass of the binary, $\eta$ is the symmetric mass ratio, ${\chi_r = \chi_s + \delta \chi_a - 76\eta\chi_s/113}$ is the reduced-spin, where ${\chi_s = (\chi_{1z} + \chi_{2z})/2}$ and ${\chi_a = (\chi_{1z} - \chi_{2z})/2}$ are the symmetric and antisymmetric dimensionless spin combinations, respectively, ${\delta = (m_1 - m_2)/M}$ is the asymmetric mass ratio and ${\chi_{\rm{eff}} = (m_1 \chi_{1z} + m_2 \chi_{2z}) / M}$ is the effective spin of the binary. $\theta_0, \theta_3, \theta_{3s}$ are the usual dimensionless chirptime coordinates known to be the most suitable coordinate system for the template placement in 3-dimensions where metric is slowly varying~\cite{Ajith:2012mn, Roy:2017qgg, Roy:2017oul}. 

For the 4-dimensional intrinsic parameter space relevant to our waveform model, we define a new coordinate \(\bar{\theta}_{3s}\) by replacing \(\chi_{\rm r}\) with \(\chi_{\rm eff}\) in \(\theta_{3s}\). 

We denote these new set of coordinates parametrizing $\intrinsic$ by $\vec{\Theta}\equiv(\theta_0, \theta_{3}, \theta_{3s}, \bar\theta_{3s})$.
The metric in the chosen coordinate system is obtained using a semi-numerical technique described in~\cite{Roy:2017oul}.

We only use the dominant spherical harmonic mode, $\tilde{h}_{2 - 2}$ to calculate the metric. As the higher harmonic modes, for which $|m| > 2$, contain more number of in-band cycles as compared to the $|m| = 2$ mode, the metric ellipsoids for these modes will be contained within the metric ellipsoid calculated using dominant harmonic mode. However, this argument is not true for $\tilde{h}_{2 - 1}$, where the metric ellipsoid will be larger, but we can still use the same metric due to relatively smaller contribution of $(2, \pm 1)$ mode to the SNR. Fig.~\ref{fig:2d_metric_slices} depicts the accuracy of the metric approximation of the match in terms of orthogonal 2-dimensional coordinate slices of the full 4-dimensional parameter space. Note that in each plot, the two coordinates that are not shown are fixed to the value corresponding to the reference point at which metric is evaluated. From Fig.~\ref{fig:2d_metric_slices}, we note that the coordinates $\theta_{3s}$ and $\bar{\theta}_{3s}$ are highly correlated, one of them can be regarded as redundant, but we need 4 parameters to uniquely map into component masses and spins. Moreover, metric obtained in this parameter space is slowly varying as desired, that's why we adopt $\bar{\theta}_{3s}$ as the $4^{\rm{th}}$ parameter.

The choice of the chosen match value to construct the 4D hyper-ellipsoid (henceforth, we call it ellipsoid) should ideally be dependent on SNR of the signal. For high SNR events a small volume in the parameter space would work because of sharply peaked likelihood function. For low SNR events the value of match at which metric is constructed should be adequate enough to capture the likelihood distribution within the intrinsic parameter space. In this work, we spray nodes in the metric ellipsoid corresponding to the match value of 0.95 for majority of the events, however for some events with higher SNR, we choose larger match values such as 0.97 or 0.98. The chosen value is adequate enough so that the metric ellipsoid encapsulates the support of the likelihood function in all cases.

We emphasize that Fig.~\ref{fig:2d_metric_slices} displays only two-dimensional cross-sections of the full four-dimensional metric ellipsoid, taken through the center. These slices do not capture the complete extent or geometry of the ellipsoid in the full parameter space and may underestimate its size or miss correlations present in higher dimensions. 
 
\begin{figure}[!hbt]
    \centering
    \includegraphics[width=\linewidth]{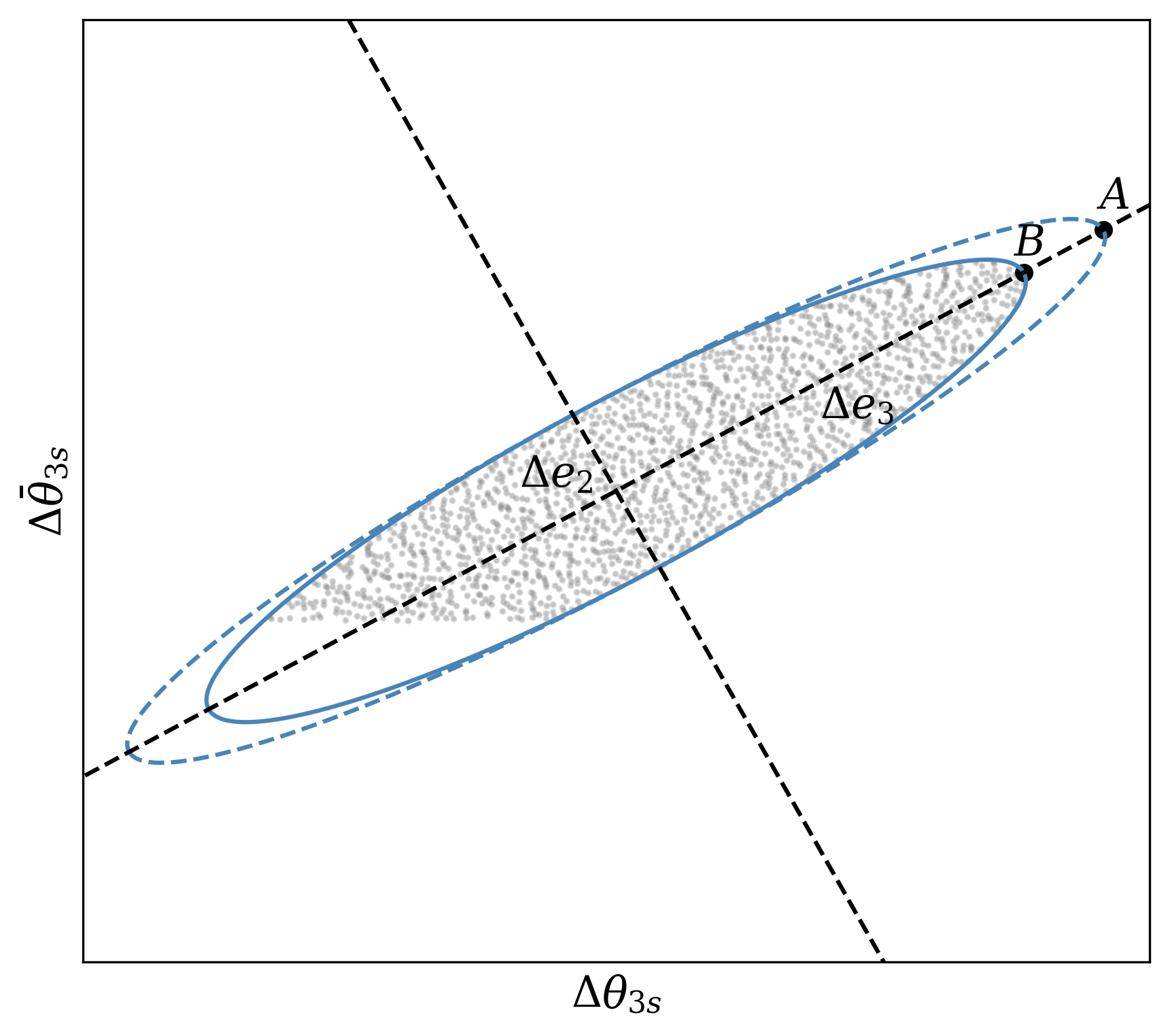}
    \caption{
    Schematic, showing the squeezing of the metric ellipsoid along its longest eigenvector direction to correct for overestimated match values near the boundary. The plot shows a 2D slice of the 4D metric ellipsoid in the $\theta_{3s} - \bar{\theta}_{3s}$ plane, with $\theta_0$ and $\theta_3$ fixed. Dashed black lines indicate the directions of the two largest eigenvectors. The dashed contour depicts the match = 0.95 ellipsoid, obtained from a semi-numerical computation of the parameter-space metric. Due to underestimated match at the ends, the ellipsoid is contracted along the longest axis, reducing its extent from point $A$ to $B$. The solid blue contour shows the adjusted (squeezed) ellipse, and gray points denote the sampling nodes. The white region inside the smaller ellipse marks unphysical parameter space. This correction confines the interpolation node placement and Bayesian sampling to regions where the match remains above threshold. It ensures that the meshfree likelihood interpolation is constructed from physically meaningful regions, improving both accuracy and efficiency of the overall parameter estimation method.
    }
    \label{fig:metric_squeezing}
\end{figure}

The metric at the center is obtained by Taylor-expanding the match function in the intrinsic parameter space around its peak and retaining terms up to quadratic order. While the metric approximation holds well for points near the expansion point, it breaks down near the tips of the ellipsoid, farther from the center. In these regions, the true match falls significantly below what the metric predicts, leading to negligible likelihood values.
This means that such points lie much farther from the center, and thus from the support of the likelihood than what the metric alone would suggest.
Placing nodes in such regions is inefficient and can degrade interpolation accuracy, since the interpolant is reliable only where the likelihood has appreciable support.
To correct for the breakdown of the metric approximation near the edges of the ellipsoid, we compress (squeeze) the ellipsoid along its longest axis, which corresponds to the largest eigenvector of the metric. The amount of compression depends on how much the true match falls below the constant match value used to define the ellipsoid. This squeezing operation is illustrated in Fig.~\ref{fig:metric_squeezing}.

After adjusting the ellipsoid, we place a fixed number of interpolation nodes within it using a Halton sequence~\cite{halton_sequence, owen2017randomized}. 
The Halton sequence produces low-discrepancy point sets that are more uniformly distributed across the parameter space than standard pseudo-random samples, making them well-suited for high-dimensional interpolation. 
In the following section, we describe how RBF interpolants are constructed from these nodes.

\subsubsection{Interpolants generation}
\label{subsec:svd_interpolants_gen}
As pointed out earlier, the inner-product $\vec z_{\ell m}^{(k)}$ corresponding to $k^{\rm{th}}$ detector represents a time-series, as it is evaluated at circular time-shifts centered around the trigger time and $\sigma_{\ell m \ell' m'}$ represents a scalar inner-product. The waveform model $\imrpxhm$ contains the following modes: $(\ell, |m|) = (2, 2), (2, 1), (3, 2), (3, 3)$ and $(4, 4)$. Therefore, there are 5 vector inner-products and 25 scalar inner-products for which interpolants are to be constructed, for each detector. However, due to the following relation,
\begin{equation}
\inp{h_{\ell -m}}{h_{\ell' -m'}}^\ast = \inp{h_{\ell' -m'}}{h_{\ell -m}}
\label{eq:conjugation_symmetry}
\end{equation}
the total independent scalar inner-products reduces to 15. In general, if $n$ number of modes are considered in the analysis, then for each detector, the number of vector and scalar inner-products for which interpolants are to be constructed would be $n$ and $n(n+1)/2$, respectively.

It is worth noting that if only the quadrupole mode ${(\ell = 2, m = 2)}$ is employed in parameter estimation for a compact binary coalescence (CBC) system having nearly equal component masses, where contributions from higher-order modes are expected to be negligible, then the right-hand side of Eq.~\eqref{eq:logl_complete} reduces to a single vector inner product and a single scalar inner product. While the primary focus of this work is on accelerating parameter estimation for systems where higher modes are relevant, the proposed framework can be readily adapted to analyses restricted to the quadrupolar mode of the gravitational-wave signal.

These vector as well as scalar inner-products are calculated at the nodes. Scalar inner-product $\sigma_{\ell m \ell' m'}$ represents a smooth scalar field in the intrinsic parameter space. Thus, at a particular point $\vec \Theta_q$ in the intrinsic parameter space, $\sigma_{\ell m \ell' m'}$ can be expressed as a linear combination of the RBFs $\phi$ centered at the interpolation nodes and monomial basis $p$~\cite{Meshfree_book}:
\begin{equation}
\label{eq:rbfcoeff}
\sigma_{\ell m \ell' m'}(\vec \Theta_q) = \sum_{i=1}^N\, a_{i}\, \phi(\|\vec \Theta_q - \vec \Theta_{i}\|_2) + \sum_{j = 1}^{M}\, b_{j}\, p_j(\vec \Theta_q)
\end{equation}

where, $\vec \Theta_i$ represents a node. The monomials terms are added in the expansion because it enhance the accuracy of RBF approximations at domain boundaries by regularizing the far-field growth of the RBF approximation~\cite{2016JCoPh, FORNBERG2002473}. The generation of interpolant for $\sigma_{\ell m \ell' m'}$ amounts to evaluating the coefficients of expansion $\{a_i\}_{i=1}^{N}$ and $\{b_j\}_{j=1}^{M}$. The $\sigma_{\ell m \ell' m'}$ calculated at the nodes provide $N$ equations in $(N+M)$ unknowns, thus extra $M$ conditions are imposed: $\sum_{j=1}^M b_{j}p_{j}(\vec\Theta_q) = 0$, in order to uniquely determine the coefficients. The number of monomial terms ($M$) to be appended depends on the chosen order $r$ of monomials. Monomials of order $r$ spans the vector space of polynomials of degree $r$, thus $M = \binom{r+d}{r}$, where, $d$ is the dimension of parameter space which in our case is $4$. The order of monomial is chosen based on the choice of RBF basis kernel. If the RBF $\phi$ is conditionally positive definite of order $r$, then the monomial terms of order atleast $(r-1)$ are required to be appended to guarantee a unique solution~\cite{Meshfree_book}.

The vectors, $\vec z_{\ell m}^{(k)}$ evaluated at the nodes are stacked (row-wise) in a matrix such that each row corresponds to a single node and subsequently, singular value decomposition (SVD) of the resultant matrix yields a set of basis vectors. Any row of the matrix can be represented in terms of basis vectors:
\begin{equation}
\vec z_{\ell m}^{(k)}(\vec \Theta_n) = \sum_{i = 1}^N\, C^{{(k)\,n}}_{i \,(\ell m)}\ \vec u^{(k)}_{i},
\label{eq:SVD}
\end{equation}
where, $C^{(k)\, n}_{i \,(\ell m)} \equiv C^{(k)}_{i \,(\ell m)}(\vec \Theta^n)$, are $N$ SVD coefficients . The set of orthonormal basis vectors $\{\vec u^{(k)}_{i}\}$ are arranged in descending order of their relative importance as determined by the spectrum of singular values. $\vec z_{\ell m}^{(k)}(\vec \Theta_n)$ can be reconstructed by considering only first $l$ basis vectors in summation Eq.~\eqref{eq:SVD}. These coefficients are also smooth scalar functions in the intrinsic parameter space and thus can be interpolated just like $\sigma_{\ell m \ell' m'}$. Therefore, in order to evaluate $z_{\ell m}^{(k)}(\vec \Theta_q)$ at any query point $\vec \Theta_q$ we generate only $l$ interpolants for $C^{{(k)\, q}}_{i \,(\ell m)}$, where, $i \in [1, l]$. As there are in total $5$ vector inner-product quantities $\vec z_{\ell m}^{(k)}$ corresponding to each mode and 15 independent scalar inner-products, $\sigma_{\ell m \ell' m'}$ for a single detector. Thus, we are equipped with a total of $(5l + 15)$ interpolants for a single detector after the end of the start-up stage.

\subsection{Online stage}
\label{subsec:online_stage}
The interpolants generated during the start-up stage can be used to calculate the $\vec z_{\ell m}$ and $\sigma_{\ell m \ell' m'}$ at any query point proposed by the sampler. However, $\vec z_{\ell m}$ is still a vector quantity and is required to be computed corresponding to the geocentric coalescence time, $t_c$ proposed by the sampler. Note that there will be distinct time-shift in each detector ($\Delta t^{(k)}$ in Eq.~\eqref{eq:logl_complete}) that depends upon its position with respect to geocenter as well as the position of the source in the sky. Therefore, we evaluate $z_{\ell m}$ at a time-stamp including the extra time-shift. Due to the finite sample rate of the time-series, the proposed time-stamp at which inner-product is to be evaluated may not correspond to any of the time samples in the series. Therefore, we fit a cubic spline with $\sim 10$ samples centered at the closest time sample and subsequently evaluate $z_{\ell m}$ at the required time-stamp using the cubic spline interpolant. Once the quantities $z_{\ell m}$ and $\sigma_{\ell m \ell' m'}$ are calculated for all modes and all the detectors, these quantities can be combined with extrinsic factors to compute the log-likelihood function Eq.~\eqref{eq:logl_complete}.

The metric ellipsoid constructed in the ${ \vec{\Theta} \equiv (\theta_0, \theta_3, \theta_{3s}, \bar{\theta}_{3s})}$ coordinates defines the region in which the interpolants are generated, and hence marks the boundary of the allowed region while sampling the posterior distribution. However, due to physical constraints: specifically, ${ \eta \leq 0.25 }$ and ${\chi_{1z}, \chi_{2z} \in [-1, 1]}$, not all points inside the ellipsoid correspond to valid physical configurations (see Figure~\ref{fig:2d_metric_slices}). As a result, the physical parameter space is bounded by nontrivial surfaces within the ellipsoid.

While sampling the posterior distribution, proposals in the intrinsic parameter space are drawn uniformly within the ellipsoid in \( \vec{\Theta} \) coordinates. Because of the nontrivial physical boundaries, a significant fraction of such proposals may fall outside the valid region, leading to low acceptance rates and inefficient exploration of the parameter space. This mismatch can also distort the effective proposal distribution and degrade posterior coverage.

To address this, we sample in the eigenbasis of the metric ellipsoid, denoted by \( \Delta e \equiv (\Delta e_0, \Delta e_1, \Delta e_2, \Delta e_3) \), where \( \Delta e_0 \) corresponds to the displacement from the center along the smallest eigenvector, \( \Delta e_1 \) along the next, and so on (see Fig.~\ref{fig:metric_squeezing}). Sampling in \( \Delta e \) coordinates mitigates the issue of nontrivial proposal geometry to a large extent. Such eigen-coordinates are also highly effective for placing templates in higher-dimensional parameter space~\cite{Sharma:2023djw}. Since the \( \Delta e \) coordinates are orthogonal and uncorrelated, the sampler achieves faster convergence and improved acceptance rates. 

The resulting posterior samples are finally mapped back to physically meaningful quantities such as chirp mass, mass ratio, and component spins. Wherever needed, these samples can be reweighted to be consistent with any desired target prior\footnote{Reweighting is performed by assigning to each posterior sample a weight proportional to the ratio of the target prior to the proposal prior evaluated at that point.}; this procedure is discussed later. Section~\ref{subsec:sampling_delta_e} has more technical details of sampling in the uncorrelated $\Delta e$ coordinates. These coordinates are not only effective for accelerated inference but are also practical for brute-force parameter estimation of gravitational-wave events, where efficient exploration of the intrinsic parameter space remains a challenge.

\begin{table}[t]
    \def\arraystretch{1.35}
    \centering
    \begin{tabular}{l l l}
    \toprule[1pt]
        Parameters     &   Range    &   Prior distribution \\     
        \midrule[1pt]
        $\mathcal{M}$  &   $[1, 6.5] \msun$           &   $\propto \mathcal{M}$ \\
        $q$            &   $[1, 10]$            & $ \propto \left [ (1 + q)/q^3 \right ]^{2/5}$     \\        
        $\chi_{1z}$    &   $[-0.998, 0.998]$    & Uniform  \\
        $\chi_{2z}$    &   $[-0.05, 0.05]$      & Uniform  \\
        $\dL$          & $[10, 500] \Mpc$            & Uniform in volume\\
        $t_c$          & $[0, 31536000\,\text{s}]$ & Uniform\\
        $\alpha$       & $[0, 2\pi]$            & Uniform\\
        $\delta$       & $[-\pi/2, \pi/2]$            & $\sin^{-1} \left [ {\text{Uniform}}[-1,1]\right ]$\\
        $\iota$        & $[0, \pi]$             & Uniform in $\cos \iota$\\
        $\psi$         & $[0, 2\pi]$            & Uniform angle\\
        $\phi_c$       & $[0, 2\pi]$            & Uniform angle \\
    \bottomrule[1pt]
    \bottomrule[1pt]
    \end{tabular}
    \caption{
    The marginalized distribution for various parameters used to generate NSBH injections. The corresponding distributions for extrinsic parameters were also used as the priors while carrying out PE analysis for each event, except for $\dL$ and $t_c$. The prior distribution for $\dL$ and $t_c$ were taken to be uniform within $[10, 1000] \Mpc$ and uniform within $[t_c^{\rm{inj}}-0.15,\;t_c^{\rm{inj}} + 0.15]s$, respectively.
    }
    \label{tab:nsbh_injs}
\end{table}

\section{Results}
\label{sec:results}
\subsection{Analyses on simulated events}
\label{subsec:sim_results}
We simulate GW data in a three-detector network comprising LIGO Hanford, LIGO Livingston and Virgo assuming stationary, Gaussian noise colored with their respective PSDs for a population of NSBH binaries. We use $\aLIGOZDHP$~\cite{aLIGO_Design} for both LIGO observatories and design sensitivity for Virgo~\cite{KAGRA:2013rdx}. The parameters of the simulated injections are drawn according to the prior distributions given in Table~\ref{tab:nsbh_injs}. Injections are drawn in such a way that the primary component mass $m_1 \in [5, 20]\, M_{\odot}$ and the secondary component mass $m_2 \in [1, 3]\, M_{\odot}$ are uniformly distributed within their chosen range. Note that they do not correspond to any astrophysical distribution. However, the choice of these ranges is inspired from the work by Biscoveanu et al.~\cite{Biscoveanu_2022} where they estimate the population properties of the NSBH mergers using the third Gravitational Wave Transients Catalog (GWTC-3)~\cite{gwtc_3} published by the LVK Collaboration. For all simulated events we use low-frequency cutoff of 15 Hz and a sample rate of 4096 Hz. 

\begin{figure}[t]
    \centering
    \includegraphics[width=\linewidth]{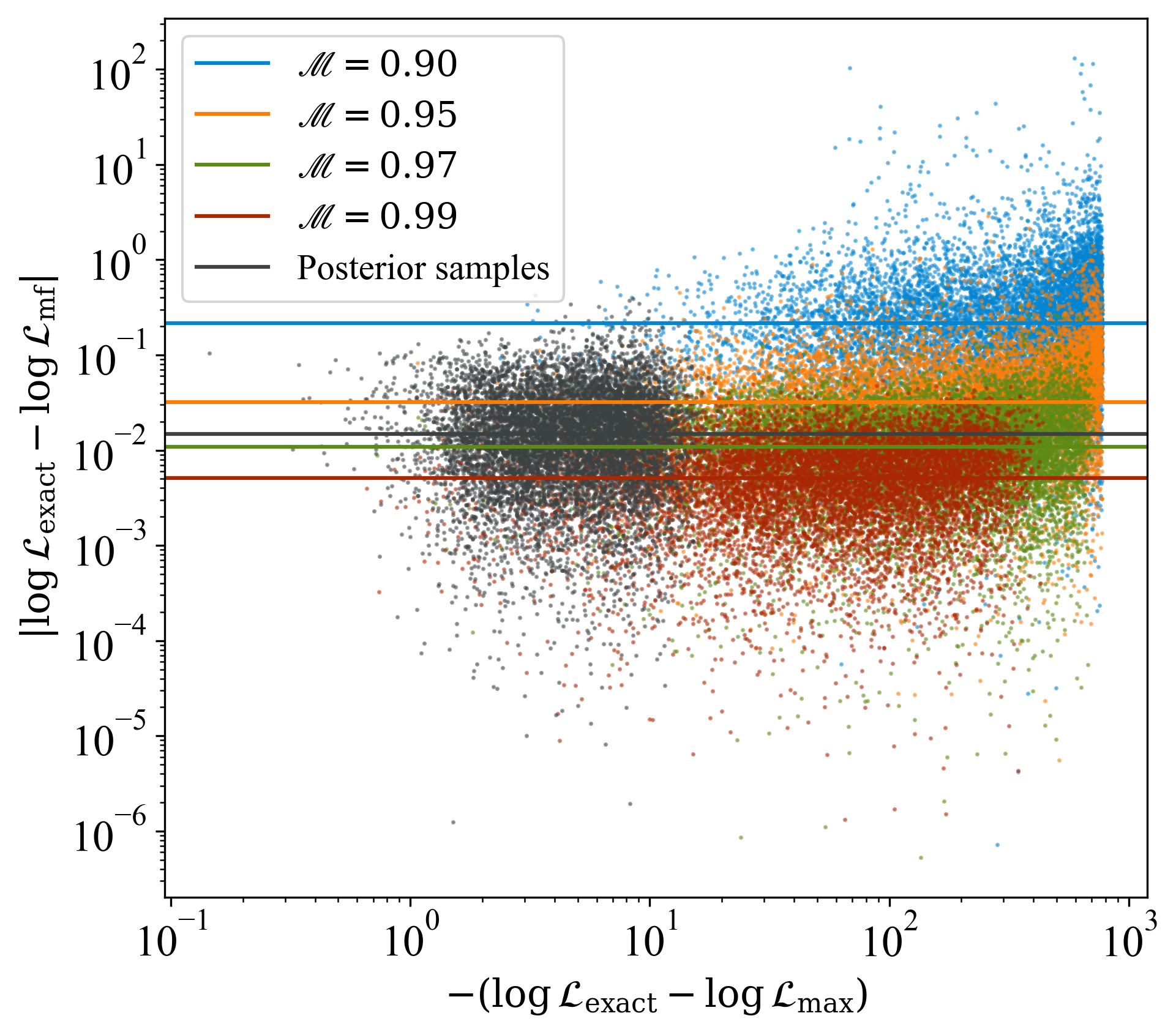}
    \caption{Absolute difference between the exact and meshfree log-likelihood values for the lightest system among 104 events, evaluated on five distinct sets of points. Four sets (colored blue, orange, green, and red), each containing $10^4$ points, correspond to the points uniformly sampled in the intrinsic parameter space and constrained to lie within a constant match ellipsoid, with all extrinsic parameters fixed to their injected values. Each set corresponds to a different match value, as indicated in the legend. The fifth set (grey) corresponds to the posterior samples obtained from a full PE run in which all parameters were varied.
    Horizontal lines indicate the median values of $|\log \mathcal{L}_{\text{exact}} - \log \mathcal{L}_{\text{mf}}|$ for each set, which are: $2.16 \times 10^{-1}$ (blue),  $3.19 \times 10^{-2}$ (orange), $1.09 \times 10^{-2}$ (green), $5.08 \times 10^{-3}$ (red) and $1.49 \times 10^{-2}$ (grey). The reference value $\log \mathcal{L}_{\text{max}} = 205.9$ denotes the maximum exact log-likelihood across the grey points.
    These results demonstrate that the meshfree likelihood achieves high accuracy--particularly in regions of high posterior density--thereby preserving the fidelity of the posterior distribution where it is most critical.}
    \label{fig:logl_difference_plot}
\end{figure}

In order to assess the accuracy of interpolated likelihood as a function of chosen match value, we compare the exact log-likelihood values, computed via direct (bruteforce) evaluation, against those obtained from the meshfree interpolation for four distinct match values: $\mathscr{M}=0.9,$ $0.95$, $0.97$, $0.99$. We selected an event with intrinsic parameters $m_1 = 5.31 \;\msun$, $m_2 = 1.22 \;\msun$, $\chi_{1z} = -0.69$, and $\chi_{2z} = -0.02$ (network SNR $\approx20$), corresponding to the lightest system in the simulated catalog for this comparison. Fig.~\ref{fig:logl_difference_plot} shows the comparison of errors in log-likelihood values evaluated at points sampled uniformly within the constant match ellipsoid corresponding to 4 different values of match. The horizontal lines indicate the corresponding median values of the log-likelihood errors. As expected, the likelihood errors increase with the size of the ellipsoid boundary. Nodes that are placed in the arbitrarily large ellipsoid (corresponding to smaller value of match) step outside the support of the steeply falling likelihood into noise-dominated regions, which reduces the fidelity of interpolation and results in large likelihood errors. We also plot the errors in log-likelihood evaluated at the posterior samples obtained from a meshfree PE analysis for this particular event. The plot indicates that the distribution of the log-likelihood errors corresponding to $\mathscr{M}\geq0.95$ are consistent throughout the parameter space, indicating that the method is accurate enough to approximate the log-likelihood values even in those regions where likelihood is significantly small. Crucially, the interpolation is significantly more accurate near the peak of the likelihood i.e., in regions that dominate the posterior support. This behavior is essential for parameter estimation, as it ensures that the meshfree approximation preserves fidelity in the parts of parameter space most relevant for inference.

We perform our PE analysis on $104$ events satisfying the criteria of network SNR greater than $20$. In our analysis, we construct the metric ellipsoid corresponding to a match value of 0.95 for most of the events. For high SNR events in the simulated catalog, we choose larger match value of 0.97 and 0.98. We place 3000 nodes in the 4D metric ellipsoid to evaluate various interpolants following the strategy outlined in section~\ref{subsubsec:node_placement}. The chosen value of the match to construct the metric ellipsoid provides a broad enough region to encapsulate the structure of likelihood function for all the events with chosen SNR threshold of 20. We use a publicly available Python package~\cite{RBF_github} for generating the RBF interpolants and we use the $\dynesty$ nested sampling package~\cite{Speagle:2019ivv} with $\mathtt{nlive} = 1000, \: \mathtt{walks} = 150, \: \mathtt{sample} = \mathtt{rwalk}$, and $\mathtt{dlogz} = 0.1$ for sampling the posterior distribution. 

To assess whether the meshfree method is able to recover the true values of the injection parameters without any bias, we generate a probability–probability (P–P) plot from the posterior samples obtained using the meshfree method. The P–P plot compares the empirical cumulative distribution function derived from the posterior samples against the theoretical cumulative distribution function. Since the prior for each simulated event is defined based on the metric computed at its reference point, the priors differ across events. To construct the P–P plot, we first reweight the posterior samples for each event using the priors used to generate the events. This reweighting is performed via importance sampling, where the weight for each posterior sample $\vec{\Lambda}_j$ is calculated as the ratio of the injection prior to the PE prior: $w_j = \frac{p_{\text{inj}}(\vec \Lambda_j)}{p_{\text{PE}}(\vec \Lambda_j)}$. 
These weights are then used to compute the effective number of samples for each event: $N_{\text{eff}} = \frac{(\sum_j\, w_j)^2}{\sum_j\, w_j^2}$.

\begin{figure}[t]
    \centering
    \includegraphics[width=\linewidth]{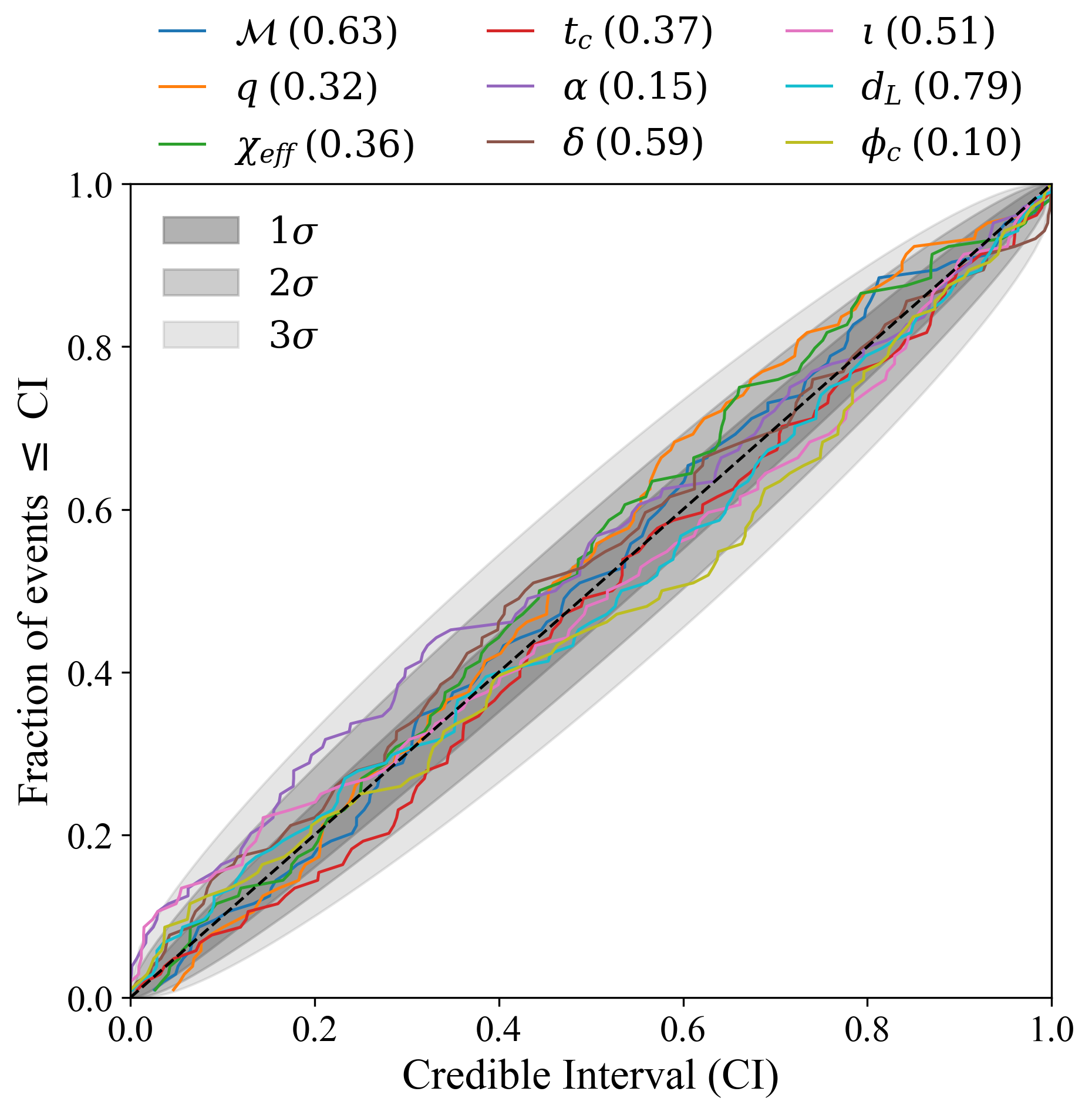}
    \caption{Probability-Probability (P-P) plot for all parameters of interest, computed using a population of 104 simulated NSBH events with network SNR $\rho > 20$. Each colored curve corresponds to a parameter, with the associated p-value listed in the legend. The black dashed line indicates perfect calibration, while the shaded gray regions denote the expected $1\sigma$, $2\sigma$, and $3\sigma$ fluctuations due to finite sample size.  
    The combined p-value, calculated using Fisher’s method~\cite{borenstein2009introduction}, is 0.424 indicating that the observed deviations in the P-P plot are statistically consistent with random fluctuations and supports the hypothesis that the posteriors are well-calibrated. All parameters are estimated in an unbiased manner by the meshfree method. 
    }
    \label{fig:pp_plot}
\end{figure}

Fig.~\ref{fig:pp_plot} shows the P–P plot, we note that all the distributions remain within the $3\sigma$ uncertainty band around the ideal calibration line (represented by the black dashed line), demonstrating that the meshfree method yields unbiased estimates of the true values. The corresponding p-values quantifying the deviation from the ideal diagonal are also reported in the figure and lie within acceptable ranges, further confirming the statistical consistency of the method.

The average wall clock time of the PE runs using meshfree method was approximately $\sim$ 2 hours 40 minutes on 64 CPUs, which corresponds to a computational cost $\sim 170$ CPU hours~\footnote{All PE runs were performed on 64 physical cores of a dual‑socket AMD EPYC 7542 2.9 GHz processor.} 
This includes both the start-up and the sampling stages. The start-up stage which includes the evaluation of optimized center to calculate the metric, computation of the metric, placement of nodes, and generation of the interpolants tend to complete within $\sim 15$ minutes (16 CPU hours) for all the systems. Due to longer in-band signal duration for lighter systems, the start-up stage tend to take longer compared to heavier systems. However, the sampling stage remains unaffected by signal length and primarily depends on the SNR of the system. As expected, the sampling time increases with increasing SNR. Overall, we observe the most substantial computational gains for the lighter systems. This is because brute-force likelihood evaluation becomes less expensive as the in-band signal duration decreases which is the case for heavier systems, while the computational cost of meshfree likelihood evaluation remains relatively unchanged.

To compute the speed-up gain using the meshfree method for the lightest system in the catalog, we perform a PE analysis using exact likelihood evaluation sampling in $\Delta e$ coordinates uniformly within the 0.95 match ellipsoid. This event has the network SNR $\approx20$. The bruteforce PE run costs $\sim 1500$ CPU hours, while meshfree PE costs $\sim 145$ CPU hours for this system, establishing the fact that meshfree likelihood approximation method is highly suited for long duration signals. 

\begin{table*}
\centering
\begin{tabular}{ccccc}
\toprule[1pt]
\toprule[1pt]
Parameter \ \ \ & Meshfree \ \ \ & \ \ \ Bruteforce ($\Delta e$) \ \ \ & \ \ \ Bruteforce (conventional) \  \ \ & \ \ \ JSD \\
\midrule
$\mchirp$ & $2.1122^{+0.0014}_{-0.0012}$ & $2.1123^{+0.0015}_{-0.0014}$ & $2.1123^{+0.0012}_{-0.0012}$ & 0.0038 \\
$q$ & $4.0773^{+0.8071}_{-0.6167}$ & $4.1367^{+0.9000}_{-0.7170}$ & $4.1993^{+0.7133}_{-0.6281}$ & 0.0217 \\
$\chi_{\rm{eff}}$ & $-0.603^{+0.107}_{-0.122}$ & $-0.605^{+0.121}_{-0.130}$ & $-0.611^{+0.104}_{-0.114}$ & 0.0082 \\
$\Delta t_{\rm{c}}$ & $0.00008^{+0.00076}_{-0.00082}$ & $0.00014^{+0.00080}_{-0.00080}$ & $0.00018^{+0.00088}_{-0.00073}$ & 0.0193 \\
$\alpha$ & $2.29^{+0.07}_{-0.09}$ & $2.30^{+0.09}_{-0.09}$ & $2.29^{+0.08}_{-0.08}$ & 0.0046 \\
$\delta$ & $-0.97^{+0.23}_{-0.13}$ & $-0.99^{+0.26}_{-0.15}$ & $-0.98^{+0.25}_{-0.16}$ & 0.0301 \\
$\cos (\iota)$ & $0.810^{+0.171}_{-1.711}$ & $0.857^{+0.131}_{-0.319}$ & $0.743^{+0.233}_{-0.599}$ & 0.1889 \\
$\dL$ & $370.6^{+89.0}_{-107.6}$ & $374.0^{+90.8}_{-121.6}$ & $329.6^{+98.5}_{-186.7}$ & 0.1010 \\
\midrule[1pt]
Computational cost (CPU hours) & 145 & 1500 & 2050 & \\
\bottomrule[1pt]
\end{tabular}
\caption{Comparison of the median values and 90\% credible intervals for intrinsic and extrinsic parameters obtained using three different parameter estimation (PE) methods for the lightest system in the simulated catalog. The meshfree results (first column) correspond to reweighted posteriors, adjusted to match the prior used in the conventional bruteforce PE run. 
The Jensen-Shannon divergence (JSD)~\cite{lin1991divergence} in the last column quantifies the agreement between the meshfree and conventional methods, based on their one-dimensional marginalized posterior distributions. The JSD is computed using logarithms with \mbox{base 2},  
and smaller JSD values indicate closer agreement, with a value of 0 corresponding to perfect overlap. 
Despite reducing computational cost by over an order of magnitude, the meshfree method yields posteriors that are in close agreement with those from the conventional PE.
}
\label{tab:PE_comparison}
\end{table*}

The speed-up provided by the meshfree approach is expected to be even more significant for 3G observatories such as the Einstein Telescope (ET)~\cite{Punturo2010_ETScienceReach,Hild2011_ETSensitivity,Maggiore2020_ETScienceCase}, which will be sensitive to frequencies as low as \mbox{5 Hz}. The corresponding increase in waveform duration, particularly for low-mass systems substantially raises the cost of likelihood evaluations in standard methods, further amplifying the benefits of interpolation-based techniques.
To assess the potential speed up we simulated the data for the lightest system in the three channels of ET assuming Gaussian, stationary noise colored with ET PSD~\cite{ET_psd} and carried out a PE analysis. 
As a simplifying assumption, we neglect the time dependence of the ET antenna pattern functions due to Earth's rotation. 
While a $\mathscr{M} = 0.95$ match ellipsoid was used earlier to sample the intrinsic parameter space, we adopt a tighter $\mathscr{M} = 0.97$ ellipsoid here, as the higher SNR ($\sim 120$) in ET is expected to produce a more sharply peaked likelihood.
The run completed in 3 hours and 6 minutes of wall-clock time on 64 CPUs. Due to the long in-band signal duration ($\sim 50$ minutes), the startup stage required 53 minutes, followed by 133 minutes of sampling. This corresponds to a total computational cost of approximately \mbox{200 CPU hours}.
A full brute-force PE analysis for this system in ET is computationally prohibitive. Instead, we estimate the runtime by measuring the average cost of a single brute-force likelihood evaluation and multiplying it by the total number of likelihood calls made during the meshfree PE run. This yields a projected computational cost of \(\mathcal{O}(10^6)\) CPU hours. The resulting speed-up from the meshfree method is \(\mathcal{O}(10^4)\), comparable to the improvements achieved by reduced order quadrature (ROQ) technique for 3G observatories~\cite{Smith:2021bqc}.

\subsection{Analysis on real GW event}
\label{subsec:lvk_gw_events}
To demonstrate the robustness of the meshfree method on real gravitational-wave strain data, we perform our analysis on \texttt{GW190814}~\cite{LIGOScientific:2020zkf}, a binary merger where higher-order mode effects arising from the highly asymmetric component masses were measured. The event showed no evidence of spin precession and is therefore well modeled using the $\imrpxhm$ waveform family. We analyze publicly available strain data released by the LIGO–Virgo–KAGRA collaboration as part of the O3 open data release~\cite{Abbott2023GWOSC}, and use the associated posterior samples, power spectral densities, and event metadata provided in the corresponding Zenodo archive~\cite{GW190814Zenodo_PE}. 

\begin{figure*}[hbtp]
    \centering
    \includegraphics[width=1\linewidth]{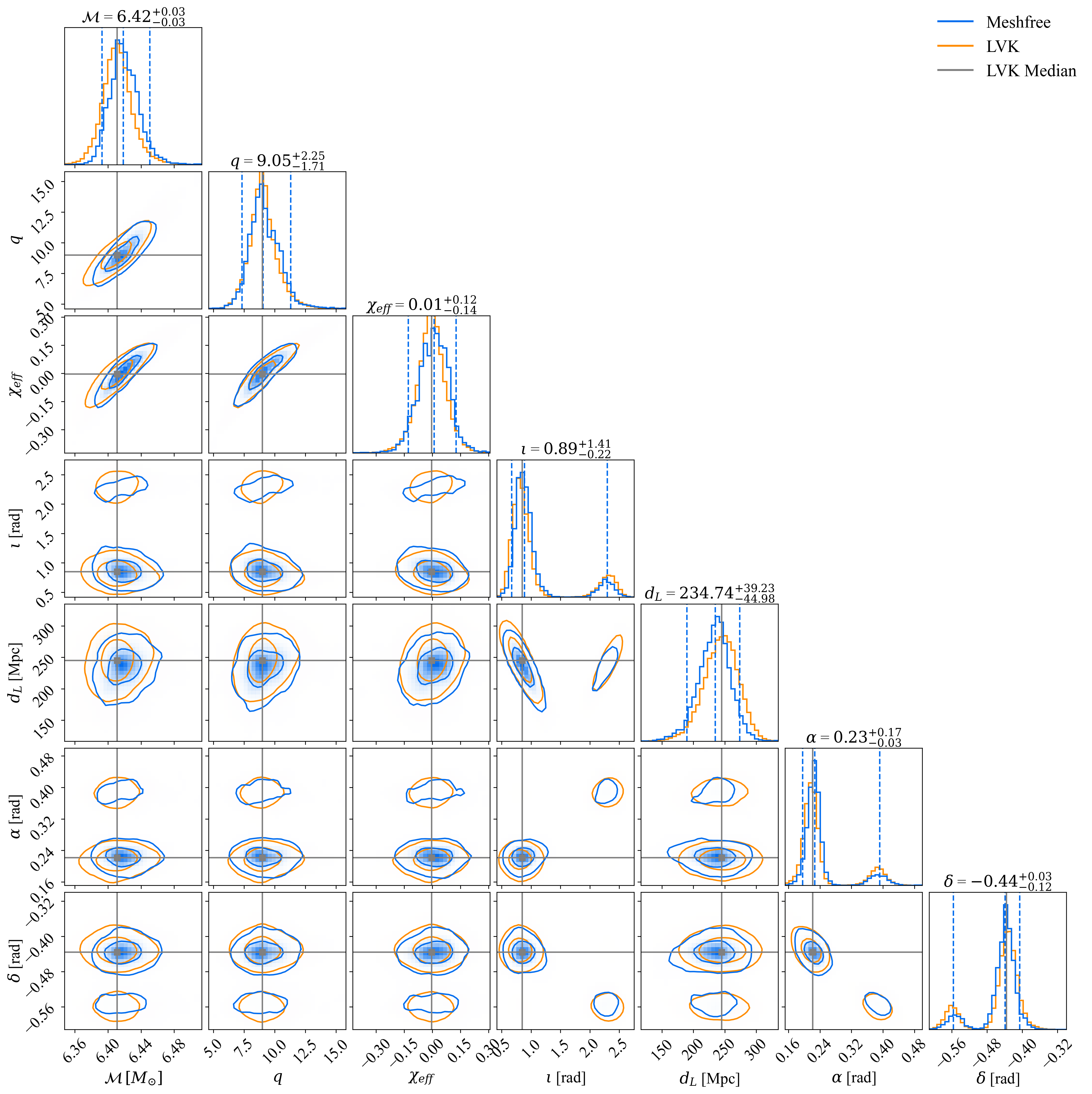}
    \caption{
    Posterior distributions for key parameters of the \texttt{GW190814} event obtained using the meshfree likelihood (blue). For reference, LVK posteriors are also shown (orange), with gray lines indicating their reported median values. The labels at the top of each column indicate the median and 90\% CI for each of the parameters corresponding to meshfree posterior. Despite using a different waveform model (\texttt{IMRPhenomXHM} was used in the meshfree inference vs.\ \texttt{IMRPhenomHM}~\cite{London:2018} in the LVK analysis), both methods show strong agreement with each other. The LVK posteriors were obtained from the metadate provided in Zenodo archive~\cite{GW190814Zenodo_PE}.
    }
    \label{fig:corner_gw190814}
\end{figure*}

To place nodes in the intrinsic parameter space, we used the best-matched template as the initial guess and evaluated the coordinates of the center to construct the metric ellipsoid (corresponding to the match $= 0.95$) by optimizing the network SNR as described in Sec.~\ref{subsubsec:node_placement}. Fig.~\ref{fig:corner_gw190814} shows the posterior distributions of different parameters obtained using meshfree likelihood along with the LVK posteriors~\cite{GW190814Zenodo_PE} and shows excellent agreement. Note that the LVK result corresponds to the  $\imrphm$~\cite{London:2018} waveform model whereas we have used $\imrpxhm$ waveform model in this analysis. In our case we sampled the posterior uniformly in $\Delta e$ coordinates within the 0.95 match ellipsoid, while LVK analysis were carried out in the $\mchirp,\; q, \chi_{1z}, \chi_{2z}$ coordinates. Thus, the two analyses differ in their prior distribution choices. Therefore, to mitigate the effect of different priors we reweight the posteriors obtained using meshfree method to make them consistent with those used in the LVK analysis.

\subsection{Sampling in \texorpdfstring{$\Delta e$}{Delta e} coordinates}
\label{subsec:sampling_delta_e}
As mentioned above, we perform sampling in the $\Delta e$ coordinates, which are obtained from the eigenvalue decomposition of the metric. By construction, these coordinates are uncorrelated, which is expected to facilitate faster convergence during sampling~\cite{Veitch:2014wba}. In contrast, sampling in the parameters: $\mathcal{M}$, $q$, $\chi_{1z}$, and $\chi_{2z}$ must navigate a more correlated distribution, thus reducing sampling efficiency compared to any uncorrelated set of parameters. 

To quantify the advantage from sampling in the $\Delta e$ coordinates, we perform parameter estimation for the lightest NSBH system in our simulated catalog using both the $\Delta e$ coordinates and the standard coordinates ($\mathcal{M}$, $q$, $\chi_{1z}$, $\chi_{2z}$) in both cases using bruteforce likelihood evaluation. For the standard coordinates, we adopt priors in $\mathcal{M}$ and $q$ that produce a uniform distribution in the component masses. The prior range for $\mathcal{M}$ was considered to be $\pm 0.075 \msun$ around the injected value and the mass ratio in $[1, 10]$. For spins, we use uniform priors with $\chi_{1z},\; \chi_{2z} \in [-0.99, 0.99]$. For the $\Delta e$ coordinates, we compute the metric at a reference point obtained by performing the SNR optimization starting from the true injection values, and the prior distributions over the intrinsic parameters were taken to be uniform within the metric ellipsoid bounded by the $0.95$ match contour.

\begin{figure}[b]
    \centering
    \includegraphics[width=\linewidth]{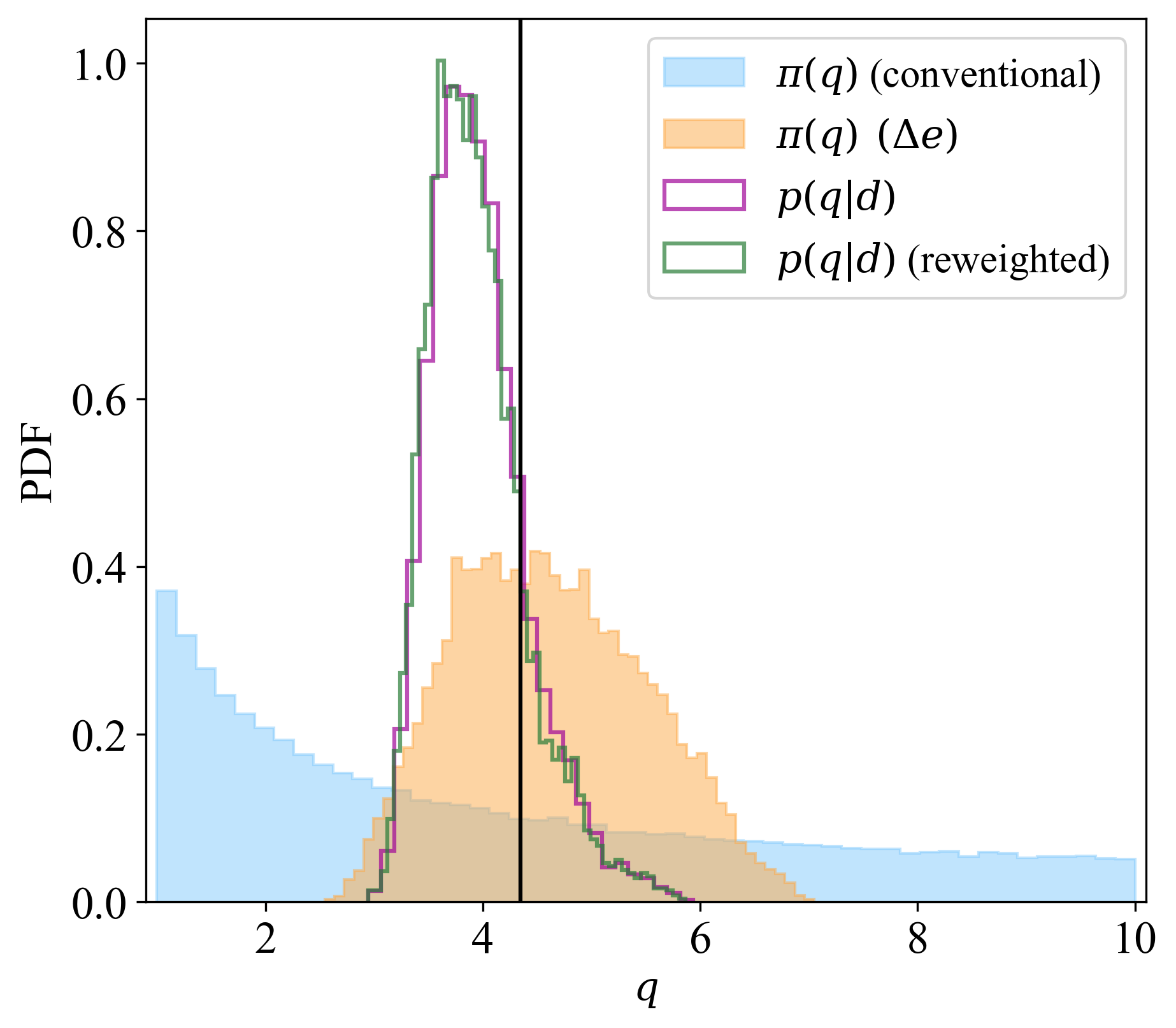}
    \caption{
    Comparison of marginalized prior distributions over mass ratio for the two sampling schemes described in Section~\ref{subsec:sampling_delta_e}. The shaded blue curve shows the prior used when sampling in conventional coordinates $\mathcal{M}_c, q, \chi_{1z}, \chi_{2z}$, while the shaded orange curve corresponds to the prior induced by uniform sampling in $\Delta e$ coordinates. Also shown are the marginalized posteriors over mass ratio obtained from $\Delta e$ sampling, both before (magenta) and after (green) reweighting. The minimal difference between the two posteriors reflects the near-uniformity of the weight distribution over the region where the posterior has support.}
    \label{fig:posterior_reweighting}
\end{figure}

\begin{figure}[t]
    \centering
    \includegraphics[width=\linewidth]{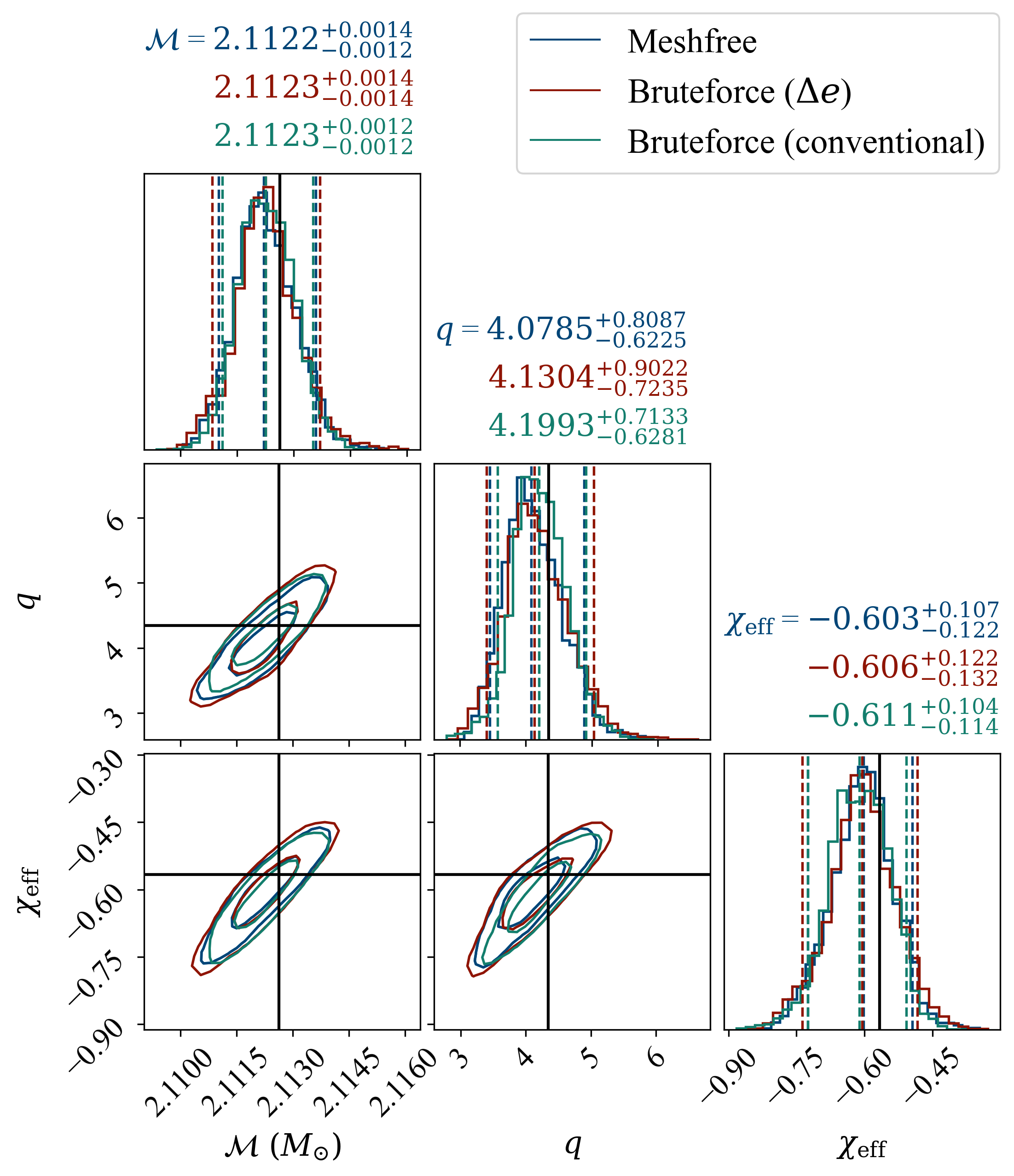}
    \caption{
    Comparison of 3D marginalized posterior distributions in intrinsic parameter space for the lightest event in the simulated catalog. Blue shows the meshfree PE run; red and green correspond to bruteforce likelihood evaluation with sampling in $\Delta e$ and $(\mchirp, q, \chi_{1z}, \chi_{2z})$ coordinates, respectively. The meshfree and bruteforce, sampled in $\Delta e$ coordinates PE posteriors are reweighted to match the priors used in $(\mchirp, q, \chi_{1z}, \chi_{2z})$ sampling. Contours in the 2D marginals indicate 50\% and 90\% credible intervals; black lines mark the true values. Column labels show median and 90\% CI from each PE run. Other parameters are omitted in this corner plot for brevity (see Table~\ref{tab:PE_comparison} for comparison of other parameters). The reweighted meshfree posteriors are identical to the standard analysis, confirming consistency.
    }
    \label{fig:corner_comparison}
\end{figure}

Parameter estimation using standard coordinates took approximately $32$ hours ($\sim 2050$ CPU hours), while the run using $\Delta e$ coordinates was completed in about $23.5$ hours ($\sim 1500$ CPU hours) on $64$ CPUs, a reduction in sampling time by a factor of ${\sim 1.36}$. This clearly demonstrates that sampling in uncorrelated coordinates significantly improves sampling efficiency. The posterior samples obtained in $\Delta e$ coordinates are then transformed into chirp mass, mass ratio and effective spin. 

Since the priors (in intrinsic parameter space) used in the meshfree method and bruteforce method with sampling in $\Delta e$ coordinates differ from those employed in the standard analysis, we reweight the posteriors obtained from sampling in the $\Delta e$ coordinates to match the priors used in the standard analysis. The reweighting procedure follows the method described in Section~\ref{subsec:sim_results}.
We would like to highlight that reweighting procedure does not drastically transform the posterior distribution because of nearly uniform distribution of these weights near the posterior support. 

For example, Fig.~\ref{fig:posterior_reweighting} shows the marginalized posterior over mass ratio obtained from sampling in $\Delta e$ coordinates with and without reweighting. The comparison of marginalized prior distributions for mass ratio corresponding to the two analyses are also shown.
Fig.~\ref{fig:corner_comparison} shows the marginalized posteriors over chirp mass, mass ratio and effective spin parameters in a corner plot obtained from sampling in $\Delta e$ coordinates (in red) and sampling in standard coordinates (in green). For comparison, the posterior obtained from meshfree method are also shown (in blue). 
The reweighted posteriors obtained by sampling in $\Delta e$ coordinates (with bruteforce likelihood evaluation) closely match those obtained from the standard analysis, confirming consistency, while being obtained approximately 26\% faster.
Similar observations have been reported for binary neutron star systems by Lee et al.~\cite{Lee2022_massspin_reparam}, who demonstrated that a Fisher-matrix based mass-spin reparameterization can enhance MCMC  efficiency by an order of magnitude. 

\section{Conclusion and Discussion}
\label{sec:conclusion}

We extend our previously developed meshfree likelihood interpolation framework to incorporate higher-order spherical harmonic modes of gravitational-wave radiation from CBCs. These modes are essential for accurately modeling asymmetric systems such as NSBH binaries. Using the aligned-spin waveform model \texttt{IMRPhenomXHM}, we demonstrate that the framework enables accurate and efficient parameter estimation for long-duration signals with significant higher-mode content. We also introduce a revised node placement strategy that improves interpolation accuracy and extends coverage in the intrinsic parameter space.

Assuming time-independent detector antenna patterns, the likelihood separates into intrinsic- and extrinsic-parameter dependent components. 
We precompute inner products at a sparse set of nodes in the intrinsic parameter space and construct interpolants using radial basis functions, employing techniques from numerical linear algebra. Each spherical harmonic mode is interpolated independently, which reduces the complexity per interpolant. Extrinsic parameters enter the likelihood through simple algebraic factors and are evaluated directly during sampling.

The interpolants are constructed within a constant-match ellipsoid in the four-dimensional intrinsic parameter space, centered near a point of locally maximal SNR identified via optimization around the best-matched template. Nodes are placed using low-dispersion Halton sequences to ensure uniform coverage. The match value that defines the ellipsoid sets a trade-off between interpolation fidelity and posterior support. In our simulations, we use match values of 0.95 for most injections and 0.97 or 0.98 for high-SNR events.

We validate the framework on 104 simulated NSBH signals in synthetic HLV data and recover unbiased posteriors. For the longest-duration signal, we achieve up to a 10$\times$ speed-up over direct likelihood evaluation. 
To assess scalability, we apply the method to the lightest simulated NSBH event in simulated ET noise and recover consistent posteriors. The analysis completes with an estimated computational saving of ${\mathcal{O}(10^4)}$ CPU-hours. This is made possible by evaluating fast surrogate models constructed using the meshfree algorithm, which avoids costly on-the-fly waveform generation and overlap calculations for long-duration signals in the ET band. The ${\sim\!10^4\times}$ speedup achieved is consistent with acceleration factors reported by Smith et al.~\cite{Smith:2021bqc} for ROQ-based inference in 3G detectors.

We also introduce a new intrinsic parameter sampling strategy that transforms coordinates to align with the eigendirections of the local metric ellipsoid. The resulting $\Delta e$ coordinates are uncorrelated by construction, improving mixing and sampler convergence. For the lightest signal in our catalog, this yields a 1.36$\times$ reduction in runtime relative to sampling in standard physical parameters. This strategy can be integrated into other gravitational-wave parameter estimation pipelines. Posterior samples obtained this way can be reweighted for  inference under arbitrary prior choices.

Although we have focused on NSBH systems with significant higher-mode content, the meshfree framework is equally applicable to symmetric BBH and BNS systems. In such cases, the likelihood simplifies further, leading to lower computational cost for parameter estimation. For instance, the lightest NSBH system in the simulated catalog (Table~\ref{tab:PE_comparison}) was analyzed in approximately 20 CPU hours by restricting both the injection and template waveform models to the quadrupole mode. This is substantially more efficient than the corresponding analysis that includes all higher modes, as also summarized in Table~\ref{tab:PE_comparison}.

While the present work assumes time-independent detector antenna pattern functions, future work will include time-dependent effects from the Earth's rotation, becoming relevant for 3G detectors. Also, while our present code allows for aligned-spin waveform models, future use will include precessing systems to more broadly extend the application of the meshfree formalism.

In addition to enabling rapid follow-up of real events, the meshfree framework could support efficient inference in large simulation campaigns to study waveform systematics  and population studies where computational cost of full Bayesian inference is a limiting factor.

\begin{acknowledgments}
We thank Charlie Hoy for a thorough reading of the manuscript and for insightful suggestions that enhanced its clarity. 
A.~Sharma thanks Divya Tahelyani for helpful feedback on the manuscript and acknowledges IIT Gandhinagar for the senior research fellowship. L.~P. is supported by the postdoctoral scholarship from Tata Institute of Fundamental Research (TIFR), Mumbai. S.~Roy is supported by the Fonds de la Recherche Scientifique -- FNRS (Belgium). We acknowledge the computational resources provided by IIT Gandhinagar and TIFR Mumbai. We also thank the high performance computing support staff at IIT Gandhinagar for their help and cooperation. 
The manuscript is based upon work supported by NSF's LIGO Laboratory, which is a major facility fully funded by the National Science Foundation (NSF), as well as the Science and Technology Facilities Council (STFC) of the United Kingdom, the Max-Planck-Society (MPS), and the State of Niedersachsen/Germany for support of the construction of Advanced LIGO and construction and operation of the GEO600 detector. Additional support for Advanced LIGO was provided by the Australian Research Council. Virgo is funded through the European Gravitational Observatory (EGO), by the French Centre National de Recherche Scientifique (CNRS), the Italian Istituto Nazionale di Fisica Nucleare (INFN) and the Dutch Nikhef, with contributions by institutions from Belgium, Germany, Greece, Hungary, Ireland, Japan, Monaco, Poland, Portugal, Spain. KAGRA is supported by the Ministry of Education, Culture, Sports, Science and Technology (MEXT), Japan Society for the Promotion of Science (JSPS) in Japan; National Research Foundation (NRF) and the Ministry of Science and ICT (MSIT) in Korea; Academia Sinica (AS) and National Science and Technology Council (NSTC) in Taiwan. 
\end{acknowledgments}

\bibliography{references}

\begin{thebibliography}{114}%
\makeatletter
\providecommand \@ifxundefined [1]{%
 \@ifx{#1\undefined}
}%
\providecommand \@ifnum [1]{%
 \ifnum #1\expandafter \@firstoftwo
 \else \expandafter \@secondoftwo
 \fi
}%
\providecommand \@ifx [1]{%
 \ifx #1\expandafter \@firstoftwo
 \else \expandafter \@secondoftwo
 \fi
}%
\providecommand \natexlab [1]{#1}%
\providecommand \enquote  [1]{``#1''}%
\providecommand \bibnamefont  [1]{#1}%
\providecommand \bibfnamefont [1]{#1}%
\providecommand \citenamefont [1]{#1}%
\providecommand \href@noop [0]{\@secondoftwo}%
\providecommand \href [0]{\begingroup \@sanitize@url \@href}%
\providecommand \@href[1]{\@@startlink{#1}\@@href}%
\providecommand \@@href[1]{\endgroup#1\@@endlink}%
\providecommand \@sanitize@url [0]{\catcode `\\12\catcode `\$12\catcode `\&12\catcode `\#12\catcode `\^12\catcode `\_12\catcode `\%12\relax}%
\providecommand \@@startlink[1]{}%
\providecommand \@@endlink[0]{}%
\providecommand \url  [0]{\begingroup\@sanitize@url \@url }%
\providecommand \@url [1]{\endgroup\@href {#1}{\urlprefix }}%
\providecommand \urlprefix  [0]{URL }%
\providecommand \Eprint [0]{\href }%
\providecommand \doibase [0]{https://doi.org/}%
\providecommand \selectlanguage [0]{\@gobble}%
\providecommand \bibinfo  [0]{\@secondoftwo}%
\providecommand \bibfield  [0]{\@secondoftwo}%
\providecommand \translation [1]{[#1]}%
\providecommand \BibitemOpen [0]{}%
\providecommand \bibitemStop [0]{}%
\providecommand \bibitemNoStop [0]{.\EOS\space}%
\providecommand \EOS [0]{\spacefactor3000\relax}%
\providecommand \BibitemShut  [1]{\csname bibitem#1\endcsname}%
\let\auto@bib@innerbib\@empty
\bibitem [{\citenamefont {Abbott}\ \emph {et~al.}(2016{\natexlab{a}})\citenamefont {Abbott} \emph {et~al.}}]{LIGOScientific:2016aoc}%
  \BibitemOpen
  \bibfield  {author} {\bibinfo {author} {\bibfnamefont {B.~P.}\ \bibnamefont {Abbott}} \emph {et~al.} (\bibinfo {collaboration} {The LIGO Scientific Collaboration, Virgo Collaboration}),\ }\bibfield  {title} {\bibinfo {title} {Observation of gravitational waves from a binary black hole merger},\ }\href {https://doi.org/10.1103/PhysRevLett.116.061102} {\bibfield  {journal} {\bibinfo  {journal} {Phys. Rev. Lett.}\ }\textbf {\bibinfo {volume} {116}},\ \bibinfo {pages} {061102} (\bibinfo {year} {2016}{\natexlab{a}})},\ \Eprint {https://arxiv.org/abs/1602.03837} {arXiv:1602.03837 [gr-qc]} \BibitemShut {NoStop}%
\bibitem [{\citenamefont {Abbott}\ \emph {et~al.}(2019)\citenamefont {Abbott} \emph {et~al.}}]{Abbott_2019}%
  \BibitemOpen
  \bibfield  {author} {\bibinfo {author} {\bibfnamefont {B.}~\bibnamefont {Abbott}} \emph {et~al.} (\bibinfo {collaboration} {The LIGO Scientific Collaboration, Virgo Collaboration}),\ }\bibfield  {title} {\bibinfo {title} {{GWTC-1: A Gravitational-Wave Transient Catalog of Compact Binary Mergers Observed by LIGO and Virgo during the First and Second Observing Runs}},\ }\bibfield  {journal} {\bibinfo  {journal} {Physical Review X}\ }\textbf {\bibinfo {volume} {9}},\ \href {https://doi.org/10.1103/physrevx.9.031040} {10.1103/physrevx.9.031040} (\bibinfo {year} {2019})\BibitemShut {NoStop}%
\bibitem [{\citenamefont {Abbott}\ \emph {et~al.}(2021{\natexlab{a}})\citenamefont {Abbott} \emph {et~al.}}]{Abbott_2021}%
  \BibitemOpen
  \bibfield  {author} {\bibinfo {author} {\bibfnamefont {R.}~\bibnamefont {Abbott}} \emph {et~al.} (\bibinfo {collaboration} {The LIGO Scientific Collaboration, Virgo Collaboration}),\ }\bibfield  {title} {\bibinfo {title} {{GWTC-2: Compact Binary Coalescences Observed by LIGO and Virgo during the First Half of the Third Observing Run}},\ }\bibfield  {journal} {\bibinfo  {journal} {Physical Review X}\ }\textbf {\bibinfo {volume} {11}},\ \href {https://doi.org/10.1103/physrevx.11.021053} {10.1103/physrevx.11.021053} (\bibinfo {year} {2021}{\natexlab{a}})\BibitemShut {NoStop}%
\bibitem [{\citenamefont {Abbott}\ \emph {et~al.}(2024)\citenamefont {Abbott} \emph {et~al.}}]{gwtc_2.1}%
  \BibitemOpen
  \bibfield  {author} {\bibinfo {author} {\bibfnamefont {R.}~\bibnamefont {Abbott}} \emph {et~al.} (\bibinfo {collaboration} {The LIGO Scientific Collaboration and the Virgo Collaboration}),\ }\bibfield  {title} {\bibinfo {title} {{GWTC-2.1: Deep extended catalog of compact binary coalescences observed by LIGO and Virgo during the first half of the third observing run}},\ }\href {https://doi.org/10.1103/PhysRevD.109.022001} {\bibfield  {journal} {\bibinfo  {journal} {Phys. Rev. D}\ }\textbf {\bibinfo {volume} {109}},\ \bibinfo {pages} {022001} (\bibinfo {year} {2024})}\BibitemShut {NoStop}%
\bibitem [{\citenamefont {Abbott}\ \emph {et~al.}(2023{\natexlab{a}})\citenamefont {Abbott} \emph {et~al.}}]{gwtc_3}%
  \BibitemOpen
  \bibfield  {author} {\bibinfo {author} {\bibfnamefont {R.}~\bibnamefont {Abbott}} \emph {et~al.} (\bibinfo {collaboration} {The LIGO Scientific Collaboration, Virgo Collaboration, and KAGRA Collaboration}),\ }\bibfield  {title} {\bibinfo {title} {{GWTC-3: Compact Binary Coalescences Observed by LIGO and Virgo during the Second Part of the Third Observing Run}},\ }\href {https://doi.org/10.1103/PhysRevX.13.041039} {\bibfield  {journal} {\bibinfo  {journal} {Phys. Rev. X}\ }\textbf {\bibinfo {volume} {13}},\ \bibinfo {pages} {041039} (\bibinfo {year} {2023}{\natexlab{a}})}\BibitemShut {NoStop}%
\bibitem [{\citenamefont {Aasi}\ \emph {et~al.}(2015)\citenamefont {Aasi} \emph {et~al.}}]{LIGOScientific:2014pky}%
  \BibitemOpen
  \bibfield  {author} {\bibinfo {author} {\bibfnamefont {J.}~\bibnamefont {Aasi}} \emph {et~al.} (\bibinfo {collaboration} {The LIGO Scientific Collaboration}),\ }\bibfield  {title} {\bibinfo {title} {{Advanced LIGO}},\ }\href {https://doi.org/10.1088/0264-9381/32/7/074001} {\bibfield  {journal} {\bibinfo  {journal} {Class. Quant. Grav.}\ }\textbf {\bibinfo {volume} {32}},\ \bibinfo {pages} {074001} (\bibinfo {year} {2015})},\ \Eprint {https://arxiv.org/abs/1411.4547} {arXiv:1411.4547 [gr-qc]} \BibitemShut {NoStop}%
\bibitem [{\citenamefont {Acernese}\ \emph {et~al.}(2015)\citenamefont {Acernese} \emph {et~al.}}]{VIRGO:2014yos}%
  \BibitemOpen
  \bibfield  {author} {\bibinfo {author} {\bibfnamefont {F.}~\bibnamefont {Acernese}} \emph {et~al.} (\bibinfo {collaboration} {Virgo Collaboration}),\ }\bibfield  {title} {\bibinfo {title} {{Advanced Virgo: a second-generation interferometric gravitational wave detector}},\ }\href {https://doi.org/10.1088/0264-9381/32/2/024001} {\bibfield  {journal} {\bibinfo  {journal} {Class. Quant. Grav.}\ }\textbf {\bibinfo {volume} {32}},\ \bibinfo {pages} {024001} (\bibinfo {year} {2015})},\ \Eprint {https://arxiv.org/abs/1408.3978} {arXiv:1408.3978 [gr-qc]} \BibitemShut {NoStop}%
\bibitem [{\citenamefont {Akutsu}\ \emph {et~al.}(2021)\citenamefont {Akutsu}, \citenamefont {Ando}, \citenamefont {Arai} \emph {et~al.}}]{Akutsu2021_KAGRA_overview}%
  \BibitemOpen
  \bibfield  {author} {\bibinfo {author} {\bibfnamefont {T.}~\bibnamefont {Akutsu}}, \bibinfo {author} {\bibfnamefont {M.}~\bibnamefont {Ando}}, \bibinfo {author} {\bibfnamefont {K.}~\bibnamefont {Arai}}, \emph {et~al.},\ }\href {https://doi.org/10.1093/ptep/ptaa125} {\emph {\bibinfo {title} {{Overview of KAGRA: Detector design and construction history}}}},\ \bibinfo {type} {Tech. Rep.}\ \bibinfo {number} {2021:05A101}\ (\bibinfo  {institution} {Progress of Theoretical and Experimental Physics, Oxford University Press},\ \bibinfo {year} {2021})\BibitemShut {NoStop}%
\bibitem [{\citenamefont {Mehta}\ \emph {et~al.}(2025)\citenamefont {Mehta}, \citenamefont {Olsen}, \citenamefont {Wadekar}, \citenamefont {Roulet}, \citenamefont {Venumadhav}, \citenamefont {Mushkin}, \citenamefont {Zackay},\ and\ \citenamefont {Zaldarriaga}}]{Mehta:2023zlk}%
  \BibitemOpen
  \bibfield  {author} {\bibinfo {author} {\bibfnamefont {A.~K.}\ \bibnamefont {Mehta}}, \bibinfo {author} {\bibfnamefont {S.}~\bibnamefont {Olsen}}, \bibinfo {author} {\bibfnamefont {D.}~\bibnamefont {Wadekar}}, \bibinfo {author} {\bibfnamefont {J.}~\bibnamefont {Roulet}}, \bibinfo {author} {\bibfnamefont {T.}~\bibnamefont {Venumadhav}}, \bibinfo {author} {\bibfnamefont {J.}~\bibnamefont {Mushkin}}, \bibinfo {author} {\bibfnamefont {B.}~\bibnamefont {Zackay}},\ and\ \bibinfo {author} {\bibfnamefont {M.}~\bibnamefont {Zaldarriaga}},\ }\bibfield  {title} {\bibinfo {title} {{New binary black hole mergers in the LIGO-Virgo O3b data}},\ }\href {https://doi.org/10.1103/PhysRevD.111.024049} {\bibfield  {journal} {\bibinfo  {journal} {Phys. Rev. D}\ }\textbf {\bibinfo {volume} {111}},\ \bibinfo {pages} {024049} (\bibinfo {year} {2025})}\BibitemShut {NoStop}%
\bibitem [{\citenamefont {Abbott}\ \emph {et~al.}(2021{\natexlab{b}})\citenamefont {Abbott} \emph {et~al.}}]{TGR_2021}%
  \BibitemOpen
  \bibfield  {author} {\bibinfo {author} {\bibfnamefont {R.}~\bibnamefont {Abbott}} \emph {et~al.} (\bibinfo {collaboration} {{The LIGO Scientific Collaboration}, {Virgo Collaboration} and the KAGRA Collaboration}),\ }\href {https://arxiv.org/abs/2112.06861} {\bibinfo {title} {{Tests of General Relativity with GWTC-3}}} (\bibinfo {year} {2021}{\natexlab{b}}),\ \Eprint {https://arxiv.org/abs/2112.06861} {arXiv:2112.06861 [gr-qc]} \BibitemShut {NoStop}%
\bibitem [{\citenamefont {Abbott}\ \emph {et~al.}(2023{\natexlab{b}})\citenamefont {Abbott} \emph {et~al.}}]{Abbott2023GWTC3pop}%
  \BibitemOpen
  \bibfield  {author} {\bibinfo {author} {\bibfnamefont {R.}~\bibnamefont {Abbott}} \emph {et~al.} (\bibinfo {collaboration} {The LIGO Scientific Collaboration, Virgo Collaboration, and KAGRA Collaboration}),\ }\bibfield  {title} {\bibinfo {title} {{Population of Merging Compact Binaries Inferred Using Gravitational‑Wave Observations of These Systems During the First Three LIGO‑Virgo Observing Runs}},\ }\href {https://doi.org/10.1103/PhysRevX.13.011048} {\bibfield  {journal} {\bibinfo  {journal} {Physical Review X}\ }\textbf {\bibinfo {volume} {13}},\ \bibinfo {pages} {011048} (\bibinfo {year} {2023}{\natexlab{b}})}\BibitemShut {NoStop}%
\bibitem [{\citenamefont {Abbott}\ \emph {et~al.}(2017)\citenamefont {Abbott} \emph {et~al.}}]{PhysRevLett.119.161101}%
  \BibitemOpen
  \bibfield  {author} {\bibinfo {author} {\bibfnamefont {B.~P.}\ \bibnamefont {Abbott}} \emph {et~al.} (\bibinfo {collaboration} {The LIGO Scientific Collaboration and Virgo Collaboration}),\ }\bibfield  {title} {\bibinfo {title} {{GW170817: Observation of Gravitational Waves from a Binary Neutron Star Inspiral}},\ }\href {https://doi.org/10.1103/PhysRevLett.119.161101} {\bibfield  {journal} {\bibinfo  {journal} {Phys. Rev. Lett.}\ }\textbf {\bibinfo {volume} {119}},\ \bibinfo {pages} {161101} (\bibinfo {year} {2017})}\BibitemShut {NoStop}%
\bibitem [{\citenamefont {Abac}\ \emph {et~al.}(2024)\citenamefont {Abac} \emph {et~al.}}]{LVK2024GW230529}%
  \BibitemOpen
  \bibfield  {author} {\bibinfo {author} {\bibfnamefont {A.~G.}\ \bibnamefont {Abac}} \emph {et~al.} (\bibinfo {collaboration} {The LIGO Scientific Collaboration, Virgo Collaboration and KAGRA Collaboration}),\ }\bibfield  {title} {\bibinfo {title} {{Observation of Gravitational Waves from the Coalescence of a 2.5–4.5~$\msun$ Compact Object and a Neutron Star}},\ }\href {https://doi.org/10.3847/2041-8213/ad5beb} {\bibfield  {journal} {\bibinfo  {journal} {The Astrophysical Journal Letters}\ }\textbf {\bibinfo {volume} {970}},\ \bibinfo {pages} {L34} (\bibinfo {year} {2024})}\BibitemShut {NoStop}%
\bibitem [{\citenamefont {{The LIGO Scientific Collaboration, Virgo Collaboration and KAGRA Collaboration}}(2025)}]{LVK2025GW231123}%
  \BibitemOpen
  \bibfield  {author} {\bibinfo {author} {\bibnamefont {{The LIGO Scientific Collaboration, Virgo Collaboration and KAGRA Collaboration}}},\ }\bibfield  {title} {\bibinfo {title} {{GW231123: a Binary Black Hole Merger with Total Mass 190–265~$\msun$}},\ }\bibfield  {journal} {\bibinfo  {journal} {arXiv preprint}\ }\href {https://doi.org/https://doi.org/10.48550/arXiv.2507.08219} {https://doi.org/10.48550/arXiv.2507.08219} (\bibinfo {year} {2025}),\ \bibinfo {note} {arXiv:2507.08219}\BibitemShut {NoStop}%
\bibitem [{\citenamefont {Abbott}\ \emph {et~al.}(2016{\natexlab{b}})\citenamefont {Abbott} \emph {et~al.}}]{Abbott2016RateBBH}%
  \BibitemOpen
  \bibfield  {author} {\bibinfo {author} {\bibfnamefont {B.~P.}\ \bibnamefont {Abbott}} \emph {et~al.} (\bibinfo {collaboration} {The LIGO Scientific Collaboration and Virgo Collaboration}),\ }\bibfield  {title} {\bibinfo {title} {{The rate of binary black hole mergers inferred from Advanced LIGO observations surrounding GW150914}},\ }\href {https://doi.org/10.3847/2041-8205/833/1/L1} {\bibfield  {journal} {\bibinfo  {journal} {Astrophysical Journal Letters}\ }\textbf {\bibinfo {volume} {833}},\ \bibinfo {pages} {L1} (\bibinfo {year} {2016}{\natexlab{b}})}\BibitemShut {NoStop}%
\bibitem [{\citenamefont {Abbott}\ \emph {et~al.}(2023{\natexlab{c}})\citenamefont {Abbott} \emph {et~al.}}]{Abbott2021PopPropsGWTC2}%
  \BibitemOpen
  \bibfield  {author} {\bibinfo {author} {\bibfnamefont {R.}~\bibnamefont {Abbott}} \emph {et~al.} (\bibinfo {collaboration} {The LIGO Scientific Collaboration and Virgo Collaboration}),\ }\bibfield  {title} {\bibinfo {title} {{Population Properties of Compact Binaries Observed by LIGO and Virgo During the First Half of the Third Observing Run}},\ }\href {https://doi.org/10.3847/2041-8213/abbe9a} {\bibfield  {journal} {\bibinfo  {journal} {Astrophysical Journal Letters}\ }\textbf {\bibinfo {volume} {951}},\ \bibinfo {pages} {L7} (\bibinfo {year} {2023}{\natexlab{c}})}\BibitemShut {NoStop}%
\bibitem [{\citenamefont {Tiwari}\ and\ \citenamefont {Fairhurst}(2021)}]{Fairhurst2023MassParam}%
  \BibitemOpen
  \bibfield  {author} {\bibinfo {author} {\bibfnamefont {V.}~\bibnamefont {Tiwari}}\ and\ \bibinfo {author} {\bibfnamefont {S.}~\bibnamefont {Fairhurst}},\ }\bibfield  {title} {\bibinfo {title} {What’s in a binary black hole’s mass parameter?},\ }\href {https://doi.org/10.3847/2041-8213/abfbe7} {\bibfield  {journal} {\bibinfo  {journal} {Astrophys. J. Lett.}\ }\textbf {\bibinfo {volume} {918}},\ \bibinfo {pages} {L31} (\bibinfo {year} {2021})}\BibitemShut {NoStop}%
\bibitem [{\citenamefont {Abbott}\ \emph {et~al.}(2021{\natexlab{c}})\citenamefont {Abbott} \emph {et~al.}}]{LIGO2021CosmicStrings}%
  \BibitemOpen
  \bibfield  {author} {\bibinfo {author} {\bibfnamefont {R.}~\bibnamefont {Abbott}} \emph {et~al.} (\bibinfo {collaboration} {The LIGO Scientific Collaboration, Virgo Collaboration and KAGRA Collaboration}),\ }\bibfield  {title} {\bibinfo {title} {{Constraints on Cosmic Strings Using Data from the Third Advanced LIGO‑Virgo Observing Run}},\ }\href {https://doi.org/10.1103/PhysRevLett.126.241102} {\bibfield  {journal} {\bibinfo  {journal} {Physical Review Letters}\ }\textbf {\bibinfo {volume} {126}},\ \bibinfo {pages} {241102} (\bibinfo {year} {2021}{\natexlab{c}})}\BibitemShut {NoStop}%
\bibitem [{\citenamefont {Abbott}\ \emph {et~al.}(2022)\citenamefont {Abbott} \emph {et~al.}}]{Abbott2022DarkPhoton}%
  \BibitemOpen
  \bibfield  {author} {\bibinfo {author} {\bibfnamefont {R.}~\bibnamefont {Abbott}} \emph {et~al.} (\bibinfo {collaboration} {The LIGO Scientific Collaboration, Virgo Collaboration and KAGRA Collaboration}),\ }\bibfield  {title} {\bibinfo {title} {{Constraints on dark photon dark matter using data from LIGO’s and Virgo’s third observing run}},\ }\href {https://doi.org/10.1103/PhysRevD.105.063030} {\bibfield  {journal} {\bibinfo  {journal} {Physical Review D}\ }\textbf {\bibinfo {volume} {105}},\ \bibinfo {pages} {063030} (\bibinfo {year} {2022})}\BibitemShut {NoStop}%
\bibitem [{\citenamefont {Aurrekoetxea}\ \emph {et~al.}(2024)\citenamefont {Aurrekoetxea}, \citenamefont {Hoy},\ and\ \citenamefont {Hannam}}]{Aurrekoetxea2024CosmicStringGW190521}%
  \BibitemOpen
  \bibfield  {author} {\bibinfo {author} {\bibfnamefont {J.~C.}\ \bibnamefont {Aurrekoetxea}}, \bibinfo {author} {\bibfnamefont {C.}~\bibnamefont {Hoy}},\ and\ \bibinfo {author} {\bibfnamefont {M.}~\bibnamefont {Hannam}},\ }\bibfield  {title} {\bibinfo {title} {{Revisiting the Cosmic String Origin of GW190521}},\ }\href {https://doi.org/10.1103/PhysRevLett.132.181401} {\bibfield  {journal} {\bibinfo  {journal} {Physical Review Letters}\ }\textbf {\bibinfo {volume} {132}},\ \bibinfo {pages} {181401} (\bibinfo {year} {2024})}\BibitemShut {NoStop}%
\bibitem [{\citenamefont {Christensen}\ and\ \citenamefont {Meyer}(2022)}]{Christensen2022RMP_PE1}%
  \BibitemOpen
  \bibfield  {author} {\bibinfo {author} {\bibfnamefont {N.}~\bibnamefont {Christensen}}\ and\ \bibinfo {author} {\bibfnamefont {R.}~\bibnamefont {Meyer}},\ }\bibfield  {title} {\bibinfo {title} {Parameter estimation with gravitational waves},\ }\href {https://doi.org/10.1103/RevModPhys.94.025001} {\bibfield  {journal} {\bibinfo  {journal} {Reviews of Modern Physics}\ }\textbf {\bibinfo {volume} {94}},\ \bibinfo {pages} {025001} (\bibinfo {year} {2022})}\BibitemShut {NoStop}%
\bibitem [{\citenamefont {Krishna}\ \emph {et~al.}(2023)\citenamefont {Krishna}, \citenamefont {Vijaykumar}, \citenamefont {Ganguly}, \citenamefont {Talbot}, \citenamefont {Biscoveanu}, \citenamefont {George}, \citenamefont {Williams},\ and\ \citenamefont {Zimmerman}}]{Krishna:2023bug}%
  \BibitemOpen
  \bibfield  {author} {\bibinfo {author} {\bibfnamefont {K.}~\bibnamefont {Krishna}}, \bibinfo {author} {\bibfnamefont {A.}~\bibnamefont {Vijaykumar}}, \bibinfo {author} {\bibfnamefont {A.}~\bibnamefont {Ganguly}}, \bibinfo {author} {\bibfnamefont {C.}~\bibnamefont {Talbot}}, \bibinfo {author} {\bibfnamefont {S.}~\bibnamefont {Biscoveanu}}, \bibinfo {author} {\bibfnamefont {R.~N.}\ \bibnamefont {George}}, \bibinfo {author} {\bibfnamefont {N.}~\bibnamefont {Williams}},\ and\ \bibinfo {author} {\bibfnamefont {A.}~\bibnamefont {Zimmerman}},\ }\bibfield  {title} {\bibinfo {title} {{Accelerated parameter estimation in Bilby with relative binning}},\ }\href@noop {} {\bibfield  {journal} {\bibinfo  {journal} {arXiv preprint}\ } (\bibinfo {year} {2023})},\ \Eprint {https://arxiv.org/abs/2312.06009} {arXiv:2312.06009 [gr-qc]} \BibitemShut {NoStop}%
\bibitem [{\citenamefont {Morisaki}\ \emph {et~al.}(2023)\citenamefont {Morisaki}, \citenamefont {Smith}, \citenamefont {Tsukada}, \citenamefont {Sachdev}, \citenamefont {Stevenson}, \citenamefont {Talbot},\ and\ \citenamefont {Zimmerman}}]{Morisaki:2023kuq}%
  \BibitemOpen
  \bibfield  {author} {\bibinfo {author} {\bibfnamefont {S.}~\bibnamefont {Morisaki}}, \bibinfo {author} {\bibfnamefont {R.}~\bibnamefont {Smith}}, \bibinfo {author} {\bibfnamefont {L.}~\bibnamefont {Tsukada}}, \bibinfo {author} {\bibfnamefont {S.}~\bibnamefont {Sachdev}}, \bibinfo {author} {\bibfnamefont {S.}~\bibnamefont {Stevenson}}, \bibinfo {author} {\bibfnamefont {C.}~\bibnamefont {Talbot}},\ and\ \bibinfo {author} {\bibfnamefont {A.}~\bibnamefont {Zimmerman}},\ }\bibfield  {title} {\bibinfo {title} {{Rapid localization and inference on compact binary coalescences with the Advanced LIGO-Virgo-KAGRA gravitational-wave detector network}},\ }\href {https://doi.org/10.1103/PhysRevD.108.123040} {\bibfield  {journal} {\bibinfo  {journal} {Phys. Rev. D}\ }\textbf {\bibinfo {volume} {108}},\ \bibinfo {pages} {123040} (\bibinfo {year} {2023})},\ \Eprint {https://arxiv.org/abs/2307.13380} {arXiv:2307.13380 [gr-qc]} \BibitemShut {NoStop}%
\bibitem [{\citenamefont {Dax}\ \emph {et~al.}(2025{\natexlab{a}})\citenamefont {Dax}, \citenamefont {Green}, \citenamefont {Gair}, \citenamefont {Gupte}, \citenamefont {P\"urrer}, \citenamefont {Raymond}, \citenamefont {Wildberger}, \citenamefont {Macke}, \citenamefont {Buonanno},\ and\ \citenamefont {Sch\"olkopf}}]{Dax2025RealTimeBNS}%
  \BibitemOpen
  \bibfield  {author} {\bibinfo {author} {\bibfnamefont {M.}~\bibnamefont {Dax}}, \bibinfo {author} {\bibfnamefont {S.~R.}\ \bibnamefont {Green}}, \bibinfo {author} {\bibfnamefont {J.}~\bibnamefont {Gair}}, \bibinfo {author} {\bibfnamefont {N.}~\bibnamefont {Gupte}}, \bibinfo {author} {\bibfnamefont {M.}~\bibnamefont {P\"urrer}}, \bibinfo {author} {\bibfnamefont {V.}~\bibnamefont {Raymond}}, \bibinfo {author} {\bibfnamefont {J.}~\bibnamefont {Wildberger}}, \bibinfo {author} {\bibfnamefont {J.~H.}\ \bibnamefont {Macke}}, \bibinfo {author} {\bibfnamefont {A.}~\bibnamefont {Buonanno}},\ and\ \bibinfo {author} {\bibfnamefont {B.}~\bibnamefont {Sch\"olkopf}},\ }\bibfield  {title} {\bibinfo {title} {{Real‑time inference for binary neutron star mergers using machine learning}},\ }\href {https://doi.org/10.1038/s41586-025-08593-z} {\bibfield  {journal} {\bibinfo  {journal} {Nature}\ }\textbf {\bibinfo {volume} {639}},\ \bibinfo {pages} {49–53} (\bibinfo {year} {2025}{\natexlab{a}})}\BibitemShut {NoStop}%
\bibitem [{\citenamefont {Green}\ and\ \citenamefont {Gair}(2021)}]{Green_2021}%
  \BibitemOpen
  \bibfield  {author} {\bibinfo {author} {\bibfnamefont {S.~R.}\ \bibnamefont {Green}}\ and\ \bibinfo {author} {\bibfnamefont {J.}~\bibnamefont {Gair}},\ }\bibfield  {title} {\bibinfo {title} {{Complete parameter inference for GW150914 using deep learning}},\ }\href {https://doi.org/10.1088/2632-2153/abfaed} {\bibfield  {journal} {\bibinfo  {journal} {Machine Learning: Science and Technology}\ }\textbf {\bibinfo {volume} {2}},\ \bibinfo {pages} {03LT01} (\bibinfo {year} {2021})}\BibitemShut {NoStop}%
\bibitem [{\citenamefont {Mushkin}\ \emph {et~al.}(2025)\citenamefont {Mushkin}, \citenamefont {Roulet}, \citenamefont {Zackay}, \citenamefont {Venumadhav}, \citenamefont {Ivashtenko}, \citenamefont {Wadekar}, \citenamefont {Mehta},\ and\ \citenamefont {Zaldarriaga}}]{Mushkin2025dotPE}%
  \BibitemOpen
  \bibfield  {author} {\bibinfo {author} {\bibfnamefont {J.}~\bibnamefont {Mushkin}}, \bibinfo {author} {\bibfnamefont {J.}~\bibnamefont {Roulet}}, \bibinfo {author} {\bibfnamefont {B.}~\bibnamefont {Zackay}}, \bibinfo {author} {\bibfnamefont {T.}~\bibnamefont {Venumadhav}}, \bibinfo {author} {\bibfnamefont {O.}~\bibnamefont {Ivashtenko}}, \bibinfo {author} {\bibfnamefont {D.}~\bibnamefont {Wadekar}}, \bibinfo {author} {\bibfnamefont {A.~K.}\ \bibnamefont {Mehta}},\ and\ \bibinfo {author} {\bibfnamefont {M.}~\bibnamefont {Zaldarriaga}},\ }\bibfield  {title} {\bibinfo {title} {{dot‑PE: Sampler‑free gravitational wave inference using matrix multiplication}},\ }\bibfield  {journal} {\bibinfo  {journal} {arXiv preprint}\ }\href {https://doi.org/10.48550/arXiv.2507.16022} {10.48550/arXiv.2507.16022} (\bibinfo {year} {2025}),\ \bibinfo {note} {arXiv:2507.16022}\BibitemShut {NoStop}%
\bibitem [{\citenamefont {Mandel}\ \emph {et~al.}(2019)\citenamefont {Mandel}, \citenamefont {Farr},\ and\ \citenamefont {Gair}}]{Mandel_2019}%
  \BibitemOpen
  \bibfield  {author} {\bibinfo {author} {\bibfnamefont {I.}~\bibnamefont {Mandel}}, \bibinfo {author} {\bibfnamefont {W.~M.}\ \bibnamefont {Farr}},\ and\ \bibinfo {author} {\bibfnamefont {J.~R.}\ \bibnamefont {Gair}},\ }\bibfield  {title} {\bibinfo {title} {Extracting distribution parameters from multiple uncertain observations with selection biases},\ }\href {https://doi.org/10.1093/mnras/stz896} {\bibfield  {journal} {\bibinfo  {journal} {Monthly Notices of the Royal Astronomical Society}\ }\textbf {\bibinfo {volume} {486}},\ \bibinfo {pages} {1086–1093} (\bibinfo {year} {2019})}\BibitemShut {NoStop}%
\bibitem [{\citenamefont {Veitch}\ \emph {et~al.}(2015{\natexlab{a}})\citenamefont {Veitch}, \citenamefont {Raymond}, \citenamefont {Farr}, \citenamefont {Farr}, \citenamefont {Graff}, \citenamefont {Vitale}, \citenamefont {et~al. (LIGO Scientific~Collaboration},\ and\ \citenamefont {Collaboration)}}]{Veitch2015LALInference}%
  \BibitemOpen
  \bibfield  {author} {\bibinfo {author} {\bibfnamefont {J.}~\bibnamefont {Veitch}}, \bibinfo {author} {\bibfnamefont {V.}~\bibnamefont {Raymond}}, \bibinfo {author} {\bibfnamefont {W.~M.}\ \bibnamefont {Farr}}, \bibinfo {author} {\bibfnamefont {B.}~\bibnamefont {Farr}}, \bibinfo {author} {\bibfnamefont {P.}~\bibnamefont {Graff}}, \bibinfo {author} {\bibfnamefont {S.}~\bibnamefont {Vitale}}, \bibinfo {author} {\bibnamefont {et~al. (LIGO Scientific~Collaboration}},\ and\ \bibinfo {author} {\bibfnamefont {V.}~\bibnamefont {Collaboration)}},\ }\bibfield  {title} {\bibinfo {title} {Parameter estimation for compact binaries with ground‑based gravitational‑wave observations using the lalinference software library},\ }\href {https://doi.org/10.1103/PhysRevD.91.042003} {\bibfield  {journal} {\bibinfo  {journal} {Physical Review D}\ }\textbf {\bibinfo {volume} {91}},\ \bibinfo {pages} {042003} (\bibinfo {year} {2015}{\natexlab{a}})}\BibitemShut {NoStop}%
\bibitem [{\citenamefont {Cornish}(2010)}]{Cornish:2010kf}%
  \BibitemOpen
  \bibfield  {author} {\bibinfo {author} {\bibfnamefont {N.~J.}\ \bibnamefont {Cornish}},\ }\bibfield  {title} {\bibinfo {title} {{Fast Fisher Matrices and Lazy Likelihoods}},\ }\href@noop {} {\bibfield  {journal} {\bibinfo  {journal} {arXiv preprint}\ } (\bibinfo {year} {2010})},\ \Eprint {https://arxiv.org/abs/1007.4820} {arXiv:1007.4820 [gr-qc]} \BibitemShut {NoStop}%
\bibitem [{\citenamefont {Zackay}\ \emph {et~al.}(2018)\citenamefont {Zackay}, \citenamefont {Dai},\ and\ \citenamefont {Venumadhav}}]{Zackay:2018qdy}%
  \BibitemOpen
  \bibfield  {author} {\bibinfo {author} {\bibfnamefont {B.}~\bibnamefont {Zackay}}, \bibinfo {author} {\bibfnamefont {L.}~\bibnamefont {Dai}},\ and\ \bibinfo {author} {\bibfnamefont {T.}~\bibnamefont {Venumadhav}},\ }\bibfield  {title} {\bibinfo {title} {{Relative Binning and Fast Likelihood Evaluation for Gravitational Wave Parameter Estimation}},\ }\href@noop {} {\bibfield  {journal} {\bibinfo  {journal} {arXiv preprint}\ } (\bibinfo {year} {2018})},\ \Eprint {https://arxiv.org/abs/1806.08792} {arXiv:1806.08792 [astro-ph.IM]} \BibitemShut {NoStop}%
\bibitem [{\citenamefont {Cornish}(2021)}]{Cornish:2021lje}%
  \BibitemOpen
  \bibfield  {author} {\bibinfo {author} {\bibfnamefont {N.~J.}\ \bibnamefont {Cornish}},\ }\bibfield  {title} {\bibinfo {title} {{Heterodyned likelihood for rapid gravitational wave parameter inference}},\ }\href {https://doi.org/10.1103/PhysRevD.104.104054} {\bibfield  {journal} {\bibinfo  {journal} {Phys. Rev. D}\ }\textbf {\bibinfo {volume} {104}},\ \bibinfo {pages} {104054} (\bibinfo {year} {2021})},\ \Eprint {https://arxiv.org/abs/2109.02728} {arXiv:2109.02728 [gr-qc]} \BibitemShut {NoStop}%
\bibitem [{\citenamefont {Leslie}\ \emph {et~al.}(2021)\citenamefont {Leslie}, \citenamefont {Dai},\ and\ \citenamefont {Pratten}}]{Leslie:2021ssu}%
  \BibitemOpen
  \bibfield  {author} {\bibinfo {author} {\bibfnamefont {N.}~\bibnamefont {Leslie}}, \bibinfo {author} {\bibfnamefont {L.}~\bibnamefont {Dai}},\ and\ \bibinfo {author} {\bibfnamefont {G.}~\bibnamefont {Pratten}},\ }\bibfield  {title} {\bibinfo {title} {{Mode-by-mode relative binning: Fast likelihood estimation for gravitational waveforms with spin-orbit precession and multiple harmonics}},\ }\href {https://doi.org/10.1103/PhysRevD.104.123030} {\bibfield  {journal} {\bibinfo  {journal} {Phys. Rev. D}\ }\textbf {\bibinfo {volume} {104}},\ \bibinfo {pages} {123030} (\bibinfo {year} {2021})},\ \Eprint {https://arxiv.org/abs/2109.09872} {arXiv:2109.09872 [astro-ph.IM]} \BibitemShut {NoStop}%
\bibitem [{\citenamefont {Narola}\ \emph {et~al.}(2024)\citenamefont {Narola}, \citenamefont {Janquart}, \citenamefont {Meijer}, \citenamefont {Haris},\ and\ \citenamefont {Van Den~Broeck}}]{Narola:2023men}%
  \BibitemOpen
  \bibfield  {author} {\bibinfo {author} {\bibfnamefont {H.}~\bibnamefont {Narola}}, \bibinfo {author} {\bibfnamefont {J.}~\bibnamefont {Janquart}}, \bibinfo {author} {\bibfnamefont {Q.}~\bibnamefont {Meijer}}, \bibinfo {author} {\bibfnamefont {K.}~\bibnamefont {Haris}},\ and\ \bibinfo {author} {\bibfnamefont {C.}~\bibnamefont {Van Den~Broeck}},\ }\bibfield  {title} {\bibinfo {title} {{Gravitational-wave parameter estimation with relative binning: Inclusion of higher-order modes and precession, and applications to lensing and third-generation detectors}},\ }\href {https://doi.org/10.1103/PhysRevD.110.084085} {\bibfield  {journal} {\bibinfo  {journal} {Phys. Rev. D}\ }\textbf {\bibinfo {volume} {110}},\ \bibinfo {pages} {084085} (\bibinfo {year} {2024})},\ \Eprint {https://arxiv.org/abs/2308.12140} {arXiv:2308.12140 [gr-qc]} \BibitemShut {NoStop}%
\bibitem [{\citenamefont {Canizares}\ \emph {et~al.}(2015)\citenamefont {Canizares}, \citenamefont {Field}, \citenamefont {Gair}, \citenamefont {Raymond}, \citenamefont {Smith},\ and\ \citenamefont {Tiglio}}]{Canizares:2014fya}%
  \BibitemOpen
  \bibfield  {author} {\bibinfo {author} {\bibfnamefont {P.}~\bibnamefont {Canizares}}, \bibinfo {author} {\bibfnamefont {S.~E.}\ \bibnamefont {Field}}, \bibinfo {author} {\bibfnamefont {J.}~\bibnamefont {Gair}}, \bibinfo {author} {\bibfnamefont {V.}~\bibnamefont {Raymond}}, \bibinfo {author} {\bibfnamefont {R.}~\bibnamefont {Smith}},\ and\ \bibinfo {author} {\bibfnamefont {M.}~\bibnamefont {Tiglio}},\ }\bibfield  {title} {\bibinfo {title} {{Accelerated gravitational-wave parameter estimation with reduced order modeling}},\ }\href {https://doi.org/10.1103/PhysRevLett.114.071104} {\bibfield  {journal} {\bibinfo  {journal} {Phys. Rev. Lett.}\ }\textbf {\bibinfo {volume} {114}},\ \bibinfo {pages} {071104} (\bibinfo {year} {2015})},\ \Eprint {https://arxiv.org/abs/1404.6284} {arXiv:1404.6284 [gr-qc]} \BibitemShut {NoStop}%
\bibitem [{\citenamefont {Smith}\ \emph {et~al.}(2016)\citenamefont {Smith}, \citenamefont {Field}, \citenamefont {Blackburn}, \citenamefont {Haster}, \citenamefont {P\"urrer}, \citenamefont {Raymond},\ and\ \citenamefont {Schmidt}}]{Smith:2016qas}%
  \BibitemOpen
  \bibfield  {author} {\bibinfo {author} {\bibfnamefont {R.}~\bibnamefont {Smith}}, \bibinfo {author} {\bibfnamefont {S.~E.}\ \bibnamefont {Field}}, \bibinfo {author} {\bibfnamefont {K.}~\bibnamefont {Blackburn}}, \bibinfo {author} {\bibfnamefont {C.-J.}\ \bibnamefont {Haster}}, \bibinfo {author} {\bibfnamefont {M.}~\bibnamefont {P\"urrer}}, \bibinfo {author} {\bibfnamefont {V.}~\bibnamefont {Raymond}},\ and\ \bibinfo {author} {\bibfnamefont {P.}~\bibnamefont {Schmidt}},\ }\bibfield  {title} {\bibinfo {title} {{Fast and accurate inference on gravitational waves from precessing compact binaries}},\ }\href {https://doi.org/10.1103/PhysRevD.94.044031} {\bibfield  {journal} {\bibinfo  {journal} {Phys. Rev. D}\ }\textbf {\bibinfo {volume} {94}},\ \bibinfo {pages} {044031} (\bibinfo {year} {2016})},\ \Eprint {https://arxiv.org/abs/1604.08253} {arXiv:1604.08253 [gr-qc]} \BibitemShut {NoStop}%
\bibitem [{\citenamefont {Morisaki}\ and\ \citenamefont {Raymond}(2020)}]{Morisaki:2020oqk}%
  \BibitemOpen
  \bibfield  {author} {\bibinfo {author} {\bibfnamefont {S.}~\bibnamefont {Morisaki}}\ and\ \bibinfo {author} {\bibfnamefont {V.}~\bibnamefont {Raymond}},\ }\bibfield  {title} {\bibinfo {title} {{Rapid Parameter Estimation of Gravitational Waves from Binary Neutron Star Coalescence using Focused Reduced Order Quadrature}},\ }\href {https://doi.org/10.1103/PhysRevD.102.104020} {\bibfield  {journal} {\bibinfo  {journal} {Phys. Rev. D}\ }\textbf {\bibinfo {volume} {102}},\ \bibinfo {pages} {104020} (\bibinfo {year} {2020})},\ \Eprint {https://arxiv.org/abs/2007.09108} {arXiv:2007.09108 [gr-qc]} \BibitemShut {NoStop}%
\bibitem [{\citenamefont {Smith}\ \emph {et~al.}(2021)\citenamefont {Smith} \emph {et~al.}}]{Smith:2021bqc}%
  \BibitemOpen
  \bibfield  {author} {\bibinfo {author} {\bibfnamefont {R.}~\bibnamefont {Smith}} \emph {et~al.},\ }\bibfield  {title} {\bibinfo {title} {{Bayesian Inference for Gravitational Waves from Binary Neutron Star Mergers in Third Generation Observatories}},\ }\href {https://doi.org/10.1103/PhysRevLett.127.081102} {\bibfield  {journal} {\bibinfo  {journal} {Phys. Rev. Lett.}\ }\textbf {\bibinfo {volume} {127}},\ \bibinfo {pages} {081102} (\bibinfo {year} {2021})},\ \Eprint {https://arxiv.org/abs/2103.12274} {arXiv:2103.12274 [gr-qc]} \BibitemShut {NoStop}%
\bibitem [{\citenamefont {Vinciguerra}\ \emph {et~al.}(2017)\citenamefont {Vinciguerra}, \citenamefont {Veitch},\ and\ \citenamefont {Mandel}}]{Vinciguerra:2017ngf}%
  \BibitemOpen
  \bibfield  {author} {\bibinfo {author} {\bibfnamefont {S.}~\bibnamefont {Vinciguerra}}, \bibinfo {author} {\bibfnamefont {J.}~\bibnamefont {Veitch}},\ and\ \bibinfo {author} {\bibfnamefont {I.}~\bibnamefont {Mandel}},\ }\bibfield  {title} {\bibinfo {title} {{Accelerating gravitational wave parameter estimation with multi-band template interpolation}},\ }\href {https://doi.org/10.1088/1361-6382/aa6d44} {\bibfield  {journal} {\bibinfo  {journal} {Class. Quant. Grav.}\ }\textbf {\bibinfo {volume} {34}},\ \bibinfo {pages} {115006} (\bibinfo {year} {2017})},\ \Eprint {https://arxiv.org/abs/1703.02062} {arXiv:1703.02062 [gr-qc]} \BibitemShut {NoStop}%
\bibitem [{\citenamefont {Morisaki}(2021)}]{Morisaki:2021ngj}%
  \BibitemOpen
  \bibfield  {author} {\bibinfo {author} {\bibfnamefont {S.}~\bibnamefont {Morisaki}},\ }\bibfield  {title} {\bibinfo {title} {{Accelerating parameter estimation of gravitational waves from compact binary coalescence using adaptive frequency resolutions}},\ }\href {https://doi.org/10.1103/PhysRevD.104.044062} {\bibfield  {journal} {\bibinfo  {journal} {Phys. Rev. D}\ }\textbf {\bibinfo {volume} {104}},\ \bibinfo {pages} {044062} (\bibinfo {year} {2021})},\ \Eprint {https://arxiv.org/abs/2104.07813} {arXiv:2104.07813 [gr-qc]} \BibitemShut {NoStop}%
\bibitem [{\citenamefont {Pankow}\ \emph {et~al.}(2015)\citenamefont {Pankow}, \citenamefont {Brady}, \citenamefont {Ochsner},\ and\ \citenamefont {O'Shaughnessy}}]{Pankow:2015cra}%
  \BibitemOpen
  \bibfield  {author} {\bibinfo {author} {\bibfnamefont {C.}~\bibnamefont {Pankow}}, \bibinfo {author} {\bibfnamefont {P.}~\bibnamefont {Brady}}, \bibinfo {author} {\bibfnamefont {E.}~\bibnamefont {Ochsner}},\ and\ \bibinfo {author} {\bibfnamefont {R.}~\bibnamefont {O'Shaughnessy}},\ }\bibfield  {title} {\bibinfo {title} {{Novel scheme for rapid parallel parameter estimation of gravitational waves from compact binary coalescences}},\ }\href {https://doi.org/10.1103/PhysRevD.92.023002} {\bibfield  {journal} {\bibinfo  {journal} {Phys. Rev. D}\ }\textbf {\bibinfo {volume} {92}},\ \bibinfo {pages} {023002} (\bibinfo {year} {2015})},\ \Eprint {https://arxiv.org/abs/1502.04370} {arXiv:1502.04370 [gr-qc]} \BibitemShut {NoStop}%
\bibitem [{\citenamefont {Lange}\ \emph {et~al.}(2018)\citenamefont {Lange}, \citenamefont {O'Shaughnessy},\ and\ \citenamefont {Rizzo}}]{Lange:2018pyp}%
  \BibitemOpen
  \bibfield  {author} {\bibinfo {author} {\bibfnamefont {J.}~\bibnamefont {Lange}}, \bibinfo {author} {\bibfnamefont {R.}~\bibnamefont {O'Shaughnessy}},\ and\ \bibinfo {author} {\bibfnamefont {M.}~\bibnamefont {Rizzo}},\ }\bibfield  {title} {\bibinfo {title} {{Rapid and accurate parameter inference for coalescing, precessing compact binaries}},\ }\href@noop {} {\bibfield  {journal} {\bibinfo  {journal} {arXiv preprint}\ } (\bibinfo {year} {2018})},\ \Eprint {https://arxiv.org/abs/1805.10457} {arXiv:1805.10457 [gr-qc]} \BibitemShut {NoStop}%
\bibitem [{\citenamefont {Wagner}\ \emph {et~al.}(2025)\citenamefont {Wagner}, \citenamefont {O'Shaughnessy}, \citenamefont {Yelikar}, \citenamefont {Manning}, \citenamefont {Fernando}, \citenamefont {Lange}, \citenamefont {Tiwari}, \citenamefont {Fernando},\ and\ \citenamefont {Williams}}]{Wagner:2025bih}%
  \BibitemOpen
  \bibfield  {author} {\bibinfo {author} {\bibfnamefont {K.~J.}\ \bibnamefont {Wagner}}, \bibinfo {author} {\bibfnamefont {R.}~\bibnamefont {O'Shaughnessy}}, \bibinfo {author} {\bibfnamefont {A.}~\bibnamefont {Yelikar}}, \bibinfo {author} {\bibfnamefont {N.}~\bibnamefont {Manning}}, \bibinfo {author} {\bibfnamefont {D.}~\bibnamefont {Fernando}}, \bibinfo {author} {\bibfnamefont {J.}~\bibnamefont {Lange}}, \bibinfo {author} {\bibfnamefont {V.}~\bibnamefont {Tiwari}}, \bibinfo {author} {\bibfnamefont {A.}~\bibnamefont {Fernando}},\ and\ \bibinfo {author} {\bibfnamefont {D.}~\bibnamefont {Williams}},\ }\bibfield  {title} {\bibinfo {title} {{Narrowing RIFT: Focused simulation-based-inference for interpreting exceptional GW sources}},\ }\href@noop {} {\bibfield  {journal} {\bibinfo  {journal} {arXiv preprint}\ } (\bibinfo {year} {2025})},\ \Eprint {https://arxiv.org/abs/2505.11655} {arXiv:2505.11655 [astro-ph.IM]} \BibitemShut {NoStop}%
\bibitem [{\citenamefont {Veitch}\ and\ \citenamefont {Del~Pozzo}(2013)}]{Veitch2013_T1300326}%
  \BibitemOpen
  \bibfield  {author} {\bibinfo {author} {\bibfnamefont {J.}~\bibnamefont {Veitch}}\ and\ \bibinfo {author} {\bibfnamefont {W.}~\bibnamefont {Del~Pozzo}},\ }\href {https://dcc.ligo.org/public/0102/T1300326/001/margphi.pdf} {\emph {\bibinfo {title} {{Analytic marginalisation of the phase parameter in gravitational-wave parameter estimation}}}},\ \bibinfo {type} {Tech. Rep.}\ \bibinfo {number} {LIGO-T1300326}\ (\bibinfo  {institution} {LIGO Scientific Collaboration},\ \bibinfo {year} {2013})\BibitemShut {NoStop}%
\bibitem [{\citenamefont {Farr}(2014)}]{Farr2014_T1400460}%
  \BibitemOpen
  \bibfield  {author} {\bibinfo {author} {\bibfnamefont {W.~M.}\ \bibnamefont {Farr}},\ }\href {https://dcc.ligo.org/public/0114/T1400460/002/margtime.pdf} {\emph {\bibinfo {title} {{Marginalisation of the time (and phase) parameter in gravitational-wave parameter estimation}}}},\ \bibinfo {type} {Tech. Rep.}\ \bibinfo {number} {LIGO-T1400460}\ (\bibinfo  {institution} {LIGO Scientific Collaboration},\ \bibinfo {year} {2014})\BibitemShut {NoStop}%
\bibitem [{\citenamefont {Singer}\ and\ \citenamefont {Price}(2016)}]{Singer:2015ema}%
  \BibitemOpen
  \bibfield  {author} {\bibinfo {author} {\bibfnamefont {L.~P.}\ \bibnamefont {Singer}}\ and\ \bibinfo {author} {\bibfnamefont {L.~R.}\ \bibnamefont {Price}},\ }\bibfield  {title} {\bibinfo {title} {{Rapid Bayesian position reconstruction for gravitational-wave transients}},\ }\href {https://doi.org/10.1103/PhysRevD.93.024013} {\bibfield  {journal} {\bibinfo  {journal} {Phys. Rev. D}\ }\textbf {\bibinfo {volume} {93}},\ \bibinfo {pages} {024013} (\bibinfo {year} {2016})},\ \Eprint {https://arxiv.org/abs/1508.03634} {arXiv:1508.03634 [gr-qc]} \BibitemShut {NoStop}%
\bibitem [{\citenamefont {Islam}\ \emph {et~al.}(2022)\citenamefont {Islam}, \citenamefont {Roulet},\ and\ \citenamefont {Venumadhav}}]{Islam:2022afg}%
  \BibitemOpen
  \bibfield  {author} {\bibinfo {author} {\bibfnamefont {T.}~\bibnamefont {Islam}}, \bibinfo {author} {\bibfnamefont {J.}~\bibnamefont {Roulet}},\ and\ \bibinfo {author} {\bibfnamefont {T.}~\bibnamefont {Venumadhav}},\ }\bibfield  {title} {\bibinfo {title} {{Factorized Parameter Estimation for Real-Time Gravitational Wave Inference}},\ }\href@noop {} {\bibfield  {journal} {\bibinfo  {journal} {arXiv preprint}\ } (\bibinfo {year} {2022})},\ \Eprint {https://arxiv.org/abs/2210.16278} {arXiv:2210.16278 [gr-qc]} \BibitemShut {NoStop}%
\bibitem [{\citenamefont {Roulet}\ \emph {et~al.}(2024)\citenamefont {Roulet}, \citenamefont {Mushkin}, \citenamefont {Wadekar}, \citenamefont {Venumadhav}, \citenamefont {Zackay},\ and\ \citenamefont {Zaldarriaga}}]{Roulet:2024hwz}%
  \BibitemOpen
  \bibfield  {author} {\bibinfo {author} {\bibfnamefont {J.}~\bibnamefont {Roulet}}, \bibinfo {author} {\bibfnamefont {J.}~\bibnamefont {Mushkin}}, \bibinfo {author} {\bibfnamefont {D.}~\bibnamefont {Wadekar}}, \bibinfo {author} {\bibfnamefont {T.}~\bibnamefont {Venumadhav}}, \bibinfo {author} {\bibfnamefont {B.}~\bibnamefont {Zackay}},\ and\ \bibinfo {author} {\bibfnamefont {M.}~\bibnamefont {Zaldarriaga}},\ }\bibfield  {title} {\bibinfo {title} {{Fast marginalization algorithm for optimizing gravitational wave detection, parameter estimation, and sky localization}},\ }\href {https://doi.org/10.1103/PhysRevD.110.044010} {\bibfield  {journal} {\bibinfo  {journal} {Phys. Rev. D}\ }\textbf {\bibinfo {volume} {110}},\ \bibinfo {pages} {044010} (\bibinfo {year} {2024})},\ \Eprint {https://arxiv.org/abs/2404.02435} {arXiv:2404.02435 [gr-qc]} \BibitemShut {NoStop}%
\bibitem [{\citenamefont {Thrane}\ and\ \citenamefont {Talbot}(2019)}]{Thrane2019PASA}%
  \BibitemOpen
  \bibfield  {author} {\bibinfo {author} {\bibfnamefont {E.}~\bibnamefont {Thrane}}\ and\ \bibinfo {author} {\bibfnamefont {C.}~\bibnamefont {Talbot}},\ }\bibfield  {title} {\bibinfo {title} {{An introduction to Bayesian inference in gravitational‑wave astronomy: Parameter estimation, model selection, and hierarchical models}},\ }\href {https://doi.org/10.1017/pasa.2019.2} {\bibfield  {journal} {\bibinfo  {journal} {Publications of the Astronomical Society of Australia}\ }\textbf {\bibinfo {volume} {36}},\ \bibinfo {pages} {e010} (\bibinfo {year} {2019})}\BibitemShut {NoStop}%
\bibitem [{\citenamefont {Fairhurst}\ \emph {et~al.}(2023{\natexlab{a}})\citenamefont {Fairhurst}, \citenamefont {Hoy}, \citenamefont {Green}, \citenamefont {Mills},\ and\ \citenamefont {Usman}}]{Fairhurst:2023idl}%
  \BibitemOpen
  \bibfield  {author} {\bibinfo {author} {\bibfnamefont {S.}~\bibnamefont {Fairhurst}}, \bibinfo {author} {\bibfnamefont {C.}~\bibnamefont {Hoy}}, \bibinfo {author} {\bibfnamefont {R.}~\bibnamefont {Green}}, \bibinfo {author} {\bibfnamefont {C.}~\bibnamefont {Mills}},\ and\ \bibinfo {author} {\bibfnamefont {S.~A.}\ \bibnamefont {Usman}},\ }\bibfield  {title} {\bibinfo {title} {{Simple parameter estimation using observable features of gravitational-wave signals}},\ }\href {https://doi.org/10.1103/PhysRevD.108.082006} {\bibfield  {journal} {\bibinfo  {journal} {Phys. Rev. D}\ }\textbf {\bibinfo {volume} {108}},\ \bibinfo {pages} {082006} (\bibinfo {year} {2023}{\natexlab{a}})},\ \Eprint {https://arxiv.org/abs/2304.03731} {arXiv:2304.03731 [gr-qc]} \BibitemShut {NoStop}%
\bibitem [{\citenamefont {Williams}\ \emph {et~al.}(2021)\citenamefont {Williams}, \citenamefont {Veitch},\ and\ \citenamefont {Messenger}}]{Williams:2021qyt}%
  \BibitemOpen
  \bibfield  {author} {\bibinfo {author} {\bibfnamefont {M.~J.}\ \bibnamefont {Williams}}, \bibinfo {author} {\bibfnamefont {J.}~\bibnamefont {Veitch}},\ and\ \bibinfo {author} {\bibfnamefont {C.}~\bibnamefont {Messenger}},\ }\bibfield  {title} {\bibinfo {title} {{Nested sampling with normalizing flows for gravitational-wave inference}},\ }\href {https://doi.org/10.1103/PhysRevD.103.103006} {\bibfield  {journal} {\bibinfo  {journal} {Phys. Rev. D}\ }\textbf {\bibinfo {volume} {103}},\ \bibinfo {pages} {103006} (\bibinfo {year} {2021})},\ \Eprint {https://arxiv.org/abs/2102.11056} {arXiv:2102.11056 [gr-qc]} \BibitemShut {NoStop}%
\bibitem [{\citenamefont {Karamanis}\ \emph {et~al.}(2022)\citenamefont {Karamanis}, \citenamefont {Nabergoj}, \citenamefont {Beutler}, \citenamefont {Peacock},\ and\ \citenamefont {Seljak}}]{Karamanis:2022ksp}%
  \BibitemOpen
  \bibfield  {author} {\bibinfo {author} {\bibfnamefont {M.}~\bibnamefont {Karamanis}}, \bibinfo {author} {\bibfnamefont {D.}~\bibnamefont {Nabergoj}}, \bibinfo {author} {\bibfnamefont {F.}~\bibnamefont {Beutler}}, \bibinfo {author} {\bibfnamefont {J.~A.}\ \bibnamefont {Peacock}},\ and\ \bibinfo {author} {\bibfnamefont {U.}~\bibnamefont {Seljak}},\ }\bibfield  {title} {\bibinfo {title} {{pocoMC: A Python package for accelerated Bayesian inference in astronomy and cosmology}},\ }\href {https://doi.org/10.21105/joss.04634} {\bibfield  {journal} {\bibinfo  {journal} {J. Open Source Softw.}\ }\textbf {\bibinfo {volume} {7}},\ \bibinfo {pages} {4634} (\bibinfo {year} {2022})},\ \Eprint {https://arxiv.org/abs/2207.05660} {arXiv:2207.05660 [astro-ph.IM]} \BibitemShut {NoStop}%
\bibitem [{\citenamefont {Wong}\ \emph {et~al.}(2023)\citenamefont {Wong}, \citenamefont {Isi},\ and\ \citenamefont {Edwards}}]{Wong:2023lgb}%
  \BibitemOpen
  \bibfield  {author} {\bibinfo {author} {\bibfnamefont {K.~W.~K.}\ \bibnamefont {Wong}}, \bibinfo {author} {\bibfnamefont {M.}~\bibnamefont {Isi}},\ and\ \bibinfo {author} {\bibfnamefont {T.~D.~P.}\ \bibnamefont {Edwards}},\ }\bibfield  {title} {\bibinfo {title} {{Fast Gravitational-wave Parameter Estimation without Compromises}},\ }\href {https://doi.org/10.3847/1538-4357/acf5cd} {\bibfield  {journal} {\bibinfo  {journal} {Astrophys. J.}\ }\textbf {\bibinfo {volume} {958}},\ \bibinfo {pages} {129} (\bibinfo {year} {2023})},\ \Eprint {https://arxiv.org/abs/2302.05333} {arXiv:2302.05333 [astro-ph.IM]} \BibitemShut {NoStop}%
\bibitem [{\citenamefont {Tiwari}\ \emph {et~al.}(2023)\citenamefont {Tiwari}, \citenamefont {Hoy}, \citenamefont {Fairhurst},\ and\ \citenamefont {MacLeod}}]{Tiwari:2023mzf}%
  \BibitemOpen
  \bibfield  {author} {\bibinfo {author} {\bibfnamefont {V.}~\bibnamefont {Tiwari}}, \bibinfo {author} {\bibfnamefont {C.}~\bibnamefont {Hoy}}, \bibinfo {author} {\bibfnamefont {S.}~\bibnamefont {Fairhurst}},\ and\ \bibinfo {author} {\bibfnamefont {D.}~\bibnamefont {MacLeod}},\ }\bibfield  {title} {\bibinfo {title} {{Fast non-Markovian sampler for estimating gravitational-wave posteriors}},\ }\href {https://doi.org/10.1103/PhysRevD.108.023001} {\bibfield  {journal} {\bibinfo  {journal} {Phys. Rev. D}\ }\textbf {\bibinfo {volume} {108}},\ \bibinfo {pages} {023001} (\bibinfo {year} {2023})},\ \Eprint {https://arxiv.org/abs/2303.01463} {arXiv:2303.01463 [astro-ph.HE]} \BibitemShut {NoStop}%
\bibitem [{\citenamefont {Williams}\ \emph {et~al.}(2023)\citenamefont {Williams}, \citenamefont {Veitch},\ and\ \citenamefont {Messenger}}]{Williams:2023ppp}%
  \BibitemOpen
  \bibfield  {author} {\bibinfo {author} {\bibfnamefont {M.~J.}\ \bibnamefont {Williams}}, \bibinfo {author} {\bibfnamefont {J.}~\bibnamefont {Veitch}},\ and\ \bibinfo {author} {\bibfnamefont {C.}~\bibnamefont {Messenger}},\ }\bibfield  {title} {\bibinfo {title} {{Importance nested sampling with normalising flows}},\ }\href {https://doi.org/10.1088/2632-2153/acd5aa} {\bibfield  {journal} {\bibinfo  {journal} {Mach. Learn. Sci. Tech.}\ }\textbf {\bibinfo {volume} {4}},\ \bibinfo {pages} {035011} (\bibinfo {year} {2023})},\ \Eprint {https://arxiv.org/abs/2302.08526} {arXiv:2302.08526 [astro-ph.IM]} \BibitemShut {NoStop}%
\bibitem [{\citenamefont {Tiwari}(2024)}]{Tiwari:2024qzr}%
  \BibitemOpen
  \bibfield  {author} {\bibinfo {author} {\bibfnamefont {V.}~\bibnamefont {Tiwari}},\ }\bibfield  {title} {\bibinfo {title} {{Varaha: A promising sampler for obtaining gravitational wave posteriors}},\ }\href@noop {} {\bibfield  {journal} {\bibinfo  {journal} {arXiv preprint}\ } (\bibinfo {year} {2024})},\ \Eprint {https://arxiv.org/abs/2405.16568} {arXiv:2405.16568 [astro-ph.HE]} \BibitemShut {NoStop}%
\bibitem [{\citenamefont {Nitz}(2024)}]{Nitz:2024nhj}%
  \BibitemOpen
  \bibfield  {author} {\bibinfo {author} {\bibfnamefont {A.~H.}\ \bibnamefont {Nitz}},\ }\bibfield  {title} {\bibinfo {title} {{Robust, Rapid, and Simple Gravitational-wave Parameter Estimation}},\ }\href@noop {} {\bibfield  {journal} {\bibinfo  {journal} {arXiv preprint}\ } (\bibinfo {year} {2024})},\ \Eprint {https://arxiv.org/abs/2410.05190} {arXiv:2410.05190 [astro-ph.IM]} \BibitemShut {NoStop}%
\bibitem [{\citenamefont {Wouters}\ \emph {et~al.}(2024)\citenamefont {Wouters}, \citenamefont {Pang}, \citenamefont {Dietrich},\ and\ \citenamefont {Van Den~Broeck}}]{Wouters:2024oxj}%
  \BibitemOpen
  \bibfield  {author} {\bibinfo {author} {\bibfnamefont {T.}~\bibnamefont {Wouters}}, \bibinfo {author} {\bibfnamefont {P.~T.~H.}\ \bibnamefont {Pang}}, \bibinfo {author} {\bibfnamefont {T.}~\bibnamefont {Dietrich}},\ and\ \bibinfo {author} {\bibfnamefont {C.}~\bibnamefont {Van Den~Broeck}},\ }\bibfield  {title} {\bibinfo {title} {{Robust parameter estimation within minutes on gravitational wave signals from binary neutron star inspirals}},\ }\href@noop {} {\bibfield  {journal} {\bibinfo  {journal} {arXiv preprint}\ } (\bibinfo {year} {2024})},\ \Eprint {https://arxiv.org/abs/2404.11397} {arXiv:2404.11397 [astro-ph.IM]} \BibitemShut {NoStop}%
\bibitem [{\citenamefont {Williams}\ \emph {et~al.}(2025)\citenamefont {Williams}, \citenamefont {Karamanis}, \citenamefont {Luo},\ and\ \citenamefont {Seljak}}]{Williams:2025szm}%
  \BibitemOpen
  \bibfield  {author} {\bibinfo {author} {\bibfnamefont {M.~J.}\ \bibnamefont {Williams}}, \bibinfo {author} {\bibfnamefont {M.}~\bibnamefont {Karamanis}}, \bibinfo {author} {\bibfnamefont {Y.}~\bibnamefont {Luo}},\ and\ \bibinfo {author} {\bibfnamefont {U.}~\bibnamefont {Seljak}},\ }\bibfield  {title} {\bibinfo {title} {{Validating Sequential Monte Carlo for Gravitational-Wave Inference}},\ }\href@noop {} {\bibfield  {journal} {\bibinfo  {journal} {arXiv preprint}\ } (\bibinfo {year} {2025})},\ \Eprint {https://arxiv.org/abs/2506.18977} {arXiv:2506.18977 [astro-ph.IM]} \BibitemShut {NoStop}%
\bibitem [{\citenamefont {Vretinaris}\ \emph {et~al.}(2025)\citenamefont {Vretinaris}, \citenamefont {Vretinaris}, \citenamefont {Mermigkas}, \citenamefont {Karamanis},\ and\ \citenamefont {Stergioulas}}]{Vretinaris:2025wdu}%
  \BibitemOpen
  \bibfield  {author} {\bibinfo {author} {\bibfnamefont {S.}~\bibnamefont {Vretinaris}}, \bibinfo {author} {\bibfnamefont {G.}~\bibnamefont {Vretinaris}}, \bibinfo {author} {\bibfnamefont {C.}~\bibnamefont {Mermigkas}}, \bibinfo {author} {\bibfnamefont {M.}~\bibnamefont {Karamanis}},\ and\ \bibinfo {author} {\bibfnamefont {N.}~\bibnamefont {Stergioulas}},\ }\bibfield  {title} {\bibinfo {title} {{Robust and fast parameter estimation for gravitational waves from binary neutron star merger remnants}},\ }\href@noop {} {\bibfield  {journal} {\bibinfo  {journal} {arXiv preprint}\ } (\bibinfo {year} {2025})},\ \Eprint {https://arxiv.org/abs/2501.11518} {arXiv:2501.11518 [gr-qc]} \BibitemShut {NoStop}%
\bibitem [{\citenamefont {Gabbard}\ \emph {et~al.}(2022)\citenamefont {Gabbard}, \citenamefont {Messenger}, \citenamefont {Heng}, \citenamefont {Tonolini},\ and\ \citenamefont {Murray-Smith}}]{Gabbard:2019rde}%
  \BibitemOpen
  \bibfield  {author} {\bibinfo {author} {\bibfnamefont {H.}~\bibnamefont {Gabbard}}, \bibinfo {author} {\bibfnamefont {C.}~\bibnamefont {Messenger}}, \bibinfo {author} {\bibfnamefont {I.~S.}\ \bibnamefont {Heng}}, \bibinfo {author} {\bibfnamefont {F.}~\bibnamefont {Tonolini}},\ and\ \bibinfo {author} {\bibfnamefont {R.}~\bibnamefont {Murray-Smith}},\ }\bibfield  {title} {\bibinfo {title} {{Bayesian parameter estimation using conditional variational autoencoders for gravitational-wave astronomy}},\ }\href {https://doi.org/10.1038/s41567-021-01425-7} {\bibfield  {journal} {\bibinfo  {journal} {Nature Phys.}\ }\textbf {\bibinfo {volume} {18}},\ \bibinfo {pages} {112} (\bibinfo {year} {2022})},\ \Eprint {https://arxiv.org/abs/1909.06296} {arXiv:1909.06296 [astro-ph.IM]} \BibitemShut {NoStop}%
\bibitem [{\citenamefont {Green}\ \emph {et~al.}(2020)\citenamefont {Green}, \citenamefont {Simpson},\ and\ \citenamefont {Gair}}]{Green:2020hst}%
  \BibitemOpen
  \bibfield  {author} {\bibinfo {author} {\bibfnamefont {S.~R.}\ \bibnamefont {Green}}, \bibinfo {author} {\bibfnamefont {C.}~\bibnamefont {Simpson}},\ and\ \bibinfo {author} {\bibfnamefont {J.}~\bibnamefont {Gair}},\ }\bibfield  {title} {\bibinfo {title} {{Gravitational-wave parameter estimation with autoregressive neural network flows}},\ }\href {https://doi.org/10.1103/PhysRevD.102.104057} {\bibfield  {journal} {\bibinfo  {journal} {Phys. Rev. D}\ }\textbf {\bibinfo {volume} {102}},\ \bibinfo {pages} {104057} (\bibinfo {year} {2020})},\ \Eprint {https://arxiv.org/abs/2002.07656} {arXiv:2002.07656 [astro-ph.IM]} \BibitemShut {NoStop}%
\bibitem [{\citenamefont {Chua}\ and\ \citenamefont {Vallisneri}(2020)}]{Chua:2019wwt}%
  \BibitemOpen
  \bibfield  {author} {\bibinfo {author} {\bibfnamefont {A.~J.~K.}\ \bibnamefont {Chua}}\ and\ \bibinfo {author} {\bibfnamefont {M.}~\bibnamefont {Vallisneri}},\ }\bibfield  {title} {\bibinfo {title} {{Learning Bayesian posteriors with neural networks for gravitational-wave inference}},\ }\href {https://doi.org/10.1103/PhysRevLett.124.041102} {\bibfield  {journal} {\bibinfo  {journal} {Phys. Rev. Lett.}\ }\textbf {\bibinfo {volume} {124}},\ \bibinfo {pages} {041102} (\bibinfo {year} {2020})},\ \Eprint {https://arxiv.org/abs/1909.05966} {arXiv:1909.05966 [gr-qc]} \BibitemShut {NoStop}%
\bibitem [{\citenamefont {Dax}\ \emph {et~al.}(2021)\citenamefont {Dax}, \citenamefont {Green}, \citenamefont {Gair}, \citenamefont {Macke}, \citenamefont {Buonanno},\ and\ \citenamefont {Sch\"olkopf}}]{Dax:2021tsq}%
  \BibitemOpen
  \bibfield  {author} {\bibinfo {author} {\bibfnamefont {M.}~\bibnamefont {Dax}}, \bibinfo {author} {\bibfnamefont {S.~R.}\ \bibnamefont {Green}}, \bibinfo {author} {\bibfnamefont {J.}~\bibnamefont {Gair}}, \bibinfo {author} {\bibfnamefont {J.~H.}\ \bibnamefont {Macke}}, \bibinfo {author} {\bibfnamefont {A.}~\bibnamefont {Buonanno}},\ and\ \bibinfo {author} {\bibfnamefont {B.}~\bibnamefont {Sch\"olkopf}},\ }\bibfield  {title} {\bibinfo {title} {{Real-Time Gravitational Wave Science with Neural Posterior Estimation}},\ }\href {https://doi.org/10.1103/PhysRevLett.127.241103} {\bibfield  {journal} {\bibinfo  {journal} {Phys. Rev. Lett.}\ }\textbf {\bibinfo {volume} {127}},\ \bibinfo {pages} {241103} (\bibinfo {year} {2021})},\ \Eprint {https://arxiv.org/abs/2106.12594} {arXiv:2106.12594 [gr-qc]} \BibitemShut {NoStop}%
\bibitem [{\citenamefont {Dax}\ \emph {et~al.}(2023)\citenamefont {Dax}, \citenamefont {Green}, \citenamefont {Gair}, \citenamefont {P\"urrer}, \citenamefont {Wildberger}, \citenamefont {Macke}, \citenamefont {Buonanno},\ and\ \citenamefont {Sch\"olkopf}}]{Dax:2022pxd}%
  \BibitemOpen
  \bibfield  {author} {\bibinfo {author} {\bibfnamefont {M.}~\bibnamefont {Dax}}, \bibinfo {author} {\bibfnamefont {S.~R.}\ \bibnamefont {Green}}, \bibinfo {author} {\bibfnamefont {J.}~\bibnamefont {Gair}}, \bibinfo {author} {\bibfnamefont {M.}~\bibnamefont {P\"urrer}}, \bibinfo {author} {\bibfnamefont {J.}~\bibnamefont {Wildberger}}, \bibinfo {author} {\bibfnamefont {J.~H.}\ \bibnamefont {Macke}}, \bibinfo {author} {\bibfnamefont {A.}~\bibnamefont {Buonanno}},\ and\ \bibinfo {author} {\bibfnamefont {B.}~\bibnamefont {Sch\"olkopf}},\ }\bibfield  {title} {\bibinfo {title} {{Neural Importance Sampling for Rapid and Reliable Gravitational-Wave Inference}},\ }\href {https://doi.org/10.1103/PhysRevLett.130.171403} {\bibfield  {journal} {\bibinfo  {journal} {Phys. Rev. Lett.}\ }\textbf {\bibinfo {volume} {130}},\ \bibinfo {pages} {171403} (\bibinfo {year} {2023})},\ \Eprint {https://arxiv.org/abs/2210.05686} {arXiv:2210.05686 [gr-qc]} \BibitemShut {NoStop}%
\bibitem [{\citenamefont {Bhardwaj}\ \emph {et~al.}(2023)\citenamefont {Bhardwaj}, \citenamefont {Alvey}, \citenamefont {Miller}, \citenamefont {Nissanke},\ and\ \citenamefont {Weniger}}]{Bhardwaj:2023xph}%
  \BibitemOpen
  \bibfield  {author} {\bibinfo {author} {\bibfnamefont {U.}~\bibnamefont {Bhardwaj}}, \bibinfo {author} {\bibfnamefont {J.}~\bibnamefont {Alvey}}, \bibinfo {author} {\bibfnamefont {B.~K.}\ \bibnamefont {Miller}}, \bibinfo {author} {\bibfnamefont {S.}~\bibnamefont {Nissanke}},\ and\ \bibinfo {author} {\bibfnamefont {C.}~\bibnamefont {Weniger}},\ }\bibfield  {title} {\bibinfo {title} {{Sequential simulation-based inference for gravitational wave signals}},\ }\href {https://doi.org/10.1103/PhysRevD.108.042004} {\bibfield  {journal} {\bibinfo  {journal} {Phys. Rev. D}\ }\textbf {\bibinfo {volume} {108}},\ \bibinfo {pages} {042004} (\bibinfo {year} {2023})},\ \Eprint {https://arxiv.org/abs/2304.02035} {arXiv:2304.02035 [gr-qc]} \BibitemShut {NoStop}%
\bibitem [{\citenamefont {Kolmus}\ \emph {et~al.}(2024)\citenamefont {Kolmus}, \citenamefont {Janquart}, \citenamefont {Baka}, \citenamefont {van Laarhoven}, \citenamefont {Van Den~Broeck},\ and\ \citenamefont {Heskes}}]{Kolmus:2024scm}%
  \BibitemOpen
  \bibfield  {author} {\bibinfo {author} {\bibfnamefont {A.}~\bibnamefont {Kolmus}}, \bibinfo {author} {\bibfnamefont {J.}~\bibnamefont {Janquart}}, \bibinfo {author} {\bibfnamefont {T.}~\bibnamefont {Baka}}, \bibinfo {author} {\bibfnamefont {T.}~\bibnamefont {van Laarhoven}}, \bibinfo {author} {\bibfnamefont {C.}~\bibnamefont {Van Den~Broeck}},\ and\ \bibinfo {author} {\bibfnamefont {T.}~\bibnamefont {Heskes}},\ }\bibfield  {title} {\bibinfo {title} {{Tuning neural posterior estimation for gravitational wave inference}},\ }\href@noop {} {\bibfield  {journal} {\bibinfo  {journal} {arXiv preprint}\ } (\bibinfo {year} {2024})},\ \Eprint {https://arxiv.org/abs/2403.02443} {arXiv:2403.02443 [astro-ph.IM]} \BibitemShut {NoStop}%
\bibitem [{\citenamefont {Dax}\ \emph {et~al.}(2025{\natexlab{b}})\citenamefont {Dax}, \citenamefont {Green}, \citenamefont {Gair}, \citenamefont {Gupte}, \citenamefont {P\"urrer}, \citenamefont {Raymond}, \citenamefont {Wildberger}, \citenamefont {Macke}, \citenamefont {Buonanno},\ and\ \citenamefont {Sch\"olkopf}}]{Dax:2024mcn}%
  \BibitemOpen
  \bibfield  {author} {\bibinfo {author} {\bibfnamefont {M.}~\bibnamefont {Dax}}, \bibinfo {author} {\bibfnamefont {S.~R.}\ \bibnamefont {Green}}, \bibinfo {author} {\bibfnamefont {J.}~\bibnamefont {Gair}}, \bibinfo {author} {\bibfnamefont {N.}~\bibnamefont {Gupte}}, \bibinfo {author} {\bibfnamefont {M.}~\bibnamefont {P\"urrer}}, \bibinfo {author} {\bibfnamefont {V.}~\bibnamefont {Raymond}}, \bibinfo {author} {\bibfnamefont {J.}~\bibnamefont {Wildberger}}, \bibinfo {author} {\bibfnamefont {J.~H.}\ \bibnamefont {Macke}}, \bibinfo {author} {\bibfnamefont {A.}~\bibnamefont {Buonanno}},\ and\ \bibinfo {author} {\bibfnamefont {B.}~\bibnamefont {Sch\"olkopf}},\ }\bibfield  {title} {\bibinfo {title} {{Real-time inference for binary neutron star mergers using machine learning}},\ }\href {https://doi.org/10.1038/s41586-025-08593-z} {\bibfield  {journal} {\bibinfo  {journal} {Nature}\ }\textbf {\bibinfo {volume} {639}},\ \bibinfo {pages} {49} (\bibinfo {year} {2025}{\natexlab{b}})},\ \Eprint
  {https://arxiv.org/abs/2407.09602} {arXiv:2407.09602 [gr-qc]} \BibitemShut {NoStop}%
\bibitem [{\citenamefont {Pathak}\ \emph {et~al.}(2023)\citenamefont {Pathak}, \citenamefont {Reza},\ and\ \citenamefont {Sengupta}}]{Pathak:2022iar}%
  \BibitemOpen
  \bibfield  {author} {\bibinfo {author} {\bibfnamefont {L.}~\bibnamefont {Pathak}}, \bibinfo {author} {\bibfnamefont {A.}~\bibnamefont {Reza}},\ and\ \bibinfo {author} {\bibfnamefont {A.~S.}\ \bibnamefont {Sengupta}},\ }\bibfield  {title} {\bibinfo {title} {{Fast likelihood evaluation using meshfree approximations for reconstructing compact binary sources}},\ }\href {https://doi.org/10.1103/PhysRevD.108.064055} {\bibfield  {journal} {\bibinfo  {journal} {Phys. Rev. D}\ }\textbf {\bibinfo {volume} {108}},\ \bibinfo {pages} {064055} (\bibinfo {year} {2023})},\ \Eprint {https://arxiv.org/abs/2210.02706} {arXiv:2210.02706 [gr-qc]} \BibitemShut {NoStop}%
\bibitem [{\citenamefont {Pathak}\ \emph {et~al.}(2024)\citenamefont {Pathak}, \citenamefont {Munishwar}, \citenamefont {Reza},\ and\ \citenamefont {Sengupta}}]{Pathak:2023ixb}%
  \BibitemOpen
  \bibfield  {author} {\bibinfo {author} {\bibfnamefont {L.}~\bibnamefont {Pathak}}, \bibinfo {author} {\bibfnamefont {S.}~\bibnamefont {Munishwar}}, \bibinfo {author} {\bibfnamefont {A.}~\bibnamefont {Reza}},\ and\ \bibinfo {author} {\bibfnamefont {A.~S.}\ \bibnamefont {Sengupta}},\ }\bibfield  {title} {\bibinfo {title} {{Prompt sky localization of compact binary sources using a meshfree approximation}},\ }\href {https://doi.org/10.1103/PhysRevD.109.024053} {\bibfield  {journal} {\bibinfo  {journal} {Phys. Rev. D}\ }\textbf {\bibinfo {volume} {109}},\ \bibinfo {pages} {024053} (\bibinfo {year} {2024})},\ \Eprint {https://arxiv.org/abs/2309.07012} {arXiv:2309.07012 [gr-qc]} \BibitemShut {NoStop}%
\bibitem [{\citenamefont {Abbott}\ \emph {et~al.}(2020{\natexlab{a}})\citenamefont {Abbott} \emph {et~al.}}]{LIGOScientific:2020zkf}%
  \BibitemOpen
  \bibfield  {author} {\bibinfo {author} {\bibfnamefont {R.}~\bibnamefont {Abbott}} \emph {et~al.} (\bibinfo {collaboration} {LIGO Scientific, Virgo}),\ }\bibfield  {title} {\bibinfo {title} {{GW190814: Gravitational Waves from the Coalescence of a 23 Solar Mass Black Hole with a 2.6 Solar Mass Compact Object}},\ }\href {https://doi.org/10.3847/2041-8213/ab960f} {\bibfield  {journal} {\bibinfo  {journal} {Astrophys. J. Lett.}\ }\textbf {\bibinfo {volume} {896}},\ \bibinfo {pages} {L44} (\bibinfo {year} {2020}{\natexlab{a}})},\ \Eprint {https://arxiv.org/abs/2006.12611} {arXiv:2006.12611 [astro-ph.HE]} \BibitemShut {NoStop}%
\bibitem [{\citenamefont {Varma}\ \emph {et~al.}(2014)\citenamefont {Varma}, \citenamefont {Ajith}, \citenamefont {Husa}, \citenamefont {Bustillo}, \citenamefont {Hannam},\ and\ \citenamefont {P\"urrer}}]{Varma:2014jxa}%
  \BibitemOpen
  \bibfield  {author} {\bibinfo {author} {\bibfnamefont {V.}~\bibnamefont {Varma}}, \bibinfo {author} {\bibfnamefont {P.}~\bibnamefont {Ajith}}, \bibinfo {author} {\bibfnamefont {S.}~\bibnamefont {Husa}}, \bibinfo {author} {\bibfnamefont {J.~C.}\ \bibnamefont {Bustillo}}, \bibinfo {author} {\bibfnamefont {M.}~\bibnamefont {Hannam}},\ and\ \bibinfo {author} {\bibfnamefont {M.}~\bibnamefont {P\"urrer}},\ }\bibfield  {title} {\bibinfo {title} {{Gravitational-wave observations of binary black holes: Effect of nonquadrupole modes}},\ }\href {https://doi.org/10.1103/PhysRevD.90.124004} {\bibfield  {journal} {\bibinfo  {journal} {Phys. Rev. D}\ }\textbf {\bibinfo {volume} {90}},\ \bibinfo {pages} {124004} (\bibinfo {year} {2014})},\ \Eprint {https://arxiv.org/abs/1409.2349} {arXiv:1409.2349 [gr-qc]} \BibitemShut {NoStop}%
\bibitem [{\citenamefont {Littenberg}\ \emph {et~al.}(2013)\citenamefont {Littenberg}, \citenamefont {Baker}, \citenamefont {Buonanno},\ and\ \citenamefont {Kelly}}]{PhysRevD.87.104003}%
  \BibitemOpen
  \bibfield  {author} {\bibinfo {author} {\bibfnamefont {T.~B.}\ \bibnamefont {Littenberg}}, \bibinfo {author} {\bibfnamefont {J.~G.}\ \bibnamefont {Baker}}, \bibinfo {author} {\bibfnamefont {A.}~\bibnamefont {Buonanno}},\ and\ \bibinfo {author} {\bibfnamefont {B.~J.}\ \bibnamefont {Kelly}},\ }\bibfield  {title} {\bibinfo {title} {Systematic biases in parameter estimation of binary black-hole mergers},\ }\href {https://doi.org/10.1103/PhysRevD.87.104003} {\bibfield  {journal} {\bibinfo  {journal} {Phys. Rev. D}\ }\textbf {\bibinfo {volume} {87}},\ \bibinfo {pages} {104003} (\bibinfo {year} {2013})}\BibitemShut {NoStop}%
\bibitem [{\citenamefont {Calder\'on~Bustillo}\ \emph {et~al.}(2016)\citenamefont {Calder\'on~Bustillo}, \citenamefont {Husa}, \citenamefont {Sintes},\ and\ \citenamefont {P\"urrer}}]{CalderonBustillo:2015lrt}%
  \BibitemOpen
  \bibfield  {author} {\bibinfo {author} {\bibfnamefont {J.}~\bibnamefont {Calder\'on~Bustillo}}, \bibinfo {author} {\bibfnamefont {S.}~\bibnamefont {Husa}}, \bibinfo {author} {\bibfnamefont {A.~M.}\ \bibnamefont {Sintes}},\ and\ \bibinfo {author} {\bibfnamefont {M.}~\bibnamefont {P\"urrer}},\ }\bibfield  {title} {\bibinfo {title} {{Impact of gravitational radiation higher order modes on single aligned-spin gravitational wave searches for binary black holes}},\ }\href {https://doi.org/10.1103/PhysRevD.93.084019} {\bibfield  {journal} {\bibinfo  {journal} {Phys. Rev. D}\ }\textbf {\bibinfo {volume} {93}},\ \bibinfo {pages} {084019} (\bibinfo {year} {2016})},\ \Eprint {https://arxiv.org/abs/1511.02060} {arXiv:1511.02060 [gr-qc]} \BibitemShut {NoStop}%
\bibitem [{\citenamefont {Chatziioannou}\ \emph {et~al.}(2019)\citenamefont {Chatziioannou}, \citenamefont {Cotesta}, \citenamefont {Ghonge} \emph {et~al.}}]{Chatziioannou:2019dsz}%
  \BibitemOpen
  \bibfield  {author} {\bibinfo {author} {\bibfnamefont {K.}~\bibnamefont {Chatziioannou}}, \bibinfo {author} {\bibfnamefont {R.}~\bibnamefont {Cotesta}}, \bibinfo {author} {\bibfnamefont {S.}~\bibnamefont {Ghonge}}, \emph {et~al.},\ }\bibfield  {title} {\bibinfo {title} {{{On the properties of the massive binary black hole merger GW170729}}},\ }\href {https://doi.org/10.1103/PhysRevD.100.104015} {\bibfield  {journal} {\bibinfo  {journal} {Phys. Rev. D}\ }\textbf {\bibinfo {volume} {100}},\ \bibinfo {pages} {104015} (\bibinfo {year} {2019})},\ \Eprint {https://arxiv.org/abs/1903.06742} {arXiv:1903.06742 [gr-qc]} \BibitemShut {NoStop}%
\bibitem [{\citenamefont {Kalaghatgi}\ \emph {et~al.}(2020)\citenamefont {Kalaghatgi}, \citenamefont {Hannam},\ and\ \citenamefont {Raymond}}]{Kalaghatgi2020IMRPhenomHM}%
  \BibitemOpen
  \bibfield  {author} {\bibinfo {author} {\bibfnamefont {C.}~\bibnamefont {Kalaghatgi}}, \bibinfo {author} {\bibfnamefont {M.}~\bibnamefont {Hannam}},\ and\ \bibinfo {author} {\bibfnamefont {V.}~\bibnamefont {Raymond}},\ }\bibfield  {title} {\bibinfo {title} {{Parameter estimation with a spinning multi‑mode waveform model: IMRPhenomHM}},\ }\href {https://doi.org/10.1103/PhysRevD.101.103004} {\bibfield  {journal} {\bibinfo  {journal} {Physical Review D}\ }\textbf {\bibinfo {volume} {101}},\ \bibinfo {pages} {103004} (\bibinfo {year} {2020})}\BibitemShut {NoStop}%
\bibitem [{\citenamefont {Abbott}\ \emph {et~al.}(2020{\natexlab{b}})\citenamefont {Abbott} \emph {et~al.}}]{LIGOScientific:2020stg}%
  \BibitemOpen
  \bibfield  {author} {\bibinfo {author} {\bibfnamefont {R.}~\bibnamefont {Abbott}} \emph {et~al.} (\bibinfo {collaboration} {LIGO Scientific, Virgo}),\ }\bibfield  {title} {\bibinfo {title} {{GW190412: Observation of a Binary-Black-Hole Coalescence with Asymmetric Masses}},\ }\href {https://doi.org/10.1103/PhysRevD.102.043015} {\bibfield  {journal} {\bibinfo  {journal} {Phys. Rev. D}\ }\textbf {\bibinfo {volume} {102}},\ \bibinfo {pages} {043015} (\bibinfo {year} {2020}{\natexlab{b}})},\ \Eprint {https://arxiv.org/abs/2004.08342} {arXiv:2004.08342 [astro-ph.HE]} \BibitemShut {NoStop}%
\bibitem [{\citenamefont {Usman}\ \emph {et~al.}(2019)\citenamefont {Usman}, \citenamefont {Mills},\ and\ \citenamefont {Fairhurst}}]{Usman:2018imj}%
  \BibitemOpen
  \bibfield  {author} {\bibinfo {author} {\bibfnamefont {S.~A.}\ \bibnamefont {Usman}}, \bibinfo {author} {\bibfnamefont {J.~C.}\ \bibnamefont {Mills}},\ and\ \bibinfo {author} {\bibfnamefont {S.}~\bibnamefont {Fairhurst}},\ }\bibfield  {title} {\bibinfo {title} {{Constraining the Inclinations of Binary Mergers from Gravitational-wave Observations}},\ }\href {https://doi.org/10.3847/1538-4357/ab0b3e} {\bibfield  {journal} {\bibinfo  {journal} {Astrophys. J.}\ }\textbf {\bibinfo {volume} {877}},\ \bibinfo {pages} {82} (\bibinfo {year} {2019})},\ \Eprint {https://arxiv.org/abs/1809.10727} {arXiv:1809.10727 [gr-qc]} \BibitemShut {NoStop}%
\bibitem [{\citenamefont {Hannam}\ \emph {et~al.}(2013)\citenamefont {Hannam}, \citenamefont {Brown}, \citenamefont {Fairhurst}, \citenamefont {Fryer},\ and\ \citenamefont {Harry}}]{Hannam:2013uu}%
  \BibitemOpen
  \bibfield  {author} {\bibinfo {author} {\bibfnamefont {M.}~\bibnamefont {Hannam}}, \bibinfo {author} {\bibfnamefont {D.~A.}\ \bibnamefont {Brown}}, \bibinfo {author} {\bibfnamefont {S.}~\bibnamefont {Fairhurst}}, \bibinfo {author} {\bibfnamefont {C.~L.}\ \bibnamefont {Fryer}},\ and\ \bibinfo {author} {\bibfnamefont {I.~W.}\ \bibnamefont {Harry}},\ }\bibfield  {title} {\bibinfo {title} {{When can gravitational-wave observations distinguish between black holes and neutron stars?}},\ }\href {https://doi.org/10.1088/2041-8205/766/1/L14} {\bibfield  {journal} {\bibinfo  {journal} {Astrophys. J. Lett.}\ }\textbf {\bibinfo {volume} {766}},\ \bibinfo {pages} {L14} (\bibinfo {year} {2013})},\ \Eprint {https://arxiv.org/abs/1301.5616} {arXiv:1301.5616 [gr-qc]} \BibitemShut {NoStop}%
\bibitem [{\citenamefont {Ohme}\ \emph {et~al.}(2013)\citenamefont {Ohme}, \citenamefont {Nielsen}, \citenamefont {Keppel},\ and\ \citenamefont {Lundgren}}]{Ohme:2013nsa}%
  \BibitemOpen
  \bibfield  {author} {\bibinfo {author} {\bibfnamefont {F.}~\bibnamefont {Ohme}}, \bibinfo {author} {\bibfnamefont {A.~B.}\ \bibnamefont {Nielsen}}, \bibinfo {author} {\bibfnamefont {D.}~\bibnamefont {Keppel}},\ and\ \bibinfo {author} {\bibfnamefont {A.}~\bibnamefont {Lundgren}},\ }\bibfield  {title} {\bibinfo {title} {{Statistical and systematic errors for gravitational-wave inspiral signals: A principal component analysis}},\ }\href {https://doi.org/10.1103/PhysRevD.88.042002} {\bibfield  {journal} {\bibinfo  {journal} {Phys. Rev. D}\ }\textbf {\bibinfo {volume} {88}},\ \bibinfo {pages} {042002} (\bibinfo {year} {2013})},\ \Eprint {https://arxiv.org/abs/1304.7017} {arXiv:1304.7017 [gr-qc]} \BibitemShut {NoStop}%
\bibitem [{\citenamefont {Foreman-Mackey}\ \emph {et~al.}(2013)\citenamefont {Foreman-Mackey}, \citenamefont {Hogg}, \citenamefont {Lang},\ and\ \citenamefont {Goodman}}]{Foreman_Mackey_2013}%
  \BibitemOpen
  \bibfield  {author} {\bibinfo {author} {\bibfnamefont {D.}~\bibnamefont {Foreman-Mackey}}, \bibinfo {author} {\bibfnamefont {D.~W.}\ \bibnamefont {Hogg}}, \bibinfo {author} {\bibfnamefont {D.}~\bibnamefont {Lang}},\ and\ \bibinfo {author} {\bibfnamefont {J.}~\bibnamefont {Goodman}},\ }\bibfield  {title} {\bibinfo {title} {{\texttt{emcee}}: The {MCMC} {H}ammer},\ }\href {https://doi.org/10.1086/670067} {\bibfield  {journal} {\bibinfo  {journal} {Publ. Astron. Soc. Pac.}\ }\textbf {\bibinfo {volume} {125}},\ \bibinfo {pages} {306} (\bibinfo {year} {2013})}\BibitemShut {NoStop}%
\bibitem [{\citenamefont {Skilling}(2006)}]{skilling2006nested}%
  \BibitemOpen
  \bibfield  {author} {\bibinfo {author} {\bibfnamefont {J.}~\bibnamefont {Skilling}},\ }\bibfield  {title} {\bibinfo {title} {{Nested sampling for general Bayesian computation}},\ }\href {https://doi.org/10.1214/06-BA127} {\bibfield  {journal} {\bibinfo  {journal} {{Bayesian Anal.}}\ }\textbf {\bibinfo {volume} {1}},\ \bibinfo {pages} {833} (\bibinfo {year} {2006})}\BibitemShut {NoStop}%
\bibitem [{\citenamefont {Garc\'\i{}a-Quir\'os}\ \emph {et~al.}(2020)\citenamefont {Garc\'\i{}a-Quir\'os}, \citenamefont {Colleoni}, \citenamefont {Husa}, \citenamefont {Estell\'es}, \citenamefont {Pratten}, \citenamefont {Ramos-Buades}, \citenamefont {Mateu-Lucena},\ and\ \citenamefont {Jaume}}]{Garcia-Quiros:2020qpx}%
  \BibitemOpen
  \bibfield  {author} {\bibinfo {author} {\bibfnamefont {C.}~\bibnamefont {Garc\'\i{}a-Quir\'os}}, \bibinfo {author} {\bibfnamefont {M.}~\bibnamefont {Colleoni}}, \bibinfo {author} {\bibfnamefont {S.}~\bibnamefont {Husa}}, \bibinfo {author} {\bibfnamefont {H.}~\bibnamefont {Estell\'es}}, \bibinfo {author} {\bibfnamefont {G.}~\bibnamefont {Pratten}}, \bibinfo {author} {\bibfnamefont {A.}~\bibnamefont {Ramos-Buades}}, \bibinfo {author} {\bibfnamefont {M.}~\bibnamefont {Mateu-Lucena}},\ and\ \bibinfo {author} {\bibfnamefont {R.}~\bibnamefont {Jaume}},\ }\bibfield  {title} {\bibinfo {title} {{Multimode frequency-domain model for the gravitational wave signal from nonprecessing black-hole binaries}},\ }\href {https://doi.org/10.1103/PhysRevD.102.064002} {\bibfield  {journal} {\bibinfo  {journal} {Phys. Rev. D}\ }\textbf {\bibinfo {volume} {102}},\ \bibinfo {pages} {064002} (\bibinfo {year} {2020})},\ \Eprint {https://arxiv.org/abs/2001.10914} {arXiv:2001.10914 [gr-qc]} \BibitemShut {NoStop}%
\bibitem [{\citenamefont {Rakhmanov}\ \emph {et~al.}(2008)\citenamefont {Rakhmanov}, \citenamefont {Romano},\ and\ \citenamefont {Whelan}}]{Rakhmanov:2008is}%
  \BibitemOpen
  \bibfield  {author} {\bibinfo {author} {\bibfnamefont {M.}~\bibnamefont {Rakhmanov}}, \bibinfo {author} {\bibfnamefont {J.~D.}\ \bibnamefont {Romano}},\ and\ \bibinfo {author} {\bibfnamefont {J.~T.}\ \bibnamefont {Whelan}},\ }\bibfield  {title} {\bibinfo {title} {{High-frequency corrections to the detector response and their effect on searches for gravitational waves}},\ }\href {https://doi.org/10.1088/0264-9381/25/18/184017} {\bibfield  {journal} {\bibinfo  {journal} {Class. Quant. Grav.}\ }\textbf {\bibinfo {volume} {25}},\ \bibinfo {pages} {184017} (\bibinfo {year} {2008})},\ \Eprint {https://arxiv.org/abs/0808.3805} {arXiv:0808.3805 [gr-qc]} \BibitemShut {NoStop}%
\bibitem [{\citenamefont {Sharma}\ \emph {et~al.}(2025)\citenamefont {Sharma}, \citenamefont {Sengupta},\ and\ \citenamefont {Mukherjee}}]{Sharma:2024sfb}%
  \BibitemOpen
  \bibfield  {author} {\bibinfo {author} {\bibfnamefont {A.}~\bibnamefont {Sharma}}, \bibinfo {author} {\bibfnamefont {A.~S.}\ \bibnamefont {Sengupta}},\ and\ \bibinfo {author} {\bibfnamefont {S.}~\bibnamefont {Mukherjee}},\ }\bibfield  {title} {\bibinfo {title} {{Accelerated parameter estimation of supermassive black hole binaries in LISA using a meshfree approximation}},\ }\href {https://doi.org/10.1103/PhysRevD.111.042009} {\bibfield  {journal} {\bibinfo  {journal} {Phys. Rev. D}\ }\textbf {\bibinfo {volume} {111}},\ \bibinfo {pages} {042009} (\bibinfo {year} {2025})},\ \Eprint {https://arxiv.org/abs/2409.14288} {arXiv:2409.14288 [gr-qc]} \BibitemShut {NoStop}%
\bibitem [{\citenamefont {Cannon}\ \emph {et~al.}(2010)\citenamefont {Cannon}, \citenamefont {Chapman}, \citenamefont {Hanna}, \citenamefont {Keppel}, \citenamefont {Searle},\ and\ \citenamefont {Weinstein}}]{GstLAL_2010}%
  \BibitemOpen
  \bibfield  {author} {\bibinfo {author} {\bibfnamefont {K.}~\bibnamefont {Cannon}}, \bibinfo {author} {\bibfnamefont {A.}~\bibnamefont {Chapman}}, \bibinfo {author} {\bibfnamefont {C.}~\bibnamefont {Hanna}}, \bibinfo {author} {\bibfnamefont {D.}~\bibnamefont {Keppel}}, \bibinfo {author} {\bibfnamefont {A.~C.}\ \bibnamefont {Searle}},\ and\ \bibinfo {author} {\bibfnamefont {A.~J.}\ \bibnamefont {Weinstein}},\ }\bibfield  {title} {\bibinfo {title} {Singular value decomposition applied to compact binary coalescence gravitational-wave signals},\ }\href {https://doi.org/10.1103/PhysRevD.82.044025} {\bibfield  {journal} {\bibinfo  {journal} {Phys. Rev. D}\ }\textbf {\bibinfo {volume} {82}},\ \bibinfo {pages} {044025} (\bibinfo {year} {2010})}\BibitemShut {NoStop}%
\bibitem [{\citenamefont {Usman}\ \emph {et~al.}(2016)\citenamefont {Usman}, \citenamefont {Nitz}, \citenamefont {Harry}, \citenamefont {Biwer}, \citenamefont {Brown}, \citenamefont {Cabero}, \citenamefont {Capano}, \citenamefont {Dal~Canton}, \citenamefont {Dent}, \citenamefont {Fairhurst} \emph {et~al.}}]{usman2016pycbc}%
  \BibitemOpen
  \bibfield  {author} {\bibinfo {author} {\bibfnamefont {S.~A.}\ \bibnamefont {Usman}}, \bibinfo {author} {\bibfnamefont {A.~H.}\ \bibnamefont {Nitz}}, \bibinfo {author} {\bibfnamefont {I.~W.}\ \bibnamefont {Harry}}, \bibinfo {author} {\bibfnamefont {C.~M.}\ \bibnamefont {Biwer}}, \bibinfo {author} {\bibfnamefont {D.~A.}\ \bibnamefont {Brown}}, \bibinfo {author} {\bibfnamefont {M.}~\bibnamefont {Cabero}}, \bibinfo {author} {\bibfnamefont {C.~D.}\ \bibnamefont {Capano}}, \bibinfo {author} {\bibfnamefont {T.}~\bibnamefont {Dal~Canton}}, \bibinfo {author} {\bibfnamefont {T.}~\bibnamefont {Dent}}, \bibinfo {author} {\bibfnamefont {S.}~\bibnamefont {Fairhurst}}, \emph {et~al.},\ }\bibfield  {title} {\bibinfo {title} {The {PyCBC} search for gravitational waves from compact binary coalescence},\ }\href {https://doi.org/10.1088/0264-9381/33/21/215004} {\bibfield  {journal} {\bibinfo  {journal} {Class. Quantum Gravity}\ }\textbf {\bibinfo {volume} {33}},\ \bibinfo {pages} {215004} (\bibinfo {year} {2016})}\BibitemShut
  {NoStop}%
\bibitem [{\citenamefont {Owen}(1996)}]{Owen:1995tm}%
  \BibitemOpen
  \bibfield  {author} {\bibinfo {author} {\bibfnamefont {B.~J.}\ \bibnamefont {Owen}},\ }\bibfield  {title} {\bibinfo {title} {{Search templates for gravitational waves from inspiraling binaries: Choice of template spacing}},\ }\href {https://doi.org/10.1103/PhysRevD.53.6749} {\bibfield  {journal} {\bibinfo  {journal} {Phys. Rev. D}\ }\textbf {\bibinfo {volume} {53}},\ \bibinfo {pages} {6749} (\bibinfo {year} {1996})},\ \Eprint {https://arxiv.org/abs/gr-qc/9511032} {arXiv:gr-qc/9511032} \BibitemShut {NoStop}%
\bibitem [{\citenamefont {Fairhurst}\ \emph {et~al.}(2023{\natexlab{b}})\citenamefont {Fairhurst}, \citenamefont {Hoy}, \citenamefont {Green}, \citenamefont {Mills},\ and\ \citenamefont {Usman}}]{Fairhurst2023simplePE}%
  \BibitemOpen
  \bibfield  {author} {\bibinfo {author} {\bibfnamefont {S.}~\bibnamefont {Fairhurst}}, \bibinfo {author} {\bibfnamefont {C.}~\bibnamefont {Hoy}}, \bibinfo {author} {\bibfnamefont {R.}~\bibnamefont {Green}}, \bibinfo {author} {\bibfnamefont {C.}~\bibnamefont {Mills}},\ and\ \bibinfo {author} {\bibfnamefont {S.~A.}\ \bibnamefont {Usman}},\ }\bibfield  {title} {\bibinfo {title} {Simple parameter estimation using observable features of gravitational-wave signals},\ }\href {https://doi.org/10.1103/PhysRevD.108.082006} {\bibfield  {journal} {\bibinfo  {journal} {Physical Review D}\ }\textbf {\bibinfo {volume} {108}},\ \bibinfo {pages} {082006} (\bibinfo {year} {2023}{\natexlab{b}})}\BibitemShut {NoStop}%
\bibitem [{\citenamefont {Dal~Canton}\ \emph {et~al.}(2021)\citenamefont {Dal~Canton}, \citenamefont {Nitz}, \citenamefont {Gadre}, \citenamefont {Cabourn~Davies}, \citenamefont {Villa-Ortega}, \citenamefont {Dent}, \citenamefont {Harry},\ and\ \citenamefont {Xiao}}]{DalCanton2021_PyCBCLive}%
  \BibitemOpen
  \bibfield  {author} {\bibinfo {author} {\bibfnamefont {T.}~\bibnamefont {Dal~Canton}}, \bibinfo {author} {\bibfnamefont {A.~H.}\ \bibnamefont {Nitz}}, \bibinfo {author} {\bibfnamefont {B.}~\bibnamefont {Gadre}}, \bibinfo {author} {\bibfnamefont {G.~S.}\ \bibnamefont {Cabourn~Davies}}, \bibinfo {author} {\bibfnamefont {V.}~\bibnamefont {Villa-Ortega}}, \bibinfo {author} {\bibfnamefont {T.}~\bibnamefont {Dent}}, \bibinfo {author} {\bibfnamefont {I.}~\bibnamefont {Harry}},\ and\ \bibinfo {author} {\bibfnamefont {L.}~\bibnamefont {Xiao}},\ }\bibfield  {title} {\bibinfo {title} {{Real-time search for compact binary mergers in Advanced LIGO and Virgo’s third observing run using PyCBC Live}},\ }\href {https://doi.org/10.3847/1538-4357/ac2f9a} {\bibfield  {journal} {\bibinfo  {journal} {The Astrophysical Journal}\ }\textbf {\bibinfo {volume} {923}},\ \bibinfo {pages} {254} (\bibinfo {year} {2021})},\ \Eprint {https://arxiv.org/abs/2008.07494} {arXiv:2008.07494 [astro-ph.HE]} \BibitemShut {NoStop}%
\bibitem [{\citenamefont {Ajith}\ \emph {et~al.}(2014)\citenamefont {Ajith}, \citenamefont {Fotopoulos}, \citenamefont {Privitera}, \citenamefont {Neunzert},\ and\ \citenamefont {Weinstein}}]{Ajith:2012mn}%
  \BibitemOpen
  \bibfield  {author} {\bibinfo {author} {\bibfnamefont {P.}~\bibnamefont {Ajith}}, \bibinfo {author} {\bibfnamefont {N.}~\bibnamefont {Fotopoulos}}, \bibinfo {author} {\bibfnamefont {S.}~\bibnamefont {Privitera}}, \bibinfo {author} {\bibfnamefont {A.}~\bibnamefont {Neunzert}},\ and\ \bibinfo {author} {\bibfnamefont {A.~J.}\ \bibnamefont {Weinstein}},\ }\bibfield  {title} {\bibinfo {title} {{Effectual template bank for the detection of gravitational waves from inspiralling compact binaries with generic spins}},\ }\href {https://doi.org/10.1103/PhysRevD.89.084041} {\bibfield  {journal} {\bibinfo  {journal} {Phys. Rev. D}\ }\textbf {\bibinfo {volume} {89}},\ \bibinfo {pages} {084041} (\bibinfo {year} {2014})},\ \Eprint {https://arxiv.org/abs/1210.6666} {arXiv:1210.6666 [gr-qc]} \BibitemShut {NoStop}%
\bibitem [{\citenamefont {Roy}\ \emph {et~al.}(2017)\citenamefont {Roy}, \citenamefont {Sengupta},\ and\ \citenamefont {Thakor}}]{Roy:2017qgg}%
  \BibitemOpen
  \bibfield  {author} {\bibinfo {author} {\bibfnamefont {S.}~\bibnamefont {Roy}}, \bibinfo {author} {\bibfnamefont {A.~S.}\ \bibnamefont {Sengupta}},\ and\ \bibinfo {author} {\bibfnamefont {N.}~\bibnamefont {Thakor}},\ }\bibfield  {title} {\bibinfo {title} {{Hybrid geometric-random template-placement algorithm for gravitational wave searches from compact binary coalescences}},\ }\href {https://doi.org/10.1103/PhysRevD.95.104045} {\bibfield  {journal} {\bibinfo  {journal} {Phys. Rev. D}\ }\textbf {\bibinfo {volume} {95}},\ \bibinfo {pages} {104045} (\bibinfo {year} {2017})},\ \Eprint {https://arxiv.org/abs/1702.06771} {arXiv:1702.06771 [gr-qc]} \BibitemShut {NoStop}%
\bibitem [{\citenamefont {Roy}\ \emph {et~al.}(2019)\citenamefont {Roy}, \citenamefont {Sengupta},\ and\ \citenamefont {Ajith}}]{Roy:2017oul}%
  \BibitemOpen
  \bibfield  {author} {\bibinfo {author} {\bibfnamefont {S.}~\bibnamefont {Roy}}, \bibinfo {author} {\bibfnamefont {A.~S.}\ \bibnamefont {Sengupta}},\ and\ \bibinfo {author} {\bibfnamefont {P.}~\bibnamefont {Ajith}},\ }\bibfield  {title} {\bibinfo {title} {{Effectual template banks for upcoming compact binary searches in Advanced-LIGO and Virgo data}},\ }\href {https://doi.org/10.1103/PhysRevD.99.024048} {\bibfield  {journal} {\bibinfo  {journal} {Phys. Rev. D}\ }\textbf {\bibinfo {volume} {99}},\ \bibinfo {pages} {024048} (\bibinfo {year} {2019})},\ \Eprint {https://arxiv.org/abs/1711.08743} {arXiv:1711.08743 [gr-qc]} \BibitemShut {NoStop}%
\bibitem [{\citenamefont {Halton}(1964)}]{halton_sequence}%
  \BibitemOpen
  \bibfield  {author} {\bibinfo {author} {\bibfnamefont {J.~H.}\ \bibnamefont {Halton}},\ }\bibfield  {title} {\bibinfo {title} {{Algorithm 247: Radical-inverse quasi-random point sequence}},\ }\href {https://doi.org/10.1145/355588.365104} {\bibfield  {journal} {\bibinfo  {journal} {Commun. ACM}\ }\textbf {\bibinfo {volume} {7}},\ \bibinfo {pages} {701–702} (\bibinfo {year} {1964})}\BibitemShut {NoStop}%
\bibitem [{\citenamefont {Owen}(2017)}]{owen2017randomized}%
  \BibitemOpen
  \bibfield  {author} {\bibinfo {author} {\bibfnamefont {A.~B.}\ \bibnamefont {Owen}},\ }\href {https://arxiv.org/abs/1706.02808} {\bibinfo {title} {{A randomized Halton algorithm in R}}} (\bibinfo {year} {2017}),\ \Eprint {https://arxiv.org/abs/1706.02808} {arXiv:1706.02808 [stat.CO]} \BibitemShut {NoStop}%
\bibitem [{\citenamefont {Fasshauer}(2007)}]{Meshfree_book}%
  \BibitemOpen
  \bibfield  {author} {\bibinfo {author} {\bibfnamefont {G.~E.}\ \bibnamefont {Fasshauer}},\ }\href {https://doi.org/10.1142/6437} {\emph {\bibinfo {title} {Meshfree Approximation Methods with MATLAB}}},\ \bibinfo {series} {Interdisciplinary Mathematical Sciences}, Vol.~\bibinfo {volume} {6}\ (\bibinfo  {publisher} {World Scientific},\ \bibinfo {address} {Singapore},\ \bibinfo {year} {2007})\BibitemShut {NoStop}%
\bibitem [{\citenamefont {{Flyer}}\ \emph {et~al.}(2016)\citenamefont {{Flyer}}, \citenamefont {{Fornberg}}, \citenamefont {{Bayona}},\ and\ \citenamefont {{Barnett}}}]{2016JCoPh}%
  \BibitemOpen
  \bibfield  {author} {\bibinfo {author} {\bibfnamefont {N.}~\bibnamefont {{Flyer}}}, \bibinfo {author} {\bibfnamefont {B.}~\bibnamefont {{Fornberg}}}, \bibinfo {author} {\bibfnamefont {V.}~\bibnamefont {{Bayona}}},\ and\ \bibinfo {author} {\bibfnamefont {G.~A.}\ \bibnamefont {{Barnett}}},\ }\bibfield  {title} {\bibinfo {title} {{On the role of polynomials in RBF-FD approximations: I. Interpolation and accuracy}},\ }\href {https://doi.org/10.1016/j.jcp.2016.05.026} {\bibfield  {journal} {\bibinfo  {journal} {Journal of Computational Physics}\ }\textbf {\bibinfo {volume} {321}},\ \bibinfo {pages} {21} (\bibinfo {year} {2016})}\BibitemShut {NoStop}%
\bibitem [{\citenamefont {Fornberg}\ \emph {et~al.}(2002)\citenamefont {Fornberg}, \citenamefont {Driscoll}, \citenamefont {Wright},\ and\ \citenamefont {Charles}}]{FORNBERG2002473}%
  \BibitemOpen
  \bibfield  {author} {\bibinfo {author} {\bibfnamefont {B.}~\bibnamefont {Fornberg}}, \bibinfo {author} {\bibfnamefont {T.}~\bibnamefont {Driscoll}}, \bibinfo {author} {\bibfnamefont {G.}~\bibnamefont {Wright}},\ and\ \bibinfo {author} {\bibfnamefont {R.}~\bibnamefont {Charles}},\ }\bibfield  {title} {\bibinfo {title} {Observations on the behavior of radial basis function approximations near boundaries},\ }\href {https://doi.org/https://doi.org/10.1016/S0898-1221(01)00299-1} {\bibfield  {journal} {\bibinfo  {journal} {Computers \& Mathematics with Applications}\ }\textbf {\bibinfo {volume} {43}},\ \bibinfo {pages} {473} (\bibinfo {year} {2002})}\BibitemShut {NoStop}%
\bibitem [{\citenamefont {Sharma}\ \emph {et~al.}(2024)\citenamefont {Sharma}, \citenamefont {Roy},\ and\ \citenamefont {Sengupta}}]{Sharma:2023djw}%
  \BibitemOpen
  \bibfield  {author} {\bibinfo {author} {\bibfnamefont {A.}~\bibnamefont {Sharma}}, \bibinfo {author} {\bibfnamefont {S.}~\bibnamefont {Roy}},\ and\ \bibinfo {author} {\bibfnamefont {A.~S.}\ \bibnamefont {Sengupta}},\ }\bibfield  {title} {\bibinfo {title} {{Template bank to search for exotic gravitational wave signals from astrophysical compact binaries}},\ }\href {https://doi.org/10.1103/PhysRevD.109.124049} {\bibfield  {journal} {\bibinfo  {journal} {Phys. Rev. D}\ }\textbf {\bibinfo {volume} {109}},\ \bibinfo {pages} {124049} (\bibinfo {year} {2024})},\ \Eprint {https://arxiv.org/abs/2311.03274} {arXiv:2311.03274 [gr-qc]} \BibitemShut {NoStop}%
\bibitem [{\citenamefont {Barsotti}\ \emph {et~al.}(2018)\citenamefont {Barsotti}, \citenamefont {Gras}, \citenamefont {Evans},\ and\ \citenamefont {Fritschel}}]{aLIGO_Design}%
  \BibitemOpen
  \bibfield  {author} {\bibinfo {author} {\bibfnamefont {L.}~\bibnamefont {Barsotti}}, \bibinfo {author} {\bibfnamefont {S.}~\bibnamefont {Gras}}, \bibinfo {author} {\bibfnamefont {M.}~\bibnamefont {Evans}},\ and\ \bibinfo {author} {\bibfnamefont {P.}~\bibnamefont {Fritschel}},\ }\href {https://dcc.ligo.org/LIGO-T1800044/public} {\emph {\bibinfo {title} {{The updated Advanced LIGO design curve}}}},\ \bibinfo {type} {Tech. Rep.}\ \bibinfo {number} {LIGO-T1800044-v5}\ (\bibinfo  {institution} {LIGO Scientific Collaboration},\ \bibinfo {year} {2018})\BibitemShut {NoStop}%
\bibitem [{\citenamefont {Abbott}\ \emph {et~al.}(2016{\natexlab{c}})\citenamefont {Abbott} \emph {et~al.}}]{KAGRA:2013rdx}%
  \BibitemOpen
  \bibfield  {author} {\bibinfo {author} {\bibfnamefont {B.~P.}\ \bibnamefont {Abbott}} \emph {et~al.} (\bibinfo {collaboration} {KAGRA Collaboration, The LIGO Scientific Collaboration and Virgo Collaboration}),\ }\bibfield  {title} {\bibinfo {title} {{{Prospects for observing and localizing gravitational-wave transients with Advanced LIGO, Advanced Virgo and KAGRA}}},\ }\href {https://doi.org/10.1007/s41114-020-00026-9} {\bibfield  {journal} {\bibinfo  {journal} {Living Rev. Rel.}\ }\textbf {\bibinfo {volume} {19}},\ \bibinfo {pages} {1} (\bibinfo {year} {2016}{\natexlab{c}})},\ \Eprint {https://arxiv.org/abs/1304.0670} {arXiv:1304.0670 [gr-qc]} \BibitemShut {NoStop}%
\bibitem [{\citenamefont {Biscoveanu}\ \emph {et~al.}(2022)\citenamefont {Biscoveanu}, \citenamefont {Landry},\ and\ \citenamefont {Vitale}}]{Biscoveanu_2022}%
  \BibitemOpen
  \bibfield  {author} {\bibinfo {author} {\bibfnamefont {S.}~\bibnamefont {Biscoveanu}}, \bibinfo {author} {\bibfnamefont {P.}~\bibnamefont {Landry}},\ and\ \bibinfo {author} {\bibfnamefont {S.}~\bibnamefont {Vitale}},\ }\bibfield  {title} {\bibinfo {title} {{Population properties and multimessenger prospects of neutron star–black hole mergers following GWTC-3}},\ }\href {https://doi.org/10.1093/mnras/stac3052} {\bibfield  {journal} {\bibinfo  {journal} {Monthly Notices of the Royal Astronomical Society}\ }\textbf {\bibinfo {volume} {518}},\ \bibinfo {pages} {5298–5312} (\bibinfo {year} {2022})}\BibitemShut {NoStop}%
\bibitem [{\citenamefont {Hines}(2015)}]{RBF_github}%
  \BibitemOpen
  \bibfield  {author} {\bibinfo {author} {\bibfnamefont {T.}~\bibnamefont {Hines}},\ }\href {https://github.com/treverhines/RBF.git} {\bibinfo {title} {Python package containing the tools necessary for radial basis function {(RBF)} applications}} (\bibinfo {year} {2015})\BibitemShut {NoStop}%
\bibitem [{\citenamefont {Speagle}(2020)}]{Speagle:2019ivv}%
  \BibitemOpen
  \bibfield  {author} {\bibinfo {author} {\bibfnamefont {J.~S.}\ \bibnamefont {Speagle}},\ }\bibfield  {title} {\bibinfo {title} {{{dynesty: a dynamic nested sampling package for estimating Bayesian posteriors and evidences}}},\ }\href {https://doi.org/10.1093/mnras/staa278} {\bibfield  {journal} {\bibinfo  {journal} {Mon. Not. Roy. Astron. Soc.}\ }\textbf {\bibinfo {volume} {493}},\ \bibinfo {pages} {3132} (\bibinfo {year} {2020})},\ \Eprint {https://arxiv.org/abs/1904.02180} {arXiv:1904.02180 [astro-ph.IM]} \BibitemShut {NoStop}%
\bibitem [{\citenamefont {Borenstein}\ \emph {et~al.}(2009)\citenamefont {Borenstein}, \citenamefont {Hedges}, \citenamefont {Higgins},\ and\ \citenamefont {Rothstein}}]{borenstein2009introduction}%
  \BibitemOpen
  \bibfield  {author} {\bibinfo {author} {\bibfnamefont {M.}~\bibnamefont {Borenstein}}, \bibinfo {author} {\bibfnamefont {L.~V.}\ \bibnamefont {Hedges}}, \bibinfo {author} {\bibfnamefont {J.~P.}\ \bibnamefont {Higgins}},\ and\ \bibinfo {author} {\bibfnamefont {H.~R.}\ \bibnamefont {Rothstein}},\ }\href {https://doi.org/10.1002/9780470743386} {\emph {\bibinfo {title} {Introduction to Meta-Analysis}}}\ (\bibinfo  {publisher} {John Wiley \& Sons},\ \bibinfo {year} {2009})\BibitemShut {NoStop}%
\bibitem [{\citenamefont {Lin}(1991)}]{lin1991divergence}%
  \BibitemOpen
  \bibfield  {author} {\bibinfo {author} {\bibfnamefont {J.}~\bibnamefont {Lin}},\ }\bibfield  {title} {\bibinfo {title} {{Divergence measures based on the Shannon entropy}},\ }\href {https://doi.org/10.1109/18.61115} {\bibfield  {journal} {\bibinfo  {journal} {IEEE Transactions on Information Theory}\ }\textbf {\bibinfo {volume} {37}},\ \bibinfo {pages} {145} (\bibinfo {year} {1991})}\BibitemShut {NoStop}%
\bibitem [{\citenamefont {Punturo}\ \emph {et~al.}(2010)\citenamefont {Punturo}, \citenamefont {Abernathy}, \citenamefont {Acernese}, \citenamefont {Allen}, \citenamefont {Andersson}, \citenamefont {Arun} \emph {et~al.}}]{Punturo2010_ETScienceReach}%
  \BibitemOpen
  \bibfield  {author} {\bibinfo {author} {\bibfnamefont {M.}~\bibnamefont {Punturo}}, \bibinfo {author} {\bibfnamefont {M.}~\bibnamefont {Abernathy}}, \bibinfo {author} {\bibfnamefont {F.}~\bibnamefont {Acernese}}, \bibinfo {author} {\bibfnamefont {B.}~\bibnamefont {Allen}}, \bibinfo {author} {\bibfnamefont {N.}~\bibnamefont {Andersson}}, \bibinfo {author} {\bibfnamefont {K.}~\bibnamefont {Arun}}, \emph {et~al.},\ }\bibfield  {title} {\bibinfo {title} {The third generation of gravitational wave observatories and their science reach},\ }\href {https://doi.org/10.1088/0264-9381/27/8/084007} {\bibfield  {journal} {\bibinfo  {journal} {Classical and Quantum Gravity}\ }\textbf {\bibinfo {volume} {27}},\ \bibinfo {pages} {084007} (\bibinfo {year} {2010})}\BibitemShut {NoStop}%
\bibitem [{\citenamefont {Hild}\ \emph {et~al.}(2011)\citenamefont {Hild}, \citenamefont {Chelkowski}, \citenamefont {Freise} \emph {et~al.}}]{Hild2011_ETSensitivity}%
  \BibitemOpen
  \bibfield  {author} {\bibinfo {author} {\bibfnamefont {S.}~\bibnamefont {Hild}}, \bibinfo {author} {\bibfnamefont {S.}~\bibnamefont {Chelkowski}}, \bibinfo {author} {\bibfnamefont {A.}~\bibnamefont {Freise}}, \emph {et~al.},\ }\bibfield  {title} {\bibinfo {title} {Sensitivity studies for third‑generation gravitational wave observatories},\ }\href {https://doi.org/10.1088/0264-9381/28/9/094013} {\bibfield  {journal} {\bibinfo  {journal} {Classical and Quantum Gravity}\ }\textbf {\bibinfo {volume} {28}},\ \bibinfo {pages} {094013} (\bibinfo {year} {2011})}\BibitemShut {NoStop}%
\bibitem [{\citenamefont {Maggiore}\ \emph {et~al.}(2020)\citenamefont {Maggiore}, \citenamefont {Van~den Broeck}, \citenamefont {Bartolo}, \citenamefont {Belgacem}, \citenamefont {Bertacca}, \citenamefont {Bizouard} \emph {et~al.}}]{Maggiore2020_ETScienceCase}%
  \BibitemOpen
  \bibfield  {author} {\bibinfo {author} {\bibfnamefont {M.}~\bibnamefont {Maggiore}}, \bibinfo {author} {\bibfnamefont {C.}~\bibnamefont {Van~den Broeck}}, \bibinfo {author} {\bibfnamefont {N.}~\bibnamefont {Bartolo}}, \bibinfo {author} {\bibfnamefont {E.}~\bibnamefont {Belgacem}}, \bibinfo {author} {\bibfnamefont {D.}~\bibnamefont {Bertacca}}, \bibinfo {author} {\bibfnamefont {M.~A.}\ \bibnamefont {Bizouard}}, \emph {et~al.},\ }\bibfield  {title} {\bibinfo {title} {{Science case for the Einstein Telescope}},\ }\href {https://doi.org/10.1088/1475-7516/2020/03/050} {\bibfield  {journal} {\bibinfo  {journal} {Journal of Cosmology and Astroparticle Physics}\ }\textbf {\bibinfo {volume} {2020}}\bibfield  {number} {\bibinfo  {number} { (03)},\ \bibinfo {pages} {050}},\ }\Eprint {https://arxiv.org/abs/1912.02622} {arXiv:1912.02622 [astro-ph.CO]} \BibitemShut {NoStop}%
\bibitem [{\citenamefont {Evans}\ \emph {et~al.}(2016)\citenamefont {Evans}, \citenamefont {Harms},\ and\ \citenamefont {Vitale}}]{ET_psd}%
  \BibitemOpen
  \bibfield  {author} {\bibinfo {author} {\bibfnamefont {M.}~\bibnamefont {Evans}}, \bibinfo {author} {\bibfnamefont {J.}~\bibnamefont {Harms}},\ and\ \bibinfo {author} {\bibfnamefont {S.}~\bibnamefont {Vitale}},\ }\href {https://dcc.ligo.org/LIGO-P1600143/public} {\emph {\bibinfo {title} {{Exploring the Sensitivity of Next Generation Gravitational Wave Detectors}}}},\ \bibinfo {type} {Tech. Rep.}\ \bibinfo {number} {LIGO-P1600143}\ (\bibinfo {year} {2016})\BibitemShut {NoStop}%
\bibitem [{\citenamefont {Abbott}\ \emph {et~al.}(2023{\natexlab{d}})\citenamefont {Abbott} \emph {et~al.}}]{Abbott2023GWOSC}%
  \BibitemOpen
  \bibfield  {author} {\bibinfo {author} {\bibfnamefont {R.}~\bibnamefont {Abbott}} \emph {et~al.} (\bibinfo {collaboration} {The LIGO Scientific Collaboration, Virgo Collaboration and KAGRA Collaboration}),\ }\bibfield  {title} {\bibinfo {title} {{Open Data from the Third Observing Run of LIGO, Virgo, KAGRA, and GEO}},\ }\href {https://doi.org/10.3847/1538-4365/acdc9f} {\bibfield  {journal} {\bibinfo  {journal} {The Astrophysical Journal Supplement Series}\ }\textbf {\bibinfo {volume} {267}},\ \bibinfo {pages} {29} (\bibinfo {year} {2023}{\natexlab{d}})}\BibitemShut {NoStop}%
\bibitem [{\citenamefont {{LIGO Scientific Collaboration and Virgo Collaboration}}(2022)}]{GW190814Zenodo_PE}%
  \BibitemOpen
  \bibfield  {author} {\bibinfo {author} {\bibnamefont {{LIGO Scientific Collaboration and Virgo Collaboration}}},\ }\href@noop {} {\bibinfo {title} {{GWTC-2.1: Deep Extended Catalog of Compact Binary Coalescences Observed by LIGO and Virgo During the First Half of the Third Observing Run} - parameter estimation data release}},\ \bibinfo {howpublished} {\url{https://doi.org/10.5281/zenodo.6513631}} (\bibinfo {year} {2022}),\ \bibinfo {note} {zenodo, version 2}\BibitemShut {NoStop}%
\bibitem [{\citenamefont {London}\ \emph {et~al.}(2018)\citenamefont {London}, \citenamefont {Khan}, \citenamefont {Fauchon-Jones}, \citenamefont {García}, \citenamefont {Hannam}, \citenamefont {Husa}, \citenamefont {Jiménez‑Forteza}, \citenamefont {Kalaghatgi},\ and\ \citenamefont {Ohme}}]{London:2018}%
  \BibitemOpen
  \bibfield  {author} {\bibinfo {author} {\bibfnamefont {L.}~\bibnamefont {London}}, \bibinfo {author} {\bibfnamefont {S.}~\bibnamefont {Khan}}, \bibinfo {author} {\bibfnamefont {E.}~\bibnamefont {Fauchon-Jones}}, \bibinfo {author} {\bibfnamefont {C.}~\bibnamefont {García}}, \bibinfo {author} {\bibfnamefont {M.}~\bibnamefont {Hannam}}, \bibinfo {author} {\bibfnamefont {S.}~\bibnamefont {Husa}}, \bibinfo {author} {\bibfnamefont {X.}~\bibnamefont {Jiménez‑Forteza}}, \bibinfo {author} {\bibfnamefont {C.}~\bibnamefont {Kalaghatgi}},\ and\ \bibinfo {author} {\bibfnamefont {F.}~\bibnamefont {Ohme}},\ }\bibfield  {title} {\bibinfo {title} {{First Higher‑Multipole Model of Gravitational Waves from Spinning and Coalescing Black‑Hole Binaries}},\ }\href {https://doi.org/10.1103/PhysRevLett.120.161102} {\bibfield  {journal} {\bibinfo  {journal} {Physical Review Letters}\ }\textbf {\bibinfo {volume} {120}},\ \bibinfo {pages} {161102} (\bibinfo {year} {2018})}\BibitemShut {NoStop}%
\bibitem [{\citenamefont {Veitch}\ \emph {et~al.}(2015{\natexlab{b}})\citenamefont {Veitch}, \citenamefont {Raymond}, \citenamefont {Farr}, \citenamefont {Farr} \emph {et~al.}}]{Veitch:2014wba}%
  \BibitemOpen
  \bibfield  {author} {\bibinfo {author} {\bibfnamefont {J.}~\bibnamefont {Veitch}}, \bibinfo {author} {\bibfnamefont {V.}~\bibnamefont {Raymond}}, \bibinfo {author} {\bibfnamefont {B.}~\bibnamefont {Farr}}, \bibinfo {author} {\bibfnamefont {W.}~\bibnamefont {Farr}}, \emph {et~al.},\ }\bibfield  {title} {\bibinfo {title} {{Parameter estimation for compact binaries with ground-based gravitational-wave observations using the LALInference software library}},\ }\href {https://doi.org/10.1103/PhysRevD.91.042003} {\bibfield  {journal} {\bibinfo  {journal} {Phys. Rev. D}\ }\textbf {\bibinfo {volume} {91}},\ \bibinfo {pages} {042003} (\bibinfo {year} {2015}{\natexlab{b}})},\ \Eprint {https://arxiv.org/abs/1409.7215} {arXiv:1409.7215 [gr-qc]} \BibitemShut {NoStop}%
\bibitem [{\citenamefont {Lee}\ \emph {et~al.}(2022)\citenamefont {Lee}, \citenamefont {Morisaki},\ and\ \citenamefont {Tagoshi}}]{Lee2022_massspin_reparam}%
  \BibitemOpen
  \bibfield  {author} {\bibinfo {author} {\bibfnamefont {E.}~\bibnamefont {Lee}}, \bibinfo {author} {\bibfnamefont {S.}~\bibnamefont {Morisaki}},\ and\ \bibinfo {author} {\bibfnamefont {H.}~\bibnamefont {Tagoshi}},\ }\bibfield  {title} {\bibinfo {title} {Mass–spin reparameterization for rapid parameter estimation of inspiral gravitational-wave signals},\ }\href {https://doi.org/10.1103/PhysRevD.105.124057} {\bibfield  {journal} {\bibinfo  {journal} {Physical Review D}\ }\textbf {\bibinfo {volume} {105}},\ \bibinfo {pages} {124057} (\bibinfo {year} {2022})},\ \Eprint {https://arxiv.org/abs/2203.05216} {arXiv:2203.05216 [gr-qc]} \BibitemShut {NoStop}%
\end{thebibliography}%

\end{document}